\documentclass[12pt,a4paper]{iopart}
\usepackage{iopams}
\usepackage{arydshl}
\usepackage{verbatim}

\newcommand{\eq}[1]{(\ref{#1})}
\newcommand{\bun}{\hat{\mathbf{b}}}
\newcommand{\eun}{\hat{\mathbf{e}}}

\newcommand{\phiave}{\langle \phi \rangle}
\newcommand{\phiwig}{\widetilde{\phi}}
\newcommand{\Phiwig}{\widetilde{\Phi}}
\newcommand{\boldr}{\mathbf{r}}
\newcommand{\bv}{\mathbf{v}}
\newcommand{\bA}{\mathbf{A}}
\newcommand{\bR}{\mathbf{R}}

\newcommand{\bB}{\mathbf{B}}
\newcommand{\bE}{\mathbf{E}}
\newcommand{\bZ}{\mathbf{Z}}
\newcommand{\bK}{\mathbf{K}}

\newcommand{\bX}{\mathbf{X}}
\newcommand{\bV}{\mathbf{V}}

\newcommand{\matrixtop}[1]{\buildrel\leftrightarrow\over{#1}}
\newcommand{\matI}{\matrixtop{\mathbf{I}}}

\newcommand{\dotcross}{ \raise 0.65ex\hbox{${\scriptstyle {{_{\displaystyle \cdot}}\atop\times}}$} }
\newcommand{\crossdot}{ \raise 0.5ex\hbox{${\scriptstyle {{_\times}\atop{\displaystyle \cdot}}}$} }
\newcommand{\rhobf}{\mbox{\boldmath$\rho$}}
\newcommand{\scrhobf}{\mbox{\scriptsize\boldmath$\rho$}}

\newcommand{\Gammabf}{\mbox{\boldmath$\Gamma$}}
\newcommand{\sumsig}{ \raise -1.3ex\hbox{${{\displaystyle \sum}\atop{\scriptstyle \sigma}}$} }
\newcounter{appnumb}

\begin{document}
\title[Phase-space Lagrangian derivation of electrostatic gyrokinetics]{Phase-space Lagrangian derivation of electrostatic gyrokinetics in general geometry}
\author{Felix I Parra$^{1,2}$}
\eads{\mailto{f.parradiaz1@physics.ox.ac.uk}}
\author{Iv\'an Calvo$^{2,3}$}
\eads{\mailto{ivan.calvo@ciemat.es}}

\vspace{1cm}

\address{$^1$Rudolf Peierls Centre for Theoretical Physics, University of Oxford, Oxford, OX1 3NP, UK}
\address{$^2$Isaac Newton Institute for Mathematical Sciences,
University of Cambridge, Cambridge, CB3 0EH, UK}
\address{$^3$Laboratorio Nacional de Fusi\'on, Asociaci\'on
EURATOM-CIEMAT, 28040 Madrid, Spain}

\begin{abstract}
Gyrokinetic theory is based on an asymptotic expansion in the
small parameter $\epsilon$, defined as the ratio of the gyroradius
and the characteristic length of variation of the magnetic field.
In this article, this ordering is strictly implemented to compute
the electrostatic gyrokinetic phase-space Lagrangian in general
magnetic geometry to order $\epsilon^2$. In particular, a new
expression for the complete second-order gyrokinetic Hamiltonian
is provided, showing that in a rigorous treatment of gyrokinetic
theory magnetic geometry and turbulence cannot be dealt with
independently. The new phase-space gyrokinetic Lagrangian gives a
Vlasov equation accurate to order $\epsilon^2$ and a Poisson
equation accurate to order $\epsilon$. The final expressions are
explicit and can be implemented into any simulation without
further computations.
\end{abstract}

\maketitle

\section{Introduction} \label{sect_intro}

Gyrokinetics \cite{catto78} has proven a very useful tool to study
turbulence in the core of fusion devices, making kinetic
simulations of turbulent fluctuations possible in reasonable
computational times \cite{dimits96, dorland00, dannert05, candy03,
chen03, peeters04}. Its main advantage is averaging over the
gyrofrequency time scale without losing the effect of the finite
size of the gyroradius that is of the order of the typical
wavelength of the turbulence. To perform this average, it is
necessary to assume certain orderings that in the electrostatic
limit can be summarized as
\begin{eqnarray} \label{orderings}
\bB(\boldr) \quad \mathrm{with} \quad \nabla \sim \frac{1}{L}
\nonumber \\ \varphi (\boldr, t)\quad \mathrm{with}\quad
\nabla_\bot \sim \frac{1}{\rho}, \quad \bun \cdot \nabla \sim
\frac{1}{L}, \quad \frac{\partial}{\partial t} \sim \omega
\nonumber \\ \frac{\omega}{\Omega} \sim \frac{\rho}{L} \sim
\frac{Z e \varphi}{Mv_t^2} \sim \epsilon \ll 1,
\end{eqnarray}
where $\varphi (\boldr, t)$ is the electrostatic potential,
$\bB(\boldr)$ is the magnetic field, $\omega$ is the
characteristic frequency of the turbulent fluctuations, $L$ is a
characteristic macroscopic scale, $v_t$, $\rho = v_t/\Omega$ and
$\Omega = ZeB/Mc$ are the thermal speed, the gyroradius and the
gyrofrequency of the species of interest, $Ze$ and $M$ are the
charge and the mass, and $e$ and $c$ are the magnitude of the
electron charge and the speed of light. Since this article is
about electrostatic gyrokinetics, we have assumed that the
magnetic field is stationary and its characteristic length of
variation is of the order of the macroscopic length $L$. The
ordering in \eq{orderings} implies that the electrostatic
potential fluctuates with some characteristic frequency $\omega$
and has a strong gradient perpendicular to the magnetic field, on
the order of the inverse of the gyroradius, whereas its gradient
parallel to the direction of the magnetic field, $\bun = \bB/B$,
is on the order of the inverse of the larger scale $L$. The
frequency $\omega$ of the turbulence is usually much smaller than
the gyrofrequency, making the gyrokinetic average over the
gyromotion valid. We have employed the small parameter $\epsilon
\sim \omega/\Omega \ll 1$ to make this explicit. In most fusion
experiments, the ratio of the gyroradius and the macroscopic
length is another small parameter that we also order as
$\epsilon$. It is easy to see that in drift wave turbulence, for
which the characteristic frequency is $\omega \sim v_t/L$, the
quantities $\omega/\Omega$ and $\rho/L$ are indeed of the same
order. More importantly, to obtain the typical gyrokinetic
formalism, it is necessary to order the electrostatic potential as
small compared with the characteristic energy of the particles.
This assumption is necessary to prove that the gyromotion of the
particles is circular to lowest order. The most common gyrokinetic
ordering assumes that the parameter $Ze \varphi/Mv_t^2$ is
comparable to $\epsilon$, as is done in \eq{orderings}. In this
way, magnetic geometry effects such as the $\nabla B$ and
curvature drifts, of order $\rho/L$, are allowed to be comparable
to the turbulent $\bE \times \bB$ drift, of order $Ze
\varphi/Mv_t^2$. Ordering these effects so that they are
comparable is very important in, for example, the core of
tokamaks, where the curvature of the magnetic field lines is
believed to be the most important drive for the turbulence
\cite{romanelli89}. The ordering in \eq{orderings} contains the
simplest assumptions that are still interesting, but it can be
extended to include components of the potential that have
perpendicular gradients of the order of the inverse of the
macroscopic length $L$ \cite{dimits92, dimits10}. The results that
we present in this article can be easily extended to some of these
more general orderings, but we leave this for future work.

There are different techniques to obtain gyrokinetics (and for
that matter, drift kinetics \cite{hazeltine73, hinton76,
helander02bk}, of which gyrokinetics is a natural extension). On
the one hand, it is possible to obtain the gyrokinetic equation by
working iteratively on the Vlasov equation \cite{lee83, wwlee83,
bernstein85, parra08}.  We will call these {\it iterative
methods}.  On the other hand, it is possible to use phase-space
Lagrangian/Hamiltonian methods that solve order by order for the
motion of the particle in a given electromagnetic field,
uncoupling the gyromotion from the slower time scales
\cite{littlejohn81, cary81, cary83, littlejohn83, littlejohn85,
dubin83, hahm88, brizard07}. Once the motion of the particle is
known, the Vlasov equation is simply obtained by its
characteristics. We will call these {\it Lagrangian methods}. Both
procedures are asymptotic expansions in the parameter $\epsilon$,
and give equivalent equations order by order, but the Lagrangian
methods have the advantage of giving the equations in a form that
exactly conserves some energy-like quantities. This property may
be very important for the global, full $f$ simulations that are
being developed \cite{heikkinen08, grandgirard06, xu07, chang08}.
To have an energy-like invariant and at the same time obtain
equations of motion and equations for the electromagnetic fields
that are the same to first order in $\epsilon$ as those obtained
with the iterative procedure, it is necessary to carry the
expansion in $\epsilon$ to higher order. For example, in a
slab~\cite{dubin83}, it is necessary to obtain the Hamiltonian to
second order in $\epsilon$. The second order piece of the
Hamiltonian, quadratic in the electrostatic potential $\varphi$,
gives second-order corrections to the equations of motion and
hence it is in principle negligible to first order. However, if
the lowest-order quasineutrality equation that contains a linear
term in $\varphi$ is employed, the second-order correction to the
Hamiltonian must be kept to obtain an energy-like invariant.

The complete calculation to order $\epsilon^2$ has not been done
for a general static magnetic field in either formalism so
far\footnote{In the particular case of a constant magnetic field
the calculation to order $\epsilon^2$ was given in
\cite{dubin83}.}. In the most common Lagrangian
formulation~\cite{brizard07}, the calculation is done in two
steps: first, the turbulent electromagnetic fields are ignored and
only the background magnetic field is considered, giving the drift
kinetic equation; in the second step, the turbulent
electromagnetic fields are added and the corresponding corrections
are calculated. Consider the case in which the magnetic field does
not vary in time, i.e. electrostatic gyrokinetics. In the first
step, the equations are expanded in the small parameter $\epsilon
\sim \rho/L$, whereas in the second step, they are expanded in
$\epsilon_\varphi \sim Ze\varphi/M v_t^2$. The expansion in
$\epsilon$ is only performed to first order because the next order
results are very tedious to calculate. The expansion in
$\epsilon_\varphi$ is continued to second order because the pieces
quadratic in $\varphi$ are needed to have an energy-like
invariant. In the expansion in $\epsilon_\varphi$, the fact that
there has been a previous expansion in $\epsilon$ is ignored, and
as a result the terms of order $\epsilon\epsilon_\varphi$ are
never calculated. The missing terms of order $\epsilon^2$ and
$\epsilon\epsilon_\varphi$ are comparable to the terms of order
$\epsilon_\varphi^2$ according to the gyrokinetic ordering in (1),
making this expansion consistent only when $\epsilon_\varphi\gg
\epsilon$. In addition, since the cross-terms that contain both
the background magnetic field and the turbulent electrostatic
potential, of order $\epsilon\epsilon_\varphi$, are always
neglected when the two-step method is presented, it is not obvious
how to calculate them following that procedure. In this article,
we present the complete phase-space Lagrangian calculation with
the standard gyrokinetic ordering (1), emphasizing the
self-consistent calculation of the terms of order $\epsilon^2$ and
$\epsilon\epsilon_\varphi$. In the gyrokinetic equations that
result from the new Lagrangian, the magnetic geometry effects and
the fluctuating potential appear together in the second-order
terms, showing that geometry and turbulence cannot be separated
and dealt with independently. Our main result is the explicit
expression for the second-order gyrokinetic Hamiltonian given in
equations~\eq{hamiltonian_o2_2}, \eq{Psi2_phi}, \eq{Psi2_phiB} and
\eq{Psi2_B}. It clearly exhibits the interplay between geometry
and turbulence inherent to gyrokinetic theory, possessing terms of
three types: terms quadratic in the electrostatic potential, terms
that include both the electrostatic potential and the magnetic
geometry, and terms that are purely geometrical.

At this point, it is fair to wonder about the motivations beyond
formal coherence to carry out the expansion consistently to second
order in $\epsilon$. Keeping the second order piece of the
Hamiltonian that is quadratic in the electrostatic potential is
necessary for the conservation of an energy-like invariant, as
already noted above. When the other second-order terms computed in
this article are included, they have two effects: (i) the
gyrokinetic Poisson's equation is modified by the effect of the
non-uniform magnetic field on the gyro-orbits, and (ii) the
equations of motion are modified to second order. Both of these
effects are not conventionally kept in gyrokinetic formulations,
but they may be crucial for conservation of momentum. Conservation
of momentum in full $f$ gyrokinetic formulations has been the
center of a recent controversy \cite{parra08, parra09b, lee09,
parra09c, parra10b, parra10c}. By assuming a gyroBohm level of
turbulent transport of momentum at long wavelengths, Catto and one
of us, FIP, have argued that to recover with a full $f$ model the
correct transport of toroidal angular momentum in a tokamak, it is
necessary to have gyrokinetic Fokker-Planck and Poisson's
equations correct to third order in $\epsilon$ in the high flow
ordering, for which the average velocity of the ions $V_i$ is of
the order of the ion thermal speed $v_{ti}$, and correct to fourth
order in the low flow ordering, for which $V_i \sim \epsilon
v_{ti} \ll v_{ti}$. In the case of slab gyrokinetics, a consistent
calculation of the transport of momentum in the low flow ordering
requires the third order Hamiltonian \cite{parra10b, parra10c}.
The requirements for a system with general geometry are still to
be sorted out, and to do so it is necessary to study the new terms
presented here and terms of even higher order.

In addition to the issues raised for full $f$ simulations, the
formulation presented here will be very useful for $\delta f$
approaches to momentum transport in tokamaks in the low flow
ordering. Reference \cite{parra10a} presents a formulation of this
problem in the electrostatic limit that requires the minimum
number of modifications to existing $\delta f$ simulations. The
most important conclusion in \cite{parra10a} is that the turbulent
pieces of the distribution function and the electrostatic
potential have to be calculated to an order higher in $\epsilon$
than usual because the contribution to momentum transport from the
lowest order pieces vanishes due to symmetry arguments
\cite{peeters05, parra10f}. These symmetry arguments do not hold
if the higher order terms of the gyrokinetic equation are
considered. To avoid calculating most of the next order
corrections to the gyrokinetic Vlasov and Poisson's equations,
reference \cite{parra10a} has to resort to a subsidiary expansion
based on the fact that in many tokamaks the poloidal component of
the magnetic field is much smaller than the toroidal component.
The new contribution to the Hamiltonian that we calculate here
gives the self-consistent higher order contributions to the
gyrokinetic Vlasov equation for the first time. Only a higher
order gyrokinetic Poisson's equation is then lacking to obtain a
complete $\delta f$ formulation in the low flow ordering that does
not require a small poloidal magnetic field; this will be the
subject of a future publication.

The rest of this article is organized as follows. In
Section~\ref{sect_order} we write the non-dimensional phase-space
Lagrangian of a particle in an electromagnetic field. The
normalization shows explicitly the standard gyrokinetic ordering
\eq{orderings}. In the first part of Section~\ref{sect_theory} we
review the phase-space Lagrangian approach to gyrokinetics to help
the understanding of the calculation that follows. In the second
half of this section we proceed to obtain the gyrokinetic
Lagrangian to second order in our expansion parameter $\epsilon$.
As mentioned above, this is our main result. In
Section~\ref{sec:GyroEqsMotion} we obtain the Vlasov equation from
this Lagrangian, and in Section~\ref{sec:GyroPoissonEq} we discuss
the consequences of this formulation for Poisson's equation. The
new Vlasov and Poisson's equations presented here are correct to
second and first order in $\epsilon$, respectively. We should
remark that in the limit where the electrostatic potential has a
scale of variation much larger than the gyroradius of the species
of interest, our gyrokinetic equations provide the highest order
guiding-center equations that we are aware of. In
Section~\ref{sec:fieldtheory} we borrow tools from classical field
theory to obtain Poisson's equation in a different way. We prove
that there is an energy-like invariant and we discuss the
stringent conditions on the equations to actually conserve it in a
simulation. We finish with a discussion of our results and the
future lines of research in Section~\ref{sec:conclusions}. The
Appendices contain the most cumbersome parts of the calculation as
well as some material included for completeness. Finally, we would
like to stress that in this article we have given all our results
in an explicit form that can be directly implemented in a computer
code.

\section{Normalized Lagrangian} \label{sect_order}

The phase-space Lagrangian for the motion of a particle of mass
$M$ and charge $Ze$ in an electromagnetic field is given by
\begin{equation}
\mathcal{L}^\bX (\boldr, \bv, \dot{\boldr}, \dot{\bv}, t) = \left
[ \frac{Ze}{c} \bA (\boldr) + M \bv \right ] \cdot \frac{d
\boldr}{dt} - H^\bX(\boldr, \bv, t), \label{lagr_orig}
\end{equation}
with the Hamiltonian
\begin{equation}
H^\bX(\boldr, \bv, t) = \frac{1}{2} M v^2 + Ze \varphi (\boldr,
t).
\end{equation}
Here $\bA$ is the vector potential that is defined such that $\bB
= \nabla \times \bA$. Notice that the phase-space Lagrangian
depends on the position of the particle $\boldr$, its velocity
$\bv$, the time derivatives of both the position and the velocity,
$\dot{\boldr} = d\boldr/dt$ and $\dot{\bv} = d\bv/dt$, and the
time $t$. For convenience, we will sometimes denote the
phase-space coordinates $\{ \boldr, \bv \}$ as $\{ X^\alpha
\}_{\alpha = 1}^6 \equiv \bX = \{\boldr, \bv\}$. We use the
superscript $^\bX$ in the Lagrangian \eq{lagr_orig} because it is
a function of the phase-space coordinates $\bX$.

The equations of motion are obtained by finding the stationary
points of the action $\sigma^\bX [\boldr(t), \bv(t)] =
\int_{t_0}^{t_1} dt\, \mathcal{L}^\bX (\boldr(t), \bv(t),
\dot{\boldr}(t), \dot{\bv}(t), t )$ with respect to variations of
the functions $\boldr(t)$ and $\bv(t)$ subject to the constraints
$\boldr(t = t_0) = \boldr_0$, $\bv(t = t_0) = \bv_0$, $\boldr(t =
t_1) = \boldr_1$ and $\bv(t = t_1) = \bv_1$. From this procedure
we obtain six equations of motion, namely
\begin{equation}
\frac{d}{dt} ( \nabla_{\dot{\boldr}} \mathcal{L}^\bX ) =
\nabla_\boldr \mathcal{L}^\bX
\end{equation}
and
\begin{equation}
\frac{d}{dt} ( \nabla_{\dot{\bv}} \mathcal{L}^\bX ) = \nabla_\bv
\mathcal{L}^\bX.
\end{equation}
This differs from the standard Lagrangian formalism where the
Lagrangian function depends only on $\boldr$, $\dot\boldr$ and
$t$. Actually, the phase-space Lagrangian formalism can be viewed
as a variational formulation of Hamilton equations (see, for
example, \cite{Goldstein}). In plasma physics, it was first
applied by Littlejohn to guiding-center dynamics in
\cite{littlejohn83}.

The Lagrangian \eq{lagr_orig} is non-dimensionalized using the
characteristic thermal velocity of the species of interest $v_t$,
the characteristic length $L^{-1} \sim |\nabla (\ln |\bA|)|$ and
the characteristic time $L/v_t$. We assume that $\epsilon =
\rho/L$ is a small parameter, with $\rho = v_t/\Omega$ and $\Omega
= ZeB_0/Mc$ the characteristic gyroradius and the characteristic
gyrofrequency of the species of interest, and $B_0 \sim |\nabla
\times \bA|$ the characteristic magnitude of the magnetic field.
We assume that the characteristic time and length scales in the
electrostatic potential are the sound gyroradius $\rho_s =
c_s/\Omega_i$ and the sound characteristic time $L/c_s$, i.e.,
$\varphi ( \boldr/\rho_s, c_s t/L)$, where $\varphi$ has
derivatives with respect to its arguments of order unity. Here
$c_s = \sqrt{T_{e0}/m_i}$ is the sound speed, $m_i$ and $\Omega_i
= eB_0/m_i c$ are the mass and the gyrofrequency of the dominant
ion species, usually singly charged, and $T_{e0}$ is the
characteristic electron temperature. The assumption on the scales
of the electrostatic potential can be easily relaxed to account
for other time and spatial scales. Since the electrostatic
potential $\varphi$ is a quantity that enters the equations of the
different species, it is normalized using parameters that do not
depend on the species, in particular the characteristic electron
temperature $T_{e0}$, the magnitude of the electron charge $e$ and
the mass of the dominant ion species $m_i$. This normalization
will be useful in Poisson's equation, where several species
appear. The new, non-dimensionalized variables are
\begin{equation}
\check{t} = \frac{v_t t}{L}, \check{\boldr} = \frac{\boldr}{L},
\check{\bv} = \frac{\bv}{v_t}, \check{\bA} = \frac{\bA}{B_0 L},
\check{\varphi} = \frac{e\varphi}{\epsilon_s T_{e0}},
\check{H}^{\check{\bX}} = \frac{H^{\check{\bX}}}{Mv_t^2},
\end{equation}
giving
\begin{equation}
\check{\mathcal{L}}^{\check{\bX}} (\check{\boldr}, \check{\bv},
\dot{\check{\boldr}}, \dot{\check{\bv}}, \check{t} ) = \left [
\frac{1}{\epsilon}\check{\bA} (\check{\boldr}) + \check{\bv}
\right ] \cdot \frac{d \check{\boldr}}{d\check{t}} -
\check{H}^{\check{\bX}} (\check{\boldr}, \check{\bv}, \check{t}),
\label{lagrange_non}
\end{equation}
with
\begin{equation}
\check{H}^{\check{\bX}} (\check{\boldr}, \check{\bv}, \check{t}) =
\frac{1}{2} \check{v}^2 + \Lambda \epsilon \check{\varphi}
(\check{\boldr}/\lambda \epsilon, \check{t}/\tau).
\label{hamilton_non}
\end{equation}
Here, $T_0 = Mv_t^2$ is the characteristic temperature of the
species of interest, $\epsilon_s = \rho_s/L$ is the ratio between
the sound gyroradius and the characteristic scale length,
\begin{equation}
\lambda = \frac{\rho_s}{\rho} = Z \sqrt{\frac{T_{e0} m_i}{T_0 M}}
\end{equation}
is the ratio between the sound gyroradius and the gyroradius of
the species of interest,
\begin{equation}
\tau = \frac{v_t}{c_s} = \sqrt{\frac{T_0 m_i}{T_{e0} M}}
\end{equation}
is the ratio between the thermal speed of the species of interest
and the sound speed, and
\begin{equation}
\Lambda = \frac{Z T_{e0}}{T_0} \lambda = Z^2 \left (
\frac{T_{e0}}{T_0} \right )^{3/2} \sqrt{\frac{m_i}{M}}.
\end{equation}

Even though the electrostatic potential is small, its
perpendicular gradient is not. This assumption has been formally
implemented by writing $\Lambda \epsilon \check{\varphi}
(\check{\boldr}/\lambda \epsilon, \check{t}/\tau)$. Here and in
what follows we assume $\Lambda \sim \lambda \sim \tau \sim 1$.
This is the maximal ordering that contains in it several
interesting regimes as subsidiary expansions in $\Lambda$,
$\lambda$ and $\tau$. The form $\Lambda \epsilon \check{\varphi}
(\check{\boldr}/\lambda \epsilon, \check{t}/\tau)$ is, however,
somewhat deceiving because the gradients along the magnetic field
lines must be small, that is, $\bun \cdot \nabla \check{\varphi}
\sim 1 \ll 1/\lambda \epsilon$, with $\bun (\check{\boldr}) : =
\check{\bB}/\check{B}$ the unit vector parallel to the magnetic
field. It is possible to formalize this condition by writing the
functions in flux coordinates $s(\check{\boldr})$,
$\psi(\check{\boldr})$ and $\alpha(\check{\boldr})$ such that
$\bun =
\partial \check{\boldr}/\partial s$ and $\check{\bB} = \nabla \alpha
\times \nabla \psi$. In these variables, the potential is given by
\begin{equation}
\check{\varphi} \equiv \check{\varphi}  ( s(\check{\boldr}),
\psi(\check{\boldr})/\lambda \epsilon,
\alpha(\check{\boldr})/\lambda \epsilon, \check{t}/\tau).
\label{phi_formal}
\end{equation}
To simplify the notation, we will often use $\check{\varphi}
(\check{\boldr}_\bot/\lambda \epsilon, \check{r}_{||},
\check{t}/\tau)$ instead of the most complete expression in
\eq{phi_formal}. Where no confusion is possible, we will write
$\check{\varphi} (\check{\boldr}, \check{t})$. In any case, we
always assume
\begin{equation}\label{eq:orderingvarphi_par}
\bun(\check{\boldr}) \cdot \nabla_{\check{\boldr}} \check{\varphi}
(\check{\boldr}, \check{t}) \sim 1
\end{equation}
and
\begin{equation}\label{eq:orderingvarphi_perp}
\nabla_{\check{\boldr}_{\perp}} \check{\varphi} (\check{\boldr},
\check{t}) := \bun(\check{\boldr}) \times (\nabla_{\check{\boldr}}
\check{\varphi} (\check{\boldr}, \check{t}) \times
\bun(\check{\boldr})) \sim \frac{1}{\lambda \epsilon}.
\end{equation}

Note that $\epsilon$ is species-dependent whereas $\epsilon_s =
\lambda \epsilon$ is not. In Sections \ref{sect_theory} and
\ref{sec:GyroEqsMotion} where we compute the gyrokinetic
phase-space Lagrangian and the equations of motion of a single
species, $\epsilon$ is the natural expansion parameter. However,
in Sections \ref{sec:GyroPoissonEq} and \ref{sec:fieldtheory},
devoted to the gyrokinetic Poisson's equation, we need to consider
several different species and $\epsilon_s$ is the appropriate,
species-independent small parameter.

Finally, a notational remark is in order. In Sections
\ref{sect_theory} and \ref{sec:GyroEqsMotion} we will be very
careful to exhibit the dependence of our results on the mass,
charge and temperature of the species through the parameters
$\Lambda$, $\lambda$ and $\tau$. Although at some places this may
seem unnecessary and awkward (and it would be if our objective
were to treat always a single species), it is very convenient to
write Poisson's equation and the gyrokinetic phase-space
Lagrangian for a mixture of species in Sections
\ref{sec:GyroPoissonEq} and \ref{sec:fieldtheory}.

From now on we will drop hats $\check{ }$ in the normalized
expressions.

\section{Phase-space Lagrangian perturbation theory} \label{sect_theory}

In this section we follow the general strategy of the applications
of Hamiltonian and phase-space Lagrangian techniques to magnetized
plasmas~\cite{littlejohn81, cary81, cary83, littlejohn83,
littlejohn85, dubin83, hahm88, brizard07}. We search order by
order in the small parameter $\epsilon$ for a change of
phase-space variables such that only one of the variables has fast
time dependence. The gyrophase $\theta$ is the fast variable that
evolves in the gyrofrequency time scale. The rest of the
phase-space variables (gyrocenter position $\bR$, parallel
velocity $u$ and magnetic moment $\mu$) evolve with the much
slower characteristic time scale $L/v_t$. To achieve this, their
time derivatives $d\bR/dt$, $du/dt$ and $d\mu/dt$ will be made
independent of the gyrophase to the order of interest. Had they
depended on $\theta$, they would necessarily show rapid time
fluctuations on top of the more physically interesting slow time
evolution. To make the time derivatives $d\bR/dt$, $du/dt$ and
$d\mu/dt$ independent of the gyrophase $\theta$, we search for a
Lagrangian that does not depend on $\theta$ (it will still depend
on its time derivative $d\theta/dt$). We first review briefly how
to perform a change of variables in a phase-space Lagrangian in
subsection~\ref{sub_change}. As we have already announced, the
objective is the gyrokinetic Lagrangian to order $\epsilon^2$ in
general magnetic geometry. Since the calculation is quite long and
complicated, we have sketched the derivation in subsection
~\ref{sub_procedure} to offer the reader a global perspective of
the formalism. This subsection also contains our own proof that
the algorithm to find the gyrokinetic change of variables can be
carried out to any order and that there exists an adiabatic
invariant $\mu$ to arbitrary order. Finally, in subsections
\ref{sub_drift} and \ref{sub_gyro} we address the calculation of
the gyrokinetic Lagrangian to second order in detail. Some of the
algebra is relegated to \ref{app_DK} and \ref{app_gyro_2}. The
results to first order are compared with the iterative method in
\cite{parra08} in \ref{app_comp}.

Before proceeding, we must mention that the phase-space Lagrangian
(or Hamiltonian) approach to gyrokinetic theory has been
geometrized \cite{littlejohn85, brizard07}. We have chosen not to
use the language of differential geometry to make the paper
accessible to a broader audience, without losing mathematical
rigor. The reader familiar with the geometrical tools will realize
that every step of our presentation can be translated into that
language in an obvious way.

\subsection{Transforming to new phase-space variables}
\label{sub_change}
Consider a transformation $T$ that can be time dependent to a new
set of gyrokinetic phase-space coordinates $\{ Z^\alpha
\}_{\alpha=1}^6\equiv\bZ$. We write\footnote{Note that in part of
the literature~\cite{brizard07} $T$ stands for the inverse of the
transformation that we call $T$.} $\bX(\bZ, t) = (\boldr(\bZ,t),
\bv(\bZ,t)) = T( \bZ,t)$. The phase-space Lagrangian
\eq{lagrange_non} can be easily written in the new set of
variables by using the chain rule, giving
\begin{equation}
\mathcal{L}^\bZ ( \bZ, \dot\bZ , t ) = \sum_{\alpha = 1}^6
\Gamma_\alpha (\bZ, t) \frac{d Z^\alpha}{d t} - H^\bZ ( \bZ, t),
\label{lagr_change}
\end{equation}
where
\begin{equation}
\fl\Gamma_\alpha(\bZ, t) = \left [ \frac{1}{\epsilon} \bA
(\boldr(\bZ, t)) + \bv(\bZ, t) \right ] \cdot \frac{\partial
\boldr(\bZ,t)}{\partial Z^\alpha}
\end{equation}
and
\begin{equation}
\fl H^\bZ ( \bZ, t) = H^\bX (\boldr(\bZ,t),\bv(\bZ,t), t)- \left [
\frac{1}{\epsilon} \bA (\boldr(\bZ, t)) + \bv(\bZ, t) \right ]
\cdot \frac{\partial \boldr(\bZ,t)}{\partial t}.
\end{equation}
By finding the stationary points of the action $\sigma^\bZ[\bZ(t)]
= \int_{t_0}^{t_1} \mathcal{L}^\bZ ( \bZ(t), \dot\bZ(t), t )d t$
with respect to variations of $\bZ (t)$ subject to the conditions
$\bZ (t = t_0) = \bZ_{0}$ and $\bZ (t = t_1) = \bZ_{1}$, we obtain
the new equations of motion
\begin{equation} \label{eqmotionZ}
\frac{d}{dt} \left ( \frac{\partial \mathcal{L}^\bZ}{\partial
\dot{Z}^\alpha} \right ) = \frac{\partial
\mathcal{L}^\bZ}{\partial Z^\alpha} , \quad \alpha=1,2,\dots,6.
\end{equation}
Note that the specific form in \eq{lagr_change} implies that the
equations of motion can be written as
\begin{equation}
\sum_{\beta = 1}^6 L_{\alpha\beta} \frac{dZ^\beta}{dt} =
\frac{\partial H^\bZ}{\partial Z^\alpha} +
\frac{\partial\Gamma_\alpha}{\partial t}, \quad
\alpha=1,2,\dots,6, \label{eqmotionwithuglyterm}
\end{equation}
with $L_{\alpha\beta}$ the $6\times 6$ antisymmetric matrix
\begin{equation}
L_{\alpha\beta} = \frac{\partial \Gamma_\beta}{\partial Z^\alpha}
- \frac{\partial \Gamma_\alpha}{\partial Z^\beta}.
\label{lagr_bra}
\end{equation}
Although our gyrokinetic change of variables has an explicit time
dependence due to the contribution of the electrostatic potential,
we will show that it is possible to choose the functions
$\Gamma_\alpha$ such that $\partial\Gamma_\alpha/\partial t\equiv
0$. In our derivation we impose then that
$\partial\Gamma_\alpha/\partial t\equiv 0$. Consequently, we drop
the last term in \eq{eqmotionwithuglyterm} and write the equations
of motion as
\begin{equation}
\sum_{\beta = 1}^6 L_{\alpha\beta} \frac{dZ^\beta}{dt} =
\frac{\partial H^\bZ}{\partial Z^\alpha}, \quad
\alpha=1,2,\dots,6.\label{eqmotion}
\end{equation}
From expression \eq{eqmotion} we define the Poisson bracket
\begin{equation}
\{ F, G \} = \sum_{\alpha,\beta = 1}^6 P^{\alpha\beta}
\frac{\partial F}{\partial Z^\alpha} \frac{\partial G}{\partial
Z^\beta}, \label{poi_bra}
\end{equation}
with $P^{\alpha\beta} = (L^{-1})^{\alpha\beta}$ the
inverse\footnote{In Section~\ref{sec:GyroEqsMotion} and
\ref{app:poissonbrackets} we show that the matrix
$L_{\alpha\beta}$ of our particular problem is indeed invertible.}
of the antisymmetric matrix defined in \eq{lagr_bra}. Then
\begin{equation} \label{ddt_poibra}
\frac{dZ^\alpha}{dt} = \{ Z^\alpha, H^\bZ \}  , \quad
\alpha=1,2,\dots,6.
\end{equation}
Noting that $L_{\alpha\beta}$ satisfies \eq{eq:ClosedForm} with
$n=3$, the proof in \ref{app_PoissonBracket} guarantees that
\eq{poi_bra} actually defines a Poisson bracket, i.e., that for
any three functions $F_1$, $F_2$ and $F_3$, the bracket satisfies
skew-symmetry
\begin{equation}\label{eq:PoissonProperty1}
\{ F_1, F_2 \} = - \{ F_2, F_1 \},
\end{equation}
the Leibniz rule
\begin{equation}\label{eq:PoissonProperty2}
\{ F_1, F_2 F_3 \} = \{ F_1, F_2 \} F_3 + \{ F_1, F_3 \} F_2,
\end{equation}
and the Jacobi identity
\begin{equation}\label{eq:PoissonProperty3}
\{ F_1, \{ F_2, F_3 \} \} + \{ F_3, \{ F_1, F_2 \} \} + \{ F_2, \{
F_3, F_1 \} \} = 0.
\end{equation}

In general, it is impossible to find a change of phase-space
variables that makes the Lagrangian as written in \eq{lagr_change}
independent of gyrophase. However, the time derivatives
$dZ^\alpha/dt$ may be gyrophase independent even if the
phase-space Lagrangian is not. This apparent discrepancy is easily
solved considering that the equations of motion remain the same if
instead of the Lagrangian \eq{lagr_change} we employ
\begin{equation}
\overline{\mathcal{L}} ( \bZ, \dot\bZ , t ) = \mathcal{L}^\bZ (
\bZ, \dot\bZ , t ) + \frac{d S}{d t}, \label{lagr_prime}
\end{equation}
where the function $S( \bZ, t )$ depends on the phase-space
variables $\bZ$ and $t$. Indeed, finding the stationary points of
the action
\begin{eqnarray}
\fl \overline{\sigma} [\bZ(t)] = \int_{t_0}^{t_1}
\overline{\mathcal{L}} ( \bZ, \dot\bZ , t )d t = \int_{t_0}^{t_1}
\mathcal{L}^\bZ ( \bZ, \dot\bZ , t ) d t+ S(\bZ_{1}, t_1) -
S(\bZ_{0}, t_0)
\end{eqnarray}
with respect to variations of $\bZ (t)$ subject to the conditions
$\bZ (t = t_0) = \bZ_{0}$ and $\bZ (t = t_1) = \bZ_{1}$ gives the
same equations of motion as finding the stationary points of the
action $\sigma^\bZ [\bZ(t)]$ because both actions differ only by
terms that are held constant. Since the Lagrangian is not unique,
we are not going to search for new phase-space variables $\bZ$
such that the phase-space Lagrangian $\mathcal{L}^\bZ$ in
\eq{lagr_change} is gyrophase independent, but such that there
exists a function $S$ for which the phase-space Lagrangian
$\overline{\mathcal{L}}$ in \eq{lagr_prime} is gyrophase
independent. This is equivalent to requiring that the time
derivatives $dZ^\alpha/dt$ be gyrophase independent. Thus, we are
searching for both the change of variables $\bZ$ and the function
$S$ such that the Lagrangian $\overline{\mathcal{L}}$ in
\eq{lagr_prime} is gyrophase independent. Explicitly,
\begin{equation} \label{gyroL_s}
\overline{\mathcal{L}} ( \bZ,\dot\bZ, t) = \sum_{\alpha = 1}^6
\overline{\Gamma}_\alpha \frac{d Z^\alpha}{d t} - \overline{H} (
\bZ, t),
\end{equation}
where
\begin{equation}
\overline{\Gamma}_\alpha (\bZ, t) = \left [ \frac{1}{\epsilon} \bA
(\boldr(\bZ, t)) + \bv(\bZ, t) \right ] \cdot \frac{\partial
\boldr(\bZ,t)}{\partial Z^\alpha} + \frac{\partial
S(\bZ,t)}{\partial Z^\alpha}
\end{equation}
and
\begin{equation}
\fl \overline{H} ( \bZ, t) = H^\bX (\boldr(\bZ,t),\bv(\bZ,t), t)-
\left [ \frac{1}{\epsilon} \bA (\boldr(\bZ, t)) + \bv(\bZ, t)
\right ] \cdot \frac{\partial \boldr(\bZ,t)}{\partial t} -
\frac{\partial S(\bZ,t)}{\partial t}
\end{equation}
are gyrophase independent. In what follows, $\bZ$ only refers to
the gyrokinetic phase-space coordinates. Notice that equations
\eq{eqmotionwithuglyterm}, \eq{lagr_bra}, \eq{eqmotion},
\eq{poi_bra} and \eq{ddt_poibra} are valid for the new Lagrangian
$\overline{\mathcal{L}}$. Simply replace $\Gamma_\alpha$ by
$\overline{\Gamma}_\alpha$ and $H^\bZ$ by $\overline{H}$.

\subsection{Obtaining the new gyrokinetic variables}
\label{sub_procedure}
As advanced in the Introduction, the detailed computation to
obtain the final form of the gyrokinetic Lagrangian to order
$\epsilon^2$ is rather involved. This is why we devote this
subsection to schematically show the steps leading to the
determination of the change of variables and the function $S$
order by order. We also give an easy proof that the algorithm can
be carried out up to arbitrary order (although in practice the
computations would become prohibitively difficult). The detailed
calculation to second order is done in subsections~\ref{sub_drift}
and \ref{sub_gyro}.

Our transformation to new phase-space coordinates will be denoted
by $T_\epsilon$,\footnote{We write $t$ explicitly in $(\boldr,
\bv) = T_\epsilon (\bR, u, \mu, \theta, t)$ because the
transformation is in general time dependent.}
\begin{equation}
(\boldr, \bv) = T_\epsilon (\bR, u, \mu, \theta, t) = T_{NP,
\epsilon} T_{P, \epsilon} (\bR, u, \mu, \theta, t),
\end{equation}
where $\bR$ is the gyrocenter position, $u$ is the gyrocenter
parallel velocity, $\mu$ is the magnetic moment and $\theta$ is
the gyrophase. For convenience, we have written the transformation
$T_\epsilon$ as the composition of two other transformations that
we call {\it non-perturbative transformation} $T_{NP, \epsilon}$
and {\it perturbative transformation} $T_{P,\epsilon}$.

First, we perform a non-perturbative change of coordinates
\begin{equation}
(\boldr, \bv) = T_{NP, \epsilon} ( \bZ_g ) = T_{NP,\epsilon}
(\bR_g, v_{||g}, \mu_g, \theta_g),
\end{equation}
where $\bR_g$, $v_{||g}$, $\mu_g$ and $\theta_g$ are lowest order
approximations to the gyrocenter position, parallel velocity,
magnetic moment and gyrophase. Their detailed definitions can be
found in subsection~\ref{sub_drift}. For simplicity, we will
sometimes use the notation $\{ Z_g^\alpha \}_{\alpha = 1}^6 \equiv
\bZ_g = \{ \bR_g, v_{||g}, \mu_g, \theta_g \}$. Physically, the
non-perturbative change of variables in subsection~\ref{sub_drift}
amounts to saying that to lowest order the gyromotion is circular.
This is a consequence of the smallness of the gyroradius and the
fact that the corrections due to the electrostatic potential are
of next order. After the change of coordinates we add the total
time derivative of a function $S_{NP} (\bR_g, \mu_g, \theta_g)$ to
the Lagrangian. The details of the calculation are given in
subsection~\ref{sub_drift}. The final result is
\begin{eqnarray}
\fl \mathcal{L}^{\bZ_g} = \left [ \frac{1}{\epsilon} \bA (\bR_g) +
v_{||g} \bun (\bR_g) + \epsilon \Gammabf_\bR^{(1)} + \epsilon^2
\Gammabf_\bR^{(2)} + \ldots \right ] \cdot \frac{d\bR_g}{dt}
\nonumber \\ + \left ( - \mu_g + \epsilon \Gamma_\theta^{(1)} +
\epsilon^2 \Gamma_\theta^{(2)} + \ldots \right )
\frac{d\theta_g}{d(t/\epsilon)} - H^{(0)} (\bR_g, v_{||g}, \mu_g)
- \epsilon H^{(1)}, \label{lagr_ini}
\end{eqnarray}
where
\begin{equation}
H^{(0)} (\bR_g, v_{||g}, \mu_g) = \frac{1}{2} v_{||g}^2 + \mu_g
B(\bR_g) \label{ham_ini}
\end{equation}
and the rest of the terms are defined in
subsection~\ref{sub_drift}. Notice that with the notation in
\eq{lagr_ini} we have made explicit the fact that
$d\theta_g/dt\sim\epsilon^{-1}$, i.e., its time variation is of
the order of the gyrofrequency time scale and hence much faster
than the evolution of the rest of the phase-space variables. The
change of variables is non-perturbative and will give
contributions to all orders in $\epsilon$. All the higher order
terms $\Gammabf_\bR^{(1)} (\bR_g, v_{||g}, \mu_g, \theta_g)$,
$\Gammabf_\bR^{(2)} (\bR_g, v_{||g}, \mu_g, \theta_g)$, ...,
$\Gamma_\theta^{(1)} (\bR_g, v_{||g}, \mu_g, \theta_g)$,
$\Gamma_\theta^{(2)} (\bR_g, v_{||g}, \mu_g, \theta_g)$, ... and
$H^{(1)} (\bR_{g\bot}/\epsilon, R_{g||}, \mu_g, \theta_g, t)$
depend on gyrophase. In the Lagrangian \eq{lagr_ini}, the
Hamiltonian has only the first order correction $H^{(1)}
(\bR_{g\bot}/\epsilon, R_{g||}, \mu_g, \theta_g, t)$, with the
higher order corrections being exactly zero. This correction
$H^{(1)} (\bR_{g\bot}/\epsilon, R_{g||}, \mu_g, \theta_g, t)$ is
the only term in the Lagrangian that has strong perpendicular
gradients because it is the only contribution that depends on the
electrostatic potential.

The gyrophase dependence in the Lagrangian \eq{lagr_ini} must be
eliminated with the definition of the gyrokinetic variables order
by order. Since we only calculate the gyrokinetic variables to
some order, we truncate the expansion in $\epsilon$ to the order
of interest. In general, we need to keep
\begin{eqnarray}
\fl \mathcal{L}^{\bZ_g} = \left [ \frac{1}{\epsilon} \bA (\bR_g) +
v_{||g} \bun (\bR_g) + \sum_{i = 1}^n \epsilon^i
\Gammabf_\bR^{(i)} (\bR_g, v_{||g}, \mu_g, \theta_g) \right ]
\cdot \frac{d\bR_g}{dt} \nonumber \\ + \left [ - \epsilon \mu_g +
\sum_{i = 1}^n \epsilon^{i+1} \Gamma_\theta^{(i)} (\bR_g, v_{||g},
\mu_g, \theta_g) \right ] \frac{d\theta_g}{dt} \nonumber \\ -
H^{(0)} (\bR_g, v_{||g}, \mu_g) - \epsilon H^{(1)}
(\bR_{g\bot}/\epsilon, R_{g||}, \mu_g, \theta_g, t) +
O(\epsilon^{n+1}, \epsilon^{n+2}). \label{lagr_n}
\end{eqnarray}
Here, we have not written explicitly $d\theta_g/d(t/\epsilon)$.
Instead, we keep the terms that are multiplying $d\theta_g/dt$ to
higher order in $\epsilon$. The notation $O(\epsilon^{n+1},
\epsilon^{n+2})$ will be extensively used in this paper and
indicates that the terms of order $\epsilon^{n+1}$ that we have
neglected are either proportional to $d\bR_g/dt$ or are in the
Hamiltonian, and that the terms of order $\epsilon^{n+2}$ that we
have neglected are proportional to $dv_{||g}/dt$, $d\mu_g/dt$ and
$d\theta_g/dt$ (the Lagrangian \eq{lagr_n} does not contain terms
proportional to $dv_{||g}/dt$ or $d\mu_g/dt$). When we perform the
expansion to obtain the gyrokinetic variables order by order, it
will be apparent that this notation is convenient because we need
to keep some terms to $O(\epsilon^n)$ and the rest to
$O(\epsilon^{n+1})$.

Employing expression \eq{lagr_n}, we find the gyrokinetic
variables by eliminating the gyrophase dependence order by order.
For now, it is enough to simply write the transformation
\begin{equation}
(\bR_g, v_{||g}, \mu_g, \theta_g) = T_{P, \epsilon} (\bZ, t) =
T_{P,\epsilon} (\bR, u, \mu, \theta, t)
\end{equation}
to the order of interest as
\begin{eqnarray}
\fl \bR_g = \bR + \sum_{i=1}^n \epsilon^{i+1} \tilde{\bR}_{i+1},
\; v_{||g} = u + \sum_{i=1}^n \epsilon^i \tilde{u}_i, \; \mu_g =
\mu + \sum_{i=1}^n \epsilon^i \tilde{\mu}_i, \; \theta_g = \theta
+ \sum_{i=1}^n \epsilon^i \tilde{\theta}_i. \label{newvar}
\end{eqnarray}
In subsection~\ref{sub_gyro} we show the connection of this
expansion with Lie transforms. Sometimes we will use the
abbreviated notation $\{Z^\alpha \}_{\alpha = 1}^6 \equiv \bZ = \{
\bR, u, \mu, \theta \}$. Notice that the variable $\bR$ is found
to an order higher than the rest. The corrections
$\tilde{\bR}_{n+1}$, $\tilde{u}_n$, $\tilde{\mu}_n$ and
$\tilde{\theta}_n$ are obtained by imposing that the Lagrangian is
gyrophase independent up to terms of order $O(\epsilon^n,
\epsilon^{n+1})$, where the terms of order $\epsilon^n$ are in the
Hamiltonian or are terms proportional to $d\bR/dt$, and the terms
of order $\epsilon^{n+1}$ are terms proportional to $du/dt$,
$d\mu/dt$ and $d\theta/dt$. To demonstrate the procedure, we show
schematically how to obtain the first corrections $\tilde{\bR}_2$,
$\tilde{u}_1$, $\tilde{\mu}_1$ and $\tilde{\theta}_1$. We then
argue that the same formalism can be extended to arbitrary order.
The proof presented here demonstrates that the gyrokinetic
variables can be consistently calculated order by order without
running into problems. Recently, this has been a controversial
issue~\cite{Sugiyama2008, Krommes2009, Sugiyama2009}.

To calculate $\tilde{\bR}_2$, $\tilde{u}_1$, $\tilde{\mu}_1$ and
$\tilde{\theta}_1$, we need to express the Lagrangian \eq{lagr_n}
in the new gyrokinetic variables to $O(\epsilon, \epsilon^2)$,
giving\footnote{Observe that in principle, the term
$\epsilon^2\bA(\bR)\cdot d\tilde{\bR}_3/dt$ should be included in
$\mathcal{L}^\bZ$ \eq{lagr_n1}. However, adding the time
derivative of $-\epsilon^2\bA(\bR)\cdot \tilde\bR_3$ eliminates
any $\tilde\bR_3$ dependence to this order. The same can be said
about \eq{lagr_nGK} and \eq{lagrangian_gyro_o2_1}, where the terms
$\epsilon^{n+1}\bA(\bR)\cdot d\tilde{\bR}_{n+2}/dt$ and
$\epsilon^3 \bA(\bR) \cdot d \tilde{\bR}_{4}/dt$ could be added.
It is very easy to show that the algorithm gives, of course,
exactly the same results with either choice.}
\begin{eqnarray}
\fl \mathcal{L}^\bZ & = \left [ \frac{1}{\epsilon} \bA (\bR) + u
\bun (\bR) + \epsilon \left ( \tilde{\bR}_2 \cdot \nabla_\bR \bA
(\bR) + \tilde{u}_1 \bun (\bR) + \Gammabf_\bR^{(1)} (\bR, u,
\mu, \theta) \right ) \right ] \cdot \frac{d \bR}{dt} \nonumber \\
\fl & + \left [ \epsilon \bA (\bR) + \epsilon^2 u \bun (\bR)
\right ] \cdot \frac{d\tilde{\bR}_2}{dt} + \left [ - \epsilon \mu
+ \epsilon^2 \left ( - \tilde{\mu}_1 + \Gamma_\theta^{(1)} (\bR,
u, \mu, \theta) \right ) \right ] \frac{d \theta}{dt} - \epsilon^2
\mu \frac{d \tilde{\theta}_1}{dt} \nonumber \\ \fl & - H^{(0)}
(\bR, u, \mu) - \epsilon \left [ u \tilde{u}_1 + \tilde{\mu}_1
B(\bR) + H^{(1)} (\bR_\bot/\epsilon, R_{||}, \mu, \theta, t)
\right ] + O(\epsilon^2, \epsilon^3). \label{lagr_n1}
\end{eqnarray}
Notice that in the functions $\Gammabf_\bR^{(1)} (\bR_g, v_{||g},
\mu_g, \theta_g)$, $\Gamma_\theta^{(1)} (\bR_g, v_{||g}, \mu_g,
\theta_g)$, $H^{(0)} (\bR_g, v_{||g}, \mu_g)$ and $H^{(1)}
(\bR_{g\bot}/\epsilon, R_{g||}, \mu_g, \theta_g, t)$ the variables
$\bR_g$, $v_{||g}$, $\mu_g$ and $\theta_g$ are replaced by the
gyrokinetic variables $\bR$, $u$, $\mu$ and $\theta$. For example,
from the definition of $H^{(0)}$ in \eq{ham_ini}, we find
\begin{equation}
H^{(0)} (\bR, u, \mu) = \frac{1}{2} u^2 + \mu B(\bR).
\end{equation}
The correction $u \tilde{u}_1 + \tilde{\mu}_1 B(\bR) + H^{(1)}
(\bR_\bot/\epsilon, R_{||}, \mu, \theta, t)$ to the Hamiltonian
contains both the correction $H^{(1)}$ and the result of Taylor
expanding $H^{(0)} (\bR_g, v_{||g}, \mu_g) = H^{(0)} (\bR +
\epsilon^2 \tilde{\bR}_2 + \ldots, u + \epsilon \tilde{u}_1 +
\ldots, \mu + \epsilon \tilde{\mu}_1 + \ldots)$ around $\bR$, $u$
and $\mu$, i.e.,
\begin{eqnarray}
\fl \tilde{u}_1 \frac{\partial H^{(0)}}{\partial u} +
\tilde{\mu}_1 \frac{\partial H^{(0)}}{\partial \mu} + H^{(1)}
(\bR_\bot/\epsilon, R_{||}, \mu, \theta, t) \nonumber \\ = u
\tilde{u}_1 + \tilde{\mu}_1 B(\bR) + H^{(1)} (\bR_\bot/\epsilon,
R_{||}, \mu, \theta, t).
\end{eqnarray}
In subsection~\ref{sub_gyro} we show that the expansion around
$\bR$ gives a term that is an order higher and can be ignored to
this order.

As indicated in subsection~\ref{sub_change}, we can always add a
time derivative to the Lagrangian \eq{lagr_n1}. By doing so, we
will get a form of the Lagrangian in which the derivatives of a
function $S^{(2)}_P (\bR_\bot/\epsilon, \bR, u, \mu, \theta, t)$
enter. By imposing that the Lagrangian is of a specific form, we
first obtain the corrections $\tilde{\bR}_2$, $\tilde{u}_1$,
$\tilde{\mu}_1$ and $\tilde{\theta}_1$ as functions of $S^{(2)}_P$
and we then find $S^{(2)}_P$ by integrating a simple differential
equation. We add to the Lagrangian \eq{lagr_n1} the total time
derivative
\begin{equation}
\fl \frac{d}{dt} \left [ \epsilon^2 S^{(2)}_P (\bR_\bot/\epsilon,
\bR, u, \mu, \theta, t) - \epsilon \bA(\bR) \cdot \tilde{\bR}_2
 - \epsilon^2 u \bun(\bR) \cdot
\tilde{\bR}_2 + \epsilon^2 \mu \tilde{\theta}_1 \right ],
\label{add_time_derivative}
\end{equation}
where we have taken into account that $S^{(2)}_P$ depends on $\bR$
in two different ways: a fast dependence due to the potential that
has wavelengths on the order of the gyroradius, and a slow
dependence due to the background magnetic field. Notice that
adding the time derivative \eq{add_time_derivative} to the
Lagrangian \eq{lagr_n1} eliminates all the terms proportional to
the time derivatives of the corrections $\tilde{\bR}_2$,
$\tilde{u}_1$, $\tilde{\mu}_1$ and $\tilde{\theta}_1$, giving
\begin{eqnarray}
\fl \overline{\mathcal{L}} = & \left [ \frac{1}{\epsilon} \bA + u
\bun + \epsilon \left ( \bB \times \tilde{\bR}_2 + \tilde{u}_1
\bun + \Gammabf_\bR^{(1)} + \nabla_{(\bR_\bot/\epsilon)} S^{(2)}_P
\right ) \right ] \cdot \frac{d\bR}{dt} \nonumber \\ \fl & +
\epsilon^2 \left ( - \bun \cdot \tilde{\bR}_2 + \frac{\partial
S^{(2)}_P}{\partial u} \right ) \frac{du}{dt} + \epsilon^2 \left (
\tilde{\theta}_1 + \frac{\partial S^{(2)}_P}{\partial \mu} \right
) \frac{d\mu}{dt} \nonumber \\ \fl & + \left [ - \epsilon \mu +
\epsilon^2 \left ( - \tilde{\mu}_1 + \Gamma_\theta^{(1)} +
\frac{\partial S^{(2)}_P}{\partial \theta} \right ) \right ]
\frac{d\theta}{dt} - H^{(0)} \nonumber \\ \fl & - \epsilon \left (
u \tilde{u}_1 + \tilde{\mu}_1 B + H^{(1)} \right ) + O(\epsilon^2,
\epsilon^3), \label{lagr_n1_2}
\end{eqnarray}
where we have used
\begin{equation}
\tilde{\bR}_2 \cdot \nabla_\bR \bA - \nabla_\bR \bA \cdot
\tilde{\bR}_2 = ( \nabla_\bR \times \bA ) \times \tilde{\bR}_2 =
\bB \times \tilde{\bR}_2,
\end{equation}
trivially deduced from the identity
\begin{equation}
[\nabla_\bR \bA - (\nabla_\bR\bA)^{\mathrm{T}}]_{ij} =
\frac{\partial A_j}{\partial R_i} - \frac{\partial A_i}{\partial
R_j} = \sum_{k=1}^3\varepsilon_{ijk}(\nabla_\bR\times\bA)_{k}.
\end{equation}
Here $\varepsilon_{ijk}$ is the Levi-Civita symbol and the
superscript ${}^\mathrm{T}$ stands for matrix transposition.
Notice that in equation \eq{lagr_n1_2} we have not made explicit
the dependence of the functions on the gyrokinetic variables, but
it is assumed that all the terms are functions of $\bR$, $u$,
$\mu$ and $\theta$. By imposing that the Lagrangian \eq{lagr_n1_2}
be equal to
\begin{eqnarray}
\fl \overline{\mathcal{L}} = & \left ( \frac{1}{\epsilon} \bA + u
\bun + \epsilon \overline{\Gammabf}_\bR^{(1)} \right ) \cdot
\frac{d\bR}{dt} - \epsilon \mu \frac{d\theta}{dt} -
\overline{H}^{(0)} - \epsilon \overline{H}^{(1)} + O(\epsilon^2,
\epsilon^3), \label{lagr_n1_2_final}
\end{eqnarray}
where $\overline{\Gammabf}_\bR^{(1)}$, $\overline{H}^{(0)} :=
H^{(0)} (\bR, u, \mu)$ and $\overline{H}^{(1)}$ are gyrophase
independent, we obtain the equations
\begin{equation}
\fl \tilde{\bR}_2 = \frac{\partial S^{(2)}_P}{\partial u} \bun
(\bR) + \frac{1}{B(\bR)} \bun (\bR) \times \left [
\Gammabf_\bR^{(1)} (\bR, u, \mu, \theta) -
\overline{\Gammabf}_\bR^{(1)} + \nabla_{(\bR_\bot/\epsilon)}
S^{(2)}_P \right ], \label{R2_1}
\end{equation}
\begin{equation}
\tilde{u}_1 =  \bun (\bR) \cdot \left [
\overline{\Gammabf}_\bR^{(1)} - \Gammabf_\bR^{(1)} (\bR, u, \mu,
\theta) \right ], \label{u1_1}
\end{equation}
\begin{equation}
\tilde{\mu}_1 = \Gamma_\theta^{(1)} (\bR, u, \mu, \theta) +
\frac{\partial S^{(2)}_P}{\partial \theta} \label{mu1_1}
\end{equation}
and
\begin{equation}
\tilde{\theta}_1 = - \frac{\partial S^{(2)}_P}{\partial \mu}.
\label{theta1_1}
\end{equation}
The corrections $\tilde{\bR}_2$, $\tilde{u}_1$, $\tilde{\mu}_1$
and $\tilde{\theta}_1$ can then be found if we obtain $S^{(2)}_P$.
To do so, we use that
\begin{equation}
u \tilde{u}_1 + B(\bR) \tilde{\mu}_1 + H^{(1)} (\bR_\bot/\epsilon,
R_{||}, \mu, \theta, t) = \overline{H}^{(1)}. \label{eq_H1GK}
\end{equation}
First, we take the gyroaverage of this equation to obtain
$\overline{H}^{(1)}$. Using the results in equations \eq{u1_1} and
\eq{mu1_1}, the gyroaverage of equation \eq{eq_H1GK} becomes
\begin{eqnarray}
\fl \overline{H}^{(1)} = u \bun(\bR) \cdot \left [
\overline{\Gammabf}_\bR^{(1)} - \left \langle \Gammabf_\bR^{(1)}
(\bR, u, \mu, \theta) \right \rangle \right ] + B(\bR) \left
\langle \Gamma_\theta^{(1)} (\bR, u, \mu, \theta) \right \rangle
\nonumber \\ + \left \langle H^{(1)} (\bR_\bot/\epsilon, R_{||},
\mu, \theta, t) \right \rangle, \label{eq_Hbar1GK}
\end{eqnarray}
where $\langle \ldots \rangle = (2\pi)^{-1} \oint d\theta\,
(\ldots)$ is the gyroaverage holding $\bR$, $u$, $\mu$ and $t$
fixed. Notice that we have been able to obtain
$\overline{H}^{(1)}$ without explicitly finding $S^{(2)}_P$.
Solving for $S^{(2)}_P$ is easy once $\overline{H}^{(1)}$ is
known. Combining equations \eq{eq_H1GK} and \eq{eq_Hbar1GK}, we
find
\begin{eqnarray}
\fl \frac{\partial S^{(2)}_P}{\partial \theta} = \frac{u}{B(\bR)}
\bun (\bR) \cdot \left [ \Gammabf_\bR^{(1)} (\bR, u, \mu, \theta)
- \left \langle \Gammabf_\bR^{(1)} (\bR, u, \mu, \theta) \right
\rangle \right ] \nonumber \\ - \left [ \Gamma_\theta^{(1)} (\bR,
u, \mu, \theta) - \left \langle \Gamma_\theta^{(1)} (\bR, u, \mu,
\theta) \right \rangle \right ] \nonumber \\ - \frac{1}{B(\bR)}
\left [ H^{(1)} (\bR_\bot/\epsilon, R_{||}, \mu, \theta, t) -
\left \langle H^{(1)} (\bR_\bot/\epsilon, R_{||}, \mu, \theta, t)
\right \rangle \right ]. \label{eq_S1GK}
\end{eqnarray}
It is then straightforward to obtain $S^{(2)}_P$ by integrating in
the gyrophase $\theta$. The specific calculation of the
corrections $\tilde{\bR}_2$, $\tilde{u}_1$, $\tilde{\mu}_1$,
$\tilde{\theta}_1$ and $\overline{H}^{(1)}$ is done in
subsection~\ref{sub_gyro}. Here we only want to demonstrate the
procedure. Notice that the final solution depends on our choice of
$\overline{\Gammabf}_\bR^{(1)}$ -- and in general on our choice of
the final expression for $\overline{\mathcal{L}}$. However, once
the choice is made, $\overline{H}^{(1)}$ is completely determined.

Once $\tilde{\bR}_2$, $\tilde{u}_1$, $\tilde{\mu}_1$,
$\tilde{\theta}_1$, $\overline{H}^{(1)}$ and $S^{(2)}_P$ have been
determined, we can prove by induction that the procedure gives the
corrections to any order. Assume that the corrections
$\tilde{\bR}_{i+1}$, $\tilde{u}_i$, $\tilde{\mu}_i$,
$\tilde{\theta}_i$ and $\overline{H}^{(i)}$ and the functions
$S^{(i+1)}_P$ have been obtained up to $i = n-1$ in such a way
that the Lagrangian is of the form
\begin{eqnarray} \label{lagr_prevorder}
\fl \overline{\mathcal{L}} = \left [ \frac{1}{\epsilon} \bA(\bR) +
u \bun (\bR) + \sum_{i=1}^{n-1} \epsilon^i
\overline{\Gammabf}_\bR^{(i)} \right ] \cdot \frac{d\bR}{dt} -
\epsilon \mu \frac{d\theta}{dt} - \overline{H}^{(0)} -
\sum_{i=1}^{n-1} \epsilon^i \overline{H}^{(i)} \nonumber \\ +
O(\epsilon^{n}, \epsilon^{n+1}).
\end{eqnarray}
Then it is possible to obtain the corrections $\tilde{\bR}_{n+1}$,
$\tilde{u}_n$, $\tilde{\mu}_n$ and $\tilde{\theta}_n$, the
function $S^{(n+1)}_P$ and the phase-space Lagrangian to order
$O(\epsilon^n, \epsilon^{n+1})$. To do so, the Lagrangian is
written to $O(\epsilon^{n}, \epsilon^{n+1})$. In general, we
cannot guess the exact form of the Lagrangian to such high order
without doing the calculation order by order, but the terms that
contain the corrections $\tilde{\bR}_{n+1}$, $\tilde{u}_n$,
$\tilde{\mu}_n$ and $\tilde{\theta}_n$ are very easy to obtain. We
find that
\begin{eqnarray}
\fl \mathcal{L}^\bZ & + \sum_{i=1}^{n-1} \frac{d}{dt} \left (
\epsilon^{i+1} S^{(i+1)}_P - \epsilon^i \bA \cdot
\tilde{\bR}_{i+1} - \epsilon^{i+1} u \bun \cdot \tilde{\bR}_{i+1}
+ \epsilon^{i+1} \mu \tilde{\theta}_i \right ) \nonumber \\ \fl &
= \left [ \frac{1}{\epsilon} \bA(\bR) + u \bun (\bR) +
\sum_{i=1}^{n-1} \epsilon^i \overline{\Gammabf}_\bR^{(i)} +
\epsilon^n \left ( \tilde{\bR}_{n+1} \cdot \nabla_\bR \bA (\bR) +
\tilde{u}_n \bun(\bR) + \tilde{\Gammabf}_\bR^{(n)} \right ) \right
] \cdot \frac{d\bR}{dt} \nonumber \\ \fl & + \left [ \epsilon^n
\bA(\bR) + \epsilon^{n+1} u \bun (\bR) \right ] \cdot
\frac{d\tilde{\bR}_{n+1}}{dt} + \epsilon^{n+1}
\tilde{\Gamma}_u^{(n)} \frac{du}{dt}+ \epsilon^{n+1}
\tilde{\Gamma}_\mu^{(n)} \frac{d\mu}{dt} \nonumber \\ \fl & +
\left [ - \epsilon \mu + \epsilon^{n+1} \left ( - \tilde{\mu}_n +
\tilde{\Gamma}_\theta^{(n)} \right ) \right ] \frac{d\theta}{dt} -
\epsilon^{n+1} \mu \frac{d\tilde{\theta}_n}{dt} -
\overline{H}^{(0)} - \sum_{i=1}^{n-1} \epsilon^i
\overline{H}^{(i)} \nonumber \\ \fl & - \epsilon^n \left [ u
\tilde{u}_n + \tilde{\mu}_n B(\bR) + \tilde{H}^{(n)} \right ] +
O(\epsilon^{n+1}, \epsilon^{n+2}). \label{lagr_nGK}
\end{eqnarray}
Here we have just separated the terms of order $O(\epsilon^n,
\epsilon^{n+1})$ into those that depend on the corrections
$\tilde{\bR}_{n+1}$, $\tilde{u}_n$, $\tilde{\mu}_n$ and
$\tilde{\theta}_n$, and the rest that we have lumped into the
terms $\tilde{\Gammabf}_\bR^{(n)}$, $\tilde{\Gamma}_u^{(n)}$,
$\tilde{\Gamma}_\mu^{(n)}$, $\tilde{\Gamma}_\theta^{(n)}$ and
$\tilde{H}^{(n)}$.

The form of the Lagrangian \eq{lagr_nGK} is very similar to the
Lagrangian \eq{lagr_n1}. We can then use the same procedure. We
add the time derivative
\begin{equation}
\fl \frac{d}{dt} \left [ \epsilon^{n+1} S^{(n+1)}_P
(\bR_\bot/\epsilon, \bR, u, \mu, \theta, t) - \epsilon^n \bA(\bR)
\cdot \tilde{\bR}_{n+1} - \epsilon^{n+1} u \bun(\bR) \cdot
\tilde{\bR}_{n+1} + \epsilon^{n+1} \mu \tilde{\theta}_n \right ]
\end{equation}
to cancel all the terms that are proportional to the time
derivatives of the corrections $\tilde{\bR}_{n+1}$, $\tilde{u}_n$,
$\tilde{\mu}_n$ and $\tilde{\theta}_n$. Then, by imposing that the
Lagrangian \eq{lagr_nGK} plus this time derivative be equal to
\begin{eqnarray}
\fl \overline{\mathcal{L}} = \left ( \frac{1}{\epsilon} \bA (\bR)
+ u \bun (\bR) + \sum_{i = 1}^n \epsilon^i
\overline{\Gammabf}_\bR^{(i)} \right ) \cdot \frac{d\bR}{dt} -
\epsilon \mu \frac{d\theta}{dt} - \overline{H}^{(0)} -
\sum_{i=1}^n \epsilon^i \overline{H}^{(i)} \nonumber \\ +
O(\epsilon^{n+1}, \epsilon^{n+2}),
\end{eqnarray}
we obtain the equations
\begin{equation}
\fl \tilde{\bR}_{n+1} = \left ( \tilde{\Gamma}_u^{(n)} +
\frac{\partial S^{(n+1)}_P}{\partial u} \right ) \bun (\bR) +
\frac{1}{B(\bR)} \bun (\bR) \times \left (
\tilde{\Gammabf}_\bR^{(n)} - \overline{\Gammabf}_\bR^{(n)} +
\nabla_{(\bR_\bot/\epsilon)} S^{(n+1)}_P \right ), \label{Rn1}
\end{equation}
\begin{equation}
\tilde{u}_n =  \bun (\bR) \cdot \left (
\overline{\Gammabf}_\bR^{(n)} - \tilde{\Gammabf}_\bR^{(n)} \right
), \label{un}
\end{equation}
\begin{equation}
\tilde{\mu}_n = \tilde{\Gamma}_\theta^{(n)} + \frac{\partial
S^{(n+1)}_P}{\partial \theta} \label{mun}
\end{equation}
and
\begin{equation}
\tilde{\theta}_n = - \tilde{\Gamma}_\mu^{(n)} - \frac{\partial
S^{(n+1)}_P}{\partial \mu}. \label{thetan}
\end{equation}
The $n$-th correction to the Hamiltonian becomes
\begin{equation}
\overline{H}^{(n)} = u \bun(\bR) \cdot \left (
\overline{\Gammabf}_\bR^{(n)} - \left \langle
\tilde{\Gammabf}_\bR^{(n)} \right \rangle \right ) + B(\bR) \left
\langle \tilde{\Gamma}_\theta^{(n)} \right \rangle + \left \langle
\tilde{H}^{(n)} \right \rangle, \label{eq_HbarnGK}
\end{equation}
and the equation for $S^{(n+1)}_P$ is
\begin{eqnarray}
\fl \frac{\partial S^{(n+1)}_P}{\partial \theta} =
\frac{u}{B(\bR)} \bun (\bR) \cdot \left (
\tilde{\Gammabf}_\bR^{(n)} - \left \langle
\tilde{\Gammabf}_\bR^{(n)} \right \rangle \right ) - \left (
\tilde{\Gamma}_\theta^{(n)} - \left \langle
\tilde{\Gamma}_\theta^{(n)} \right \rangle \right ) \nonumber \\ -
\frac{1}{B(\bR)} \left ( \tilde{H}^{(n)} - \left \langle
\tilde{H}^{(n)} \right \rangle \right ).
\end{eqnarray}

In subsections \ref{sub_drift} and \ref{sub_gyro} we obtain the
phase-space Lagrangian to $O(\epsilon^2, \epsilon^3)$.
Specifically, in subsection~\ref{sub_drift} and \ref{app_DK} we
derive equation \eq{lagr_ini}. In subsection \ref{sub_gyro} and
\ref{app_gyro_2} we use the perturbation procedure explained here
to go from equation \eq{lagr_ini} to the final result.

\subsection{Non-perturbative change of variables}
\label{sub_drift}
We perform a change of variables $(\boldr, \bv) = T_{NP, \epsilon}
(\bZ_g) = T_{NP,\epsilon} (\bR_g, v_{||g}, \mu_g, \theta_g)$
defined by
\begin{equation}\label{eq:changer}
\boldr = \bR_g  + \epsilon \rhobf ( \bR_g, \mu_g, \theta_g),
\end{equation}
and
\begin{equation}\label{eq:changev}
\bv = v_{||g} \bun(\bR_g) + \rhobf ( \bR_g, \mu_g, \theta_g)
\times \bB (\bR_g),
\end{equation}
with the gyroradius vector defined as
\begin{equation}\label{eq:defrho}
\rhobf ( \bR_g, \mu_g, \theta_g ) = - \sqrt{\frac{2
\mu_g}{B(\bR_g)}} \left [ \sin \theta_g \eun_1 (\bR_g) - \cos
\theta_g \eun_2 (\bR_g) \right ].
\end{equation}
The unit vectors $\eun_1 (\boldr)$ and $\eun_2 (\boldr)$ are
orthogonal to each other and to $\bun = \bB/B$, and satisfy
$\eun_1 \times \eun_2 = \bun$ at every location $\boldr$.
Physically, $\bR_g$ is the guiding center position, $v_{||g}$ the
velocity parallel to the magnetic field at the guiding center
position, $\mu_g$ the lowest order magnetic moment, and $\theta_g$
the lowest order gyrophase. For a homogeneous static magnetic
field and in the absence of electric field, the change of
coordinates defined by \eq{eq:changer}, \eq{eq:changev} and
\eq{eq:defrho} exactly eliminates the gyrophase dependence. Note
in passing that it is a well defined change of coordinates. These
formulae explicitly give $\{\boldr,\bv\}$ as a function of $\bZ_g
= \{ \bR_g, v_{||g}, \mu_g, \theta_g \}$ and it is easy to see
that the transformation is invertible for small $\epsilon$: it is
clearly invertible for $\epsilon=0$ and the transformation is
continuous in $\epsilon$.

We substitute the relations $\bX (\bZ_g)$, given in
\eq{eq:changer} and \eq{eq:changev}, and
\begin{equation}
\frac{dX^\alpha}{dt} (\bZ_g, \dot{\bZ}_g) = \sum_{\beta = 1}^6
\frac{\partial X^\alpha (\bZ_g)}{\partial Z_g^\beta} {\dot
Z}_g^\beta, \quad \alpha = 1, 2, \ldots, 6
\end{equation}
into the non-dimensionalized Lagrangian \eq{lagrange_non},
$\mathcal{L}^\bX (\bX, \dot{\bX}, t)$. The resulting Lagrangian
that we denote as $\mathcal{L}^\bX ( \bX ( \bZ_g), \dot{\bX} (
\bZ_g, \dot{\bZ}_g ), t )$ differs from the Lagrangian
$\mathcal{L}^{\bZ_g}$ in \eq{lagr_ini} and \eq{lagrange_drift_4}
by the time derivative of a function $S_{NP}$ and even though both
Lagrangians give the same equations of motion, we have decided to
stress the difference. The Lagrangian $\mathcal{L}^\bX ( \bX (
\bZ_g), \dot{\bX} ( \bZ_g, \dot{\bZ}_g ), t )$ is
\begin{eqnarray}
\fl \mathcal{L}^\bX ( \bX ( \bZ_g), \dot{\bX} ( \bZ_g, \dot{\bZ}_g
), t ) = \left [ \frac{1}{\epsilon}\bA ( \bR_g + \epsilon \rhobf )
+ v_{||g} \bun_g + \rhobf \times \bB_g \right ] \cdot
\frac{d}{dt} \left ( \bR_g + \epsilon \rhobf \right ) \nonumber \\
- H^{(0)} - \epsilon H^{(1)}, \label{lagrange_drift_1}
\end{eqnarray}
with $H^{(0)}(\bR_g, v_{||g}, \mu_g)$ defined in \eq{ham_ini} and
\begin{eqnarray} \label{H1}
\fl H^{(1)} (\bR_{g\bot}/\epsilon, R_{g||}, \mu_g, \theta_g, t) =
\Lambda \langle \phi \rangle ( \bR_{g\bot}/\lambda \epsilon,
R_{g||}, \mu_g/\lambda^2, t/\tau) \nonumber \\ + \Lambda
\tilde{\phi} ( \bR_{g\bot}/\lambda \epsilon, R_{g||},
\mu_g/\lambda^2, \theta_g + \pi \Theta (-\lambda), t/\tau),
\end{eqnarray}
where $\Theta(x)$ is the Heaviside step function, with $\Theta(x)
= 1$ for $x > 0$ and $\Theta(x) = 0$ for $x < 0$. Any magnetic
quantity with subindex $g$ is evaluated at $\bR_g$, e.g.,
$\bB_g:=\bB(\bR_g)$. We write $\rhobf\equiv \rhobf ( \bR_g, \mu_g,
\theta_g)$ when no confusion is possible.

We have defined a new function $\phi(\bR_g,\mu_g,\theta_g,t)$
(notice the difference in the font between $\phi$ and $\varphi$)
given by
\begin{equation} \label{defphi}
\phi( \bR_{g}, \mu_g, \theta_g,t) := \varphi (
\bR_{g}+\epsilon\rhobf (\bR_g, \mu_g, \theta_g), t ).
\end{equation}
Then $\phiave$ is the gyroaverage of $\phi$,
\begin{equation}
\langle \phi \rangle(\bR_g,\mu_g,t) = \frac{1}{2\pi} \int_0^{2\pi}
d\theta_g\, \phi ( \bR_{g}, \mu_g, \theta_g, t),
\end{equation}
and $\phiwig$ the gyrophase dependent piece,
\begin{eqnarray}
\tilde\phi (\bR_g,\mu_g,\theta_g,t) = \phi
(\bR_g,\mu_g,\theta_g,t) - \langle \phi \rangle(\bR_g,\mu_g,t).
\end{eqnarray}
We now prove that the notation in \eq{H1},
\begin{equation}
\phi ( \bR_{g}, \mu_g,\theta_g, t) \equiv \phi (
\bR_{g\bot}/\lambda \epsilon, R_{g||}, \mu_g/\lambda^2, \theta_g +
\pi \Theta (- \lambda), t/\tau),
\end{equation}
is appropriate. First, we show that $\mu_g$ is always divided by
$\lambda^2$ and that the sign of $\lambda$ determines the phase of
$\theta_g$, and later we demonstrate that if conditions
\eq{eq:orderingvarphi_par} and \eq{eq:orderingvarphi_perp} are
satisfied, then
\begin{equation}\label{eq:orderingphi_par}
\bun(\bR_g) \cdot \nabla_{\bR_g} {\phi}({\bR}_g,\mu_g,\theta_g,t)
\sim 1
\end{equation}
and
\begin{equation}\label{eq:orderingphi_perp}
\nabla_{\bR_{g\perp}}{\phi} ({\bR}_g, \mu_g, \theta_g, t) :=
\bun(\bR_g) \times (\nabla_{\bR_g} {\phi} ({\bR}_g, \mu_g,
\theta_g, t) \times \bun(\bR_g)) \sim \frac{1}{\lambda \epsilon}
\end{equation}
are also satisfied. To show that $\mu_g$ always appears divided by
$\lambda^2$ and that we need to add $\pi$ to $\theta_g$ when
$\lambda$ is negative, it is enough to realize that $\phi$ depends
on $\mu_g$ and $\theta_g$ through the dependence of $\varphi$ on
$\boldr/\lambda \epsilon = \bR/\lambda \epsilon + \rhobf(\bR_g,
\mu_g, \theta_g)/\lambda$ and that $\rhobf$ as defined in
\eq{eq:defrho} only depends on $\mu_g$ through the multiplying
term $\sqrt{\mu_g}$. It is then obvious that $\boldr/\lambda
\epsilon = \bR_g/\lambda \epsilon + \rhobf(\bR_g, \mu_g/\lambda^2,
\theta_g + \pi \Theta (-\lambda))$. To prove that
\eq{eq:orderingvarphi_par} and \eq{eq:orderingvarphi_perp} imply
\eq{eq:orderingphi_par} and \eq{eq:orderingphi_perp}, we employ
\begin{eqnarray}
\fl \nabla_{\bR_g} \phi (\bR_g, \mu_g, \theta_g, t) =
\nabla_{\bR_g} \varphi (\bR_g + \epsilon \rhobf (\bR_g, \mu_g,
\theta_g), t) \nonumber \\ = \nabla_\boldr \varphi (\boldr, t) +
\epsilon \nabla_{\bR_g} \rhobf (\bR_g, \mu_g, \theta_g) \cdot
\nabla_\boldr \varphi (\boldr, t),
\end{eqnarray}
with $\epsilon \nabla_{\bR_g} \rhobf \cdot \nabla_\boldr \varphi
\sim \epsilon \nabla_{\bR_g} \rhobf \cdot \nabla_{\boldr_\bot}
\varphi \sim \lambda^{-1} \sim 1$. Then, using equations
\eq{eq:orderingvarphi_par} and \eq{eq:orderingvarphi_perp}, it is
easy to see that equations \eq{eq:orderingphi_par} and
\eq{eq:orderingphi_perp} are correct. Note that when we write
$H^{(1)} (\bR_{g\bot}/\epsilon, R_{g||}, \mu_g, \theta_g, t)$ in
\eq{H1} we are emphasizing the dependence on $\epsilon$ because
the asymptotic procedure is based on expanding in $\epsilon \ll
1$. The dependence on $\Lambda$, $\lambda$ and $\tau$ is only
written explicitly in the function $\phi$.

We now show how to simplify \eq{lagrange_drift_1}. Employing
\begin{equation}
\nabla_{\bR_g} \rhobf = - \frac{\nabla_{\bR_g} B_g}{2B_g} \rhobf -
(\nabla_{\bR_g} \bun_g \cdot \rhobf) \bun_g + \nabla_{\bR_g}
\eun_{2g} \cdot \eun_{1g} (\rhobf \times \bun_g) \label{gradrho},
\end{equation}
\begin{equation}
\frac{\partial \rhobf}{\partial \mu_g} = \frac{1}{2\mu_g} \rhobf
\label{drhodmu}
\end{equation}
and
\begin{equation}
\frac{\partial \rhobf}{\partial\theta_g} = - \rhobf \times \bun_g,
\label{drhodtheta}
\end{equation}
we write the Lagrangian in \eq{lagrange_drift_1} as
\begin{eqnarray}
\fl \mathcal{L}^\bX & (\bX (\bZ_g), \dot{\bX} (\bZ_g,
\dot{\bZ}_g), t) = \Bigg [ \frac{1}{\epsilon} \bA (\bR_g +
\epsilon \rhobf) + v_{||g} \bun_g + \rhobf \times \bB_g +
\nabla_{\bR_g} \rhobf \cdot \bA (\bR_g + \epsilon \rhobf) +
\nonumber \\ \fl & + \epsilon \Bigg ( 2\mu_g \nabla_{\bR_g}
\eun_{2g} \cdot \eun_{1g} - v_{||g} \nabla_{\bR_g} \bun_g \cdot
\rhobf \Bigg ) \Bigg ] \cdot \frac{d \bR_g}{dt} + \frac{1}{2\mu_g}
\bA (\bR_g + \epsilon \rhobf) \cdot \rhobf \frac{d\mu_g}{dt}
\nonumber \\ \fl & + \left [ \bA (\bR_g + \epsilon \rhobf) \cdot
\frac{\partial \rhobf}{\partial \theta_g} - 2 \epsilon \mu_g
\right ] \frac{d\theta_g}{dt} - H^{(0)} - \epsilon H^{(1)} .
\label{lagrange_drift_2}
\end{eqnarray}
To obtain \eq{gradrho} we have used $\nabla_{\bR_g} \eun_{1g} = -
(\nabla_{\bR_g} \bun \cdot \eun_{1g}) \bun_g - (\nabla_{\bR_g}
\eun_{2g} \cdot \eun_{1g}) \eun_{2g}$ and $\nabla_{\bR_g}
\eun_{2g} = - (\nabla_{\bR_g} \bun \cdot \eun_{2g}) \bun_g +
(\nabla_{\bR_g} \eun_{2g} \cdot \eun_{1g}) \eun_{1g}$. To simplify
the Lagrangian \eq{lagrange_drift_2}, we add the time derivative
of
\begin{equation} \label{S_NP}
S_{NP} (\bR_g, \mu_g, \theta_g) = - \int_0^{\mu_g}
\frac{d\mu_g^\prime}{2\mu_g^\prime} \, \bA ( \bR_g + \epsilon
\rhobf ( \bR_g, \mu_g^\prime, \theta_g ) ) \cdot \rhobf ( \bR_g,
\mu_g^\prime, \theta_g ).
\end{equation}
As a result we find
\begin{eqnarray}
\fl \mathcal{L}^{\bZ_g} = & \Bigg [ \frac{1}{\epsilon} \bA (\bR_g
+ \epsilon \rhobf) + v_{||g} \bun_g + \rhobf \times \bB_g +
\nabla_{\bR_g} \rhobf \cdot \bA (\bR_g + \epsilon \rhobf) +
\nabla_{\bR_g} S_{NP} \nonumber \\ \fl & + \epsilon \Bigg ( 2\mu_g
\nabla_{\bR_g} \eun_{2g} \cdot \eun_{1g} - v_{||g} \nabla_{\bR_g}
\bun_g \cdot \rhobf \Bigg ) \Bigg ] \cdot \frac{d \bR_g}{dt}
\nonumber \\ \fl & + \left [ \bA (\bR_g + \epsilon \rhobf) \cdot
\frac{\partial \rhobf}{\partial \theta_g} - 2 \epsilon \mu_g +
\frac{\partial S_{NP}}{\partial \theta_g} \right ]
\frac{d\theta_g}{dt} - H^{(0)} - \epsilon H^{(1)}.
\label{lagrange_drift_3}
\end{eqnarray}
In \ref{app_DK} we prove that
\begin{eqnarray} \label{dS_NPdR}
\fl \nabla_{\bR_g} S_{NP} & = - \frac{1}{\epsilon} \bA (\bR_g +
\epsilon \rhobf) + \frac{1}{\epsilon} \bA_g - \rhobf \times \bB_g
- \nabla_{\bR_g} \rhobf \cdot \bA (\bR_g + \epsilon \rhobf) -
\epsilon \mu_g \nabla_{\bR_g} \eun_{2g} \cdot \eun_{1g} \nonumber \\
\fl & - \int_0^{\mu_g} \frac{d\mu_g^\prime}{2\mu_g^\prime} \,
\Bigg \{ \rhobf^\prime \times [ \bB (\bR_g + \epsilon
\rhobf^\prime ) - \bB_g ] + \epsilon [ ( \rhobf^\prime \times
\bun_g ) \cdot \bB (\bR_g + \epsilon \rhobf^\prime) ]
\nabla_{\bR_g} \bun_g \cdot \rhobf^\prime \nonumber \\ \fl & +
\frac{2\epsilon \mu_g^\prime}{B_g} [ \bun_g \cdot \bB (\bR_g +
\epsilon \rhobf^\prime) - B_g ] \nabla_{\bR_g} \eun_{2g} \cdot
\eun_{1g} \Bigg \}
\end{eqnarray}
and
\begin{equation} \label{dS_NPdtheta}
\fl \frac{\partial S_{NP}}{\partial \theta_g} = - \frac{\partial
\rhobf}{\partial \theta_g} \cdot \bA (\bR_g + \epsilon \rhobf ) +
\epsilon \mu_g +  \frac{\epsilon}{B_g} \int_0^{\mu_g}
d\mu_g^\prime \, [  \bun_g \cdot \bB (\bR_g + \epsilon
\rhobf^\prime) - B_g ],
\end{equation}
where we use the abbreviated notation $\rhobf^\prime \equiv \rhobf
(\bR_g, \mu_g^\prime, \theta_g)$. Substituting equations
\eq{dS_NPdR} and \eq{dS_NPdtheta} into the Lagrangian
\eq{lagrange_drift_3} finally gives
\begin{eqnarray}
\fl \mathcal{L}^{\bZ_g} = \Bigg ( \frac{1}{\epsilon} \bA_g +
v_{||g} \bun_g + \epsilon \Delta \Gammabf_\bR \Bigg ) \cdot
\frac{d \bR_g}{dt} + \left ( - \epsilon \mu_g + \epsilon^2 \Delta
\Gamma_\theta \right ) \frac{d\theta_g}{dt} - H^{(0)} - \epsilon
H^{(1)}, \label{lagrange_drift_4}
\end{eqnarray}
with
\begin{eqnarray}
\fl \Delta & \Gammabf_\bR = \mu_g \nabla_{\bR_g} \eun_{2g} \cdot
\eun_{1g} - v_{||g} \nabla_{\bR_g} \bun_g \cdot \rhobf_g -
\int_0^{\mu_g} \frac{d\mu_g^\prime}{2\mu_g^\prime} \, \Bigg \{
\frac{1}{\epsilon} \rhobf^\prime \times [ \bB (\bR_g + \epsilon
\rhobf^\prime ) - \bB_g ] \nonumber \\
\fl & + [ ( \rhobf^\prime \times \bun_g ) \cdot \bB (\bR_g +
\epsilon \rhobf^\prime) ] \nabla_{\bR_g} \bun_g \cdot
\rhobf^\prime + \frac{2 \mu_g^\prime}{B_g} [ \bun_g \cdot \bB
(\bR_g + \epsilon \rhobf^\prime) - B_g ] \nabla_{\bR_g} \eun_{2g}
\cdot \eun_{1g} \Bigg \}
\end{eqnarray}
and
\begin{equation}
\Delta \Gamma_\theta = \frac{1}{\epsilon B_g} \int_0^{\mu_g}
d\mu_g^\prime \, [  \bun_g \cdot \bB (\bR_g + \epsilon
\rhobf^\prime) - B_g ].
\end{equation}

It is easy to write the Lagrangian \eq{lagrange_drift_4} order by
order. We use
\begin{equation}
\bB (\bR_g + \epsilon \rhobf) = \bB_g + \epsilon \rhobf \cdot
\nabla_{\bR_g} \bB_g + \frac{\epsilon^2}{2} \rhobf \rhobf :
\nabla_{\bR_g} \nabla_{\bR_g} \bB_g + O( \epsilon^3 ),
\end{equation}
where our double-dot convention is $\mathbf{a} \mathbf{b}:
\matrixtop{\mathbf{M}} = \mathbf{b} \cdot \matrixtop{\mathbf{M}}
\cdot \mathbf{a}$, to obtain
\begin{eqnarray}
\fl \mathcal{L}^{\bZ_g} = \left ( \frac{1}{\epsilon} \bA_g +
v_{||g} \bun_g + \epsilon \Gammabf_\bR^{(1)} + \epsilon^2
\Gammabf_\bR^{(2)} \right ) \cdot \frac{d\bR_g}{dt} + \left ( -
\epsilon \mu_g + \epsilon^2 \Gamma_\theta^{(1)} + \epsilon^3
\Gamma_\theta^{(2)} \right ) \frac{d\theta_g}{dt} \nonumber \\
- H^{(0)} - \epsilon H^{(1)} + O(\epsilon^3, \epsilon^4),
\label{lagrangian_drift_5}
\end{eqnarray}
where
\begin{eqnarray}
\fl \Gammabf_\bR^{(1)} = \mu_g \nabla_{\bR_g} \eun_{2g} \cdot
\eun_{1g} - v_{||g} \nabla_{\bR_g} \bun_g \cdot \rhobf -
\frac{1}{2} (\rhobf \cdot \nabla_{\bR_g} B_g) \rhobf \times \bun_g
\nonumber \\ + \frac{1}{2} [ \rhobf \cdot \nabla_{\bR_g} \bun_g
\cdot (\rhobf \times \bun_g) ] \bB_g, \label{GammaR_o1}
\end{eqnarray}
\begin{eqnarray}
\fl \Gammabf_\bR^{(2)} =  \frac{1}{6} \rhobf \rhobf :
\nabla_{\bR_g} \nabla_{\bR_g} \bB_g \times \rhobf  - \frac{B_g}{3}
[\rhobf \cdot \nabla_{\bR_g} \bun_g \cdot (\rhobf \times \bun_g)]
\nabla_{\bR_g} \bun_g \cdot \rhobf \nonumber \\ -
\frac{2\mu_g}{3B_g} (\rhobf \cdot \nabla_{\bR_g} B_g)
\nabla_{\bR_g} \eun_{2g} \cdot \eun_{1g}, \label{GammaR_o2}
\end{eqnarray}
\begin{equation}
\fl \Gamma_\theta^{(1)} = \frac{2\mu_g}{3B_g} \rhobf \cdot
\nabla_{\bR_g} B_g \label{Gammatheta_o1}
\end{equation}
and
\begin{equation}
\fl \Gamma_\theta^{(2)} = \frac{\mu_g}{4B_g} \rhobf
\rhobf : \nabla_{\bR_g} \nabla_{\bR_g} \bB_g \cdot \bun_g.
\label{Gammatheta_o2}
\end{equation}

\subsection{Perturbative change of variables} \label{sub_gyro}

In this subsection we find a new set of coordinates
$\{\bR,u,\mu,\theta\}$ that makes the Lagrangian
\eq{lagrangian_drift_5} gyrophase independent. We employ the
procedure described in subsection~\ref{sub_procedure}. The
transformation $(\bR_g, v_{||g}, \mu_g, \theta_g)= T_{P,\epsilon}
(\bR, u, \mu, \theta, t)$ is customarily written in the form of a
Lie transform~\cite{cary81, brizard07},
\begin{equation}
T_{P,\epsilon} = T_1 T_2 T_3 \dots, \label{Tgyro}
\end{equation}
where
\begin{equation}
T_n = \exp \Bigg [ \epsilon^{n+1} \bR_{n+1} \cdot \nabla_\bR +
\epsilon^n \left( u_n \frac{\partial}{\partial u} +  \mu_n
\frac{\partial}{\partial \mu} + \theta_n \frac{\partial}{\partial
\theta}\right) \Bigg ].
\end{equation}
Instead of this form, we use the form in \eq{newvar} that we find
more convenient. The connection between the two arrangements is
trivial. To first order we find
\begin{equation}
\tilde{\bR}_2 = \bR_2, \; \tilde{u}_1 = u_1, \; \tilde{\mu}_1 =
\mu_1,\; \tilde{\theta}_1 = \theta_1.
\end{equation}
To second order, the relation is
\begin{equation}
\tilde{\bR}_3 = \bR_3 + \frac{1}{2} \bR_2 \cdot
\nabla_{(\bR_\bot/\epsilon)} \bR_2 + \frac{u_1}{2} \frac{\partial
\bR_2}{\partial u} + \frac{\mu_1}{2} \frac{\partial
\bR_2}{\partial \mu} + \frac{\theta_1}{2} \frac{\partial
\bR_2}{\partial \theta}, \label{tildeR3}
\end{equation}
\begin{equation}
\tilde{u}_2 = u_2 + \frac{1}{2} \bR_2 \cdot
\nabla_{(\bR_\bot/\epsilon)} u_1 + \frac{u_1}{2} \frac{\partial
u_1}{\partial u} + \frac{\mu_1}{2} \frac{\partial u_1}{\partial
\mu} + \frac{\theta_1}{2} \frac{\partial u_1}{\partial \theta},
\end{equation}
\begin{equation}
\tilde{\mu}_2 = \mu_2 + \frac{1}{2} \bR_2 \cdot
\nabla_{(\bR_\bot/\epsilon)} \mu_1 + \frac{u_1}{2} \frac{\partial
\mu_1}{\partial u} + \frac{\mu_1}{2} \frac{\partial
\mu_1}{\partial \mu} + \frac{\theta_1}{2} \frac{\partial
\mu_1}{\partial \theta}
\end{equation}
and
\begin{equation}
\tilde{\theta}_2 = \theta_2 + \frac{1}{2} \bR_2 \cdot
\nabla_{(\bR_\bot/\epsilon)} \theta_1 + \frac{u_1}{2}
\frac{\partial \theta_1}{\partial u} + \frac{\mu_1}{2}
\frac{\partial \theta_1}{\partial \mu} + \frac{\theta_1}{2}
\frac{\partial \theta_1}{\partial \theta}.
\end{equation}

In subsection \ref{ss_gyro_1}, the corrections $\bR_2$, $u_1$,
$\mu_1$ and $\theta_1$ are calculated following the procedure in
subsection~\ref{sub_procedure}, and the Lagrangian is obtained to
$O(\epsilon, \epsilon^2)$. In subsection \ref{ss_gyro_2} the
Lagrangian is obtained to next order. It is possible to do so
without explicitly obtaining $\tilde{\bR}_3$, $\tilde{u}_2$,
$\tilde{\mu}_2$ and $\tilde{\theta}_2$.

\subsubsection{Perturbative change of variables to first order.}
\label{ss_gyro_1}
We obtain the first-order gyrokinetic correction to the
Hamiltonian, $\overline{H}^{(1)}$, by employing equation
\eq{eq_Hbar1GK}. We need to know that $\langle \Gamma_\theta^{(1)}
(\bR, u, \mu, \theta) \rangle = 0$, $\langle H^{(1)}
(\bR_\bot/\epsilon, R_{||}, \mu, \theta, t) \rangle = \Lambda
\phiave (\bR_\bot/\lambda \epsilon, R_{||}, \mu/\lambda^2,
t/\tau)$ and
\begin{equation}
\langle \Gammabf_\bR^{(1)} (\bR, u, \mu, \theta) \rangle = \mu
\nabla_\bR \eun_2 \cdot \eun_1 + \frac{\mu}{2B} \bun \times
\nabla_\bR B - \frac{\mu}{2} \bun \bun \cdot \nabla_\bR \times
\bun,
\end{equation}
where we have used that
\begin{equation} \label{rhorhoave}
\langle \rhobf \rhobf \rangle = \frac{\mu}{B} ( \matI - \bun \bun
),
\end{equation}
with $\matI$ the unit matrix. For the remainder of the section,
whenever we do not write explicitly the arguments of the
functions, it will be understood that they are evaluated at
$(\bR,u,\mu,\theta)$, i.e. $\bun \equiv \bun(\bR)$,
$\phiave\equiv\phiave (\bR_\bot/\lambda \epsilon, R_{||},
\mu/\lambda^2, t/\tau)$, and so on. Substituting the values of
$\langle \Gammabf_\bR^{(1)} (\bR, u, \mu, \theta) \rangle$,
$\langle \Gamma_\theta^{(1)} (\bR, u, \mu, \theta) \rangle$ and
$\langle H^{(1)} (\bR_\bot/\epsilon, R_{||}, \mu, \theta, t)
\rangle$ into equation \eq{eq_Hbar1GK}, we find
\begin{equation}
\overline{H}^{(1)} = \Lambda \phiave + u \bun \cdot
\overline{\Gammabf}_\bR^{(1)} - u \mu \bun \cdot \nabla_\bR \eun_2
\cdot \eun_1 + \frac{u \mu}{2} \bun \cdot \nabla_\bR \times \bun .
\label{Hbar1_2}
\end{equation}
Notice that we have the freedom to choose
$\overline{\Gammabf}_\bR^{(1)}$ as we wish. Our choice will affect
the corrections $\bR_2$ and $u_1$, and the final form of
$\overline{H}^{(1)}$. To coincide with previous derivations in the
literature~\cite{brizard07}, we choose
\begin{equation}
\overline{\Gammabf}_\bR^{(1)} = \mu \nabla_\bR \eun_2 \cdot \eun_1
- \frac{\mu}{2} \bun \bun \cdot \nabla_\bR \times \bun,
\label{GammaRbar1_final}
\end{equation}
giving
\begin{equation}
\overline{H}^{(1)} = \Lambda \phiave. \label{Hbar1_final}
\end{equation}
In equation \eq{GammaRbar1_final}, we have chosen
$\overline{\Gammabf}_\bR^{(1)} = \mu \nabla_\bR \eun_2 \cdot
\eun_1 + \ldots$ instead of $\overline{\Gammabf}_\bR^{(1)} = \mu
\bun \bun \cdot \nabla_\bR \eun_2 \cdot \eun_1 + \ldots $ to
manifestly show that the equations of motion are independent of
the choice of $\eun_1$ and $\eun_2$ \cite{littlejohn88}.

The function $S^{(2)}_P$ can be obtained by solving equation
\eq{eq_S1GK}. Substituting equations \eq{H1}, \eq{GammaR_o1} and
\eq{Gammatheta_o1} into \eq{eq_S1GK} gives
\begin{eqnarray}
\fl \frac{\partial S^{(2)}_P}{\partial \theta} = - \frac{u^2}{B}
\bun \cdot \nabla_\bR \bun \cdot \rhobf + \frac{u}{4} \left [
\rhobf (\rhobf \times \bun) + (\rhobf \times \bun) \rhobf \right ]
: \nabla_\bR \bun - \frac{2\mu}{3B} \rhobf \cdot \nabla_\bR B -
\frac{\Lambda \phiwig}{B}, \label{dS2dtheta}
\end{eqnarray}
where we have used that
\begin{equation} \label{rhorhowig}
\rhobf \rhobf - \langle \rhobf \rhobf \rangle = \frac{1}{2} \left
[ \rhobf \rhobf - (\rhobf \times \bun) ( \rhobf \times \bun)
\right ].
\end{equation}
Integrating equation \eq{dS2dtheta} in the gyrophase gives
\begin{eqnarray}
\fl S^{(2)}_P = - \frac{u^2}{B} \bun \cdot \nabla_\bR \bun \cdot
(\rhobf \times \bun) - \frac{u}{8} \left [ \rhobf \rhobf - (\rhobf
\times \bun) (\rhobf \times \bun) \right ]: \nabla_\bR \bun
\nonumber \\ - \frac{2\mu}{3B} (\rhobf \times \bun) \cdot
\nabla_\bR B  - \frac{\Lambda \Phiwig}{B}, \label{S2}
\end{eqnarray}
where the function $\Phiwig$ is the integral
\begin{eqnarray}
\fl \Phiwig (\bR_\bot/\lambda \epsilon, R_{||}, \mu/\lambda^2,
\theta + \pi \Theta (- \lambda), t/\tau) \nonumber \\ =
\int^\theta d\theta^\prime\, \phiwig (\bR_\bot/\lambda \epsilon,
R_{||}, \mu/\lambda^2, \theta^\prime + \pi \Theta (- \lambda),
t/\tau)
\end{eqnarray}
such that $\langle \Phiwig \rangle = 0$. Here we have used that
$\rhobf = \partial (\rhobf \times \bun)/\partial \theta$ and
$\rhobf (\rhobf \times \bun) + (\rhobf \times \bun) \rhobf = -
(1/2)\partial [\rhobf \rhobf - (\rhobf \times \bun) (\rhobf \times
\bun) ]/\partial \theta$.

Using $S^{(2)}_P$ in the expressions \eq{R2_1}, \eq{u1_1},
\eq{mu1_1} and \eq{theta1_1}, the first order corrections to the
gyrokinetic variables become
\begin{eqnarray}
\fl \bR_2 = - \frac{2u}{B} \bun \bun \cdot \nabla_\bR \bun \cdot
(\rhobf \times \bun) - \frac{1}{8} \bun \left [ \rhobf \rhobf -
(\rhobf \times \bun) (\rhobf \times \bun) \right ]: \nabla_\bR
\bun - \frac{u}{B} \bun \times \nabla_\bR \bun \cdot \rhobf
\nonumber \\ - \frac{1}{2B} \rhobf \rhobf \cdot \nabla_\bR B -
\frac{\Lambda}{\lambda B^2} \bun \times \nabla_{(\bR_\bot/\lambda
\epsilon)} \Phiwig, \label{R2}
\end{eqnarray}
\begin{equation}
\fl u_1 = u \bun \cdot \nabla_\bR \bun \cdot \rhobf - \frac{B}{4}
\left [ \rhobf (\rhobf \times \bun) + (\rhobf \times \bun) \rhobf
\right ] : \nabla_\bR \bun, \label{u1}
\end{equation}
\begin{equation}
\fl \mu_1 = - \frac{u^2}{B} \bun \cdot \nabla_\bR \bun \cdot
\rhobf + \frac{u}{4} \left [ \rhobf (\rhobf \times \bun) + (\rhobf
\times \bun) \rhobf \right ]: \nabla_\bR \bun - \frac{\Lambda
\phiwig}{B} \label{mu1}
\end{equation}
and
\begin{eqnarray}
\fl \theta_1 = \frac{u^2}{2\mu B} \bun \cdot \nabla_\bR \bun \cdot
(\rhobf \times \bun) + \frac{u}{8\mu} \left [ \rhobf \rhobf -
(\rhobf \times \bun) (\rhobf \times \bun) \right ]: \nabla_\bR
\bun \nonumber \\ + \frac{1}{B} (\rhobf \times \bun) \cdot
\nabla_\bR B + \frac{\Lambda}{\lambda^2 B} \frac{\partial
\Phiwig}{\partial (\mu/\lambda^2)}. \label{theta1}
\end{eqnarray}
In \ref{app_comp} we show that this result is equivalent to the
result obtained with the iterative method in \cite{parra08}.

\subsubsection{Perturbative change of variables to second order.}
\label{ss_gyro_2}
In this subsection we apply the change of variables \eq{newvar} to
the Lagrangian \eq{lagrangian_drift_5} to $O(\epsilon^2,
\epsilon^3)$. The idea is to write an expression similar to
\eq{lagr_nGK} with $n = 2$ so that we can use the technique
demonstrated in subsection~\ref{sub_procedure}.

To $O(\epsilon^2, \epsilon^3)$, the Lagrangian
\eq{lagrangian_drift_5} becomes
\begin{eqnarray}
\fl \mathcal{L}^\bZ & = \Bigg [ \frac{1}{\epsilon} \bA + u \bun +
\epsilon \left ( \Gammabf_\bR^{(1)} + \bR_2 \cdot \nabla_\bR \bA +
u_1 \bun \right ) + \epsilon^2 \Bigg ( \Gammabf_\bR^{(2)} +
\tilde{\bR}_3 \cdot \nabla_\bR \bA + \tilde{u}_2 \bun \nonumber \\
\fl & + u \bR_2 \cdot \nabla_\bR \bun + u_1 \frac{\partial
\Gammabf_\bR^{(1)}}{\partial u} + \mu_1 \frac{\partial
\Gammabf_\bR^{(1)}}{\partial \mu} + \theta_1 \frac{\partial
\Gammabf_\bR^{(1)}}{\partial \theta} \Bigg ) \Bigg ] \cdot
\frac{d\bR}{dt} \nonumber \\ \fl & + \Bigg [ \epsilon \bA +
\epsilon^2 u \bun + \epsilon^3 \left ( \Gammabf_\bR^{(1)} + \bR_2
\cdot \nabla_\bR \bA + u_1 \bun \right ) \Bigg ] \cdot
\frac{d\bR_2}{dt} + \Bigg ( \epsilon^2 \bA + \epsilon^3 u \bun
\Bigg ) \cdot \frac{d\tilde{\bR}_3}{dt} \nonumber \\ \fl & + \Bigg
[ - \epsilon \mu + \epsilon^2 \left ( - \mu_1 +
\Gamma_\theta^{(1)} \right ) + \epsilon^3 \Bigg ( - \tilde{\mu}_2
+ \Gamma_\theta^{(2)} + \mu_1 \frac{\partial
\Gamma_\theta^{(1)}}{\partial \mu} + \theta_1 \frac{\partial
\Gamma_\theta^{(1)}}{\partial \theta} \Bigg ) \Bigg ]
\frac{d\theta}{dt} \nonumber \\ \fl & + \Bigg [ - \epsilon^2 \mu +
\epsilon^3 \left ( - \mu_1 + \Gamma_\theta^{(1)} \right ) \Bigg ]
\frac{d\theta_1}{dt} - \epsilon^3 \mu \frac{d\tilde{\theta}_2}{dt}
- H^{(0)} - \epsilon \left ( u u_1 + \mu_1 B + H^{(1)} \right )
\nonumber \\ \fl & - \epsilon^2 \Bigg ( u \tilde{u}_2 +
\frac{u_1^2}{2} + \tilde{\mu}_2 B + \mu \bR_2 \cdot \nabla_\bR B +
\bR_2 \cdot \nabla_{(\bR_\bot/\epsilon)} H^{(1)} + \mu_1
\frac{\partial H^{(1)}}{\partial \mu} + \theta_1 \frac{\partial
H^{(1)}}{\partial \theta} \Bigg ) \nonumber \\ \fl & +
O(\epsilon^3, \epsilon^4), \label{lagrangian_gyro_o2_1}
\end{eqnarray}
where we have used that $\partial \Gamma_\theta^{(1)} /\partial u
= 0$, that $\partial H^{(1)}/\partial u = 0$ and that
$\Gammabf_\bR^{(1)}$ and $\Gamma_\theta^{(1)}$ only depend slowly
on $\bR$.

We have seen in subsection \ref{sub_procedure} that to first order
we need to add to the Lagrangian \eq{lagrangian_gyro_o2_1} the
time derivative
\begin{eqnarray}
\frac{d}{dt} \left ( \epsilon^2 S^{(2)}_P - \epsilon \bA \cdot
\bR_2 - \epsilon^2 u \bun \cdot \bR_2 + \epsilon^2 \mu \theta_1
\right ),
\end{eqnarray}
giving as a result the Lagrangian
\begin{eqnarray}
\fl \mathcal{L}^\bZ & + \frac{d}{dt} \left ( \epsilon^2 S^{(2)}_P
- \epsilon \bA \cdot \bR_2 - \epsilon^2 u \bun \cdot \bR_2 +
\epsilon^2 \mu \theta_1 \right ) \nonumber \\ \fl & = \left [
\frac{1}{\epsilon} \bA + u \bun + \epsilon
\overline{\Gammabf}_\bR^{(1)} + \epsilon^2 \left ( \tilde{\bR}_3
\cdot \nabla_\bR \bA + \tilde{u}_2 \bun +
\tilde{\Gammabf}_\bR^{(2)} \right ) \right ] \cdot \frac{d\bR}{dt}
+ \left ( \epsilon^2 \bA + \epsilon^3 u \bun \right ) \cdot
\frac{d\tilde{\bR}_3}{dt} \nonumber \\ \fl & + \epsilon^3
\tilde{\Gamma}_u^{(2)} \frac{du}{dt} + \epsilon^3
\tilde{\Gamma}_\mu^{(2)} \frac{d\mu}{dt} + \left [ - \epsilon \mu
+ \epsilon^3 \left ( - \tilde{\mu}_2 + \tilde{\Gamma}_\theta^{(2)}
\right ) \right ] \frac{d\theta}{dt} - \epsilon^3 \mu
\frac{d\tilde{\theta}_2}{dt} -
\overline{H}^{(0)} - \epsilon \overline{H}^{(1)} \nonumber \\
\fl & - \epsilon^2 \left [ u \tilde{u}_2 + \tilde{\mu}_2 B +
\tilde{H}^{(2)} \right ] + O(\epsilon^3, \epsilon^4),
\label{lagrangian_gyro_o2_2}
\end{eqnarray}
with
\begin{eqnarray} \label{Gammat_R_o2}
\fl \tilde{\Gammabf}_\bR^{(2)} = \Gammabf_\bR^{(2)} + u
(\nabla_\bR \times \bun) \times \bR_2 + u_1 \frac{\partial
\Gammabf_\bR^{(1)}}{\partial u} + \mu_1 \frac{\partial
\Gammabf_\bR^{(1)}}{\partial \mu} + \theta_1 \frac{\partial
\Gammabf_\bR^{(1)}}{\partial \theta} + \nabla_\bR S^{(2)}_P
\nonumber\\ + \nabla_{(\bR_\bot/\epsilon)} \bR_2 \cdot \left (
\Gammabf_\bR^{(1)} + \bR_2 \cdot \nabla_\bR \bA + u_1 \bun \right
) + \left ( - \mu_1 + \Gamma_\theta^{(1)} \right )
\nabla_{(\bR_\bot/\epsilon)} \theta_1,
\end{eqnarray}
\begin{equation}
\fl \tilde{\Gamma}_u^{(2)} = \left ( \Gammabf_\bR^{(1)} + \bR_2
\cdot \nabla_\bR \bA + u_1 \bun \right ) \cdot \frac{\partial
\bR_2}{\partial u} + \left ( - \mu_1 + \Gamma_\theta^{(1)} \right
) \frac{\partial \theta_1}{\partial u},
\end{equation}
\begin{equation}
\fl \tilde{\Gamma}_\mu^{(2)} = \left ( \Gammabf_\bR^{(1)} + \bR_2
\cdot \nabla_\bR \bA + u_1 \bun \right ) \cdot \frac{\partial
\bR_2}{\partial \mu} + \left ( - \mu_1 + \Gamma_\theta^{(1)}
\right ) \frac{\partial \theta_1}{\partial \mu},
\end{equation}
\begin{eqnarray} \label{Gammat_theta_o2}
\fl \tilde{\Gamma}_\theta^{(2)} = \Gamma_\theta^{(2)} + \mu_1
\frac{\partial \Gamma_\theta^{(1)}}{\partial \mu} + \theta_1
\frac{\partial \Gamma_\theta^{(1)}}{\partial \theta} + \left (
\Gammabf_\bR^{(1)} + \bR_2 \cdot \nabla_\bR \bA + u_1 \bun \right
) \cdot \frac{\partial \bR_2}{\partial \theta} \nonumber \\ +
\left ( - \mu_1 + \Gamma_\theta^{(1)} \right ) \frac{\partial
\theta_1}{\partial \theta}
\end{eqnarray}
and
\begin{equation} \label{Ht_o2}
\fl \tilde{H}^{(2)} = \frac{u_1^2}{2} + \mu \bR_2 \cdot \nabla_\bR
B + \bR_2 \cdot \nabla_{(\bR_\bot/\epsilon)} H^{(1)} + \mu_1
\frac{\partial H^{(1)}}{\partial \mu} + \theta_1 \frac{\partial
H^{(1)}}{\partial \theta} - \frac{\partial S^{(2)}_P}{\partial t}.
\end{equation}
Notice that in equation \eq{Gammat_R_o2} we are using
\begin{equation}
\bR_2 \cdot \nabla_\bR \bun - \nabla_\bR \bun \cdot \bR_2 =
(\nabla_\bR \times \bun) \times \bR_2,
\end{equation}
and we have taken into account that $S^{(2)}_P$ depends on $\bR$
in two different ways. On the one hand, there is the dependence on
$\bR_\bot/\epsilon$, that was the only dependence that was taken
into account in subsection~\ref{ss_gyro_1}. This dependence gives
the strong gradient $\nabla_{(\bR_\bot/\epsilon)} S^{(2)}_P = -
(\Lambda/\lambda B) \nabla_{(\bR_\bot/\lambda \epsilon)} \Phiwig$.
On the other hand there is a slow dependence on $\bR$ that gives
the gradient
\begin{eqnarray}
\fl \nabla_\bR S^{(2)}_P = \nabla_\bR \Bigg [ - \frac{u^2}{B} \bun
\cdot \nabla_\bR \bun \cdot (\rhobf \times \bun) - \frac{u}{8}
\left ( \rhobf \rhobf - (\rhobf \times \bun) (\rhobf \times \bun)
\right ): \nabla_\bR \bun \nonumber \\ - \frac{2\mu}{3B} (\rhobf
\times \bun) \cdot \nabla_\bR B \Bigg ] + \frac{\Lambda
\Phiwig}{B^2} \nabla_\bR B - \frac{\Lambda}{B} \bun \bun \cdot
\nabla_\bR \tilde\Phi. \label{nablaS2_slow}
\end{eqnarray}

In subsection~\ref{sub_procedure} we showed that by adding the
total time derivative
\begin{eqnarray}
\frac{d}{dt} \left ( \epsilon^3 S^{(3)}_P (\bR_\bot/\epsilon, \bR,
u, \mu, \theta, t) - \epsilon^2 \bA \cdot \tilde{\bR}_3 -
\epsilon^3 u \bun \cdot \tilde{\bR}_3 + \epsilon^3 \mu
\tilde{\theta}_2 \right ) \label{add_lagrange_4}
\end{eqnarray}
to the Lagrangian \eq{lagrangian_gyro_o2_2} and making the result
equal to
\begin{equation}\label{eq:lagrangiangyroorderepsilon2}
\fl \overline{\mathcal{L}} = \left [ \frac{1}{\epsilon} \bA(\bR) +
u \bun (\bR) + \epsilon \overline{\Gammabf}_\bR^{(1)} \right ]
\cdot \frac{d\bR}{dt} - \epsilon \mu \frac{d\theta}{dt} -
\overline{H}^{(0)} - \epsilon \overline{H}^{(1)} - \epsilon^2
\overline{H}^{(2)} + O(\epsilon^3, \epsilon^4),
\end{equation}
where we have explicitly set $\overline{\Gammabf}^{(2)}_\bR = 0$,
we obtain the equations for the corrections
\begin{equation}
\fl \tilde{\bR}_3 = \left ( \tilde{\Gamma}_u^{(2)} +
\frac{\partial S^{(3)}_P}{\partial u} \right ) \bun (\bR) +
\frac{1}{B(\bR)} \bun (\bR) \times \left (
\tilde{\Gammabf}_\bR^{(2)} + \nabla_{(\bR_\bot/\epsilon)}
S^{(3)}_P \right ), \label{R3}
\end{equation}
\begin{equation}
\tilde{u}_2 = - \bun (\bR) \cdot \tilde{\Gammabf}_\bR^{(2)},
\label{u2}
\end{equation}
\begin{equation}
\tilde{\mu}_2 = \tilde{\Gamma}_\theta^{(2)} + \frac{\partial
S^{(3)}_P}{\partial \theta} \label{mu2}
\end{equation}
and
\begin{equation}
\tilde{\theta}_2 = - \tilde{\Gamma}_\mu^{(2)} - \frac{\partial
S^{(3)}_P}{\partial \mu}. \label{theta2}
\end{equation}
The correction to the Hamiltonian is
\begin{equation}
\overline{H}^{(2)} = - u \bun(\bR) \cdot \left \langle
\tilde{\Gammabf}_\bR^{(2)} \right \rangle + B(\bR) \left \langle
\tilde{\Gamma}_\theta^{(2)} \right \rangle + \left \langle
\tilde{H}^{(2)} \right \rangle, \label{Hbar2_1}
\end{equation}
and the equation for $S^{(3)}_P$ is
\begin{eqnarray}
\fl \frac{\partial S^{(3)}_P}{\partial \theta} = \frac{u}{B(\bR)}
\bun (\bR) \cdot \left ( \tilde{\Gammabf}_\bR^{(2)} - \left
\langle \tilde{\Gammabf}_\bR^{(2)} \right \rangle \right ) - \left
( \tilde{\Gamma}_\theta^{(2)} - \left \langle
\tilde{\Gamma}_\theta^{(2)} \right \rangle \right ) \nonumber \\ -
\frac{1}{B(\bR)} \left ( \tilde{H}^{(2)} - \left \langle
\tilde{H}^{(2)} \right \rangle \right ).
\end{eqnarray}
In this article we do not solve for $S^{(3)}_P$ and hence we are
not able to write explicitly the corrections $\tilde{\bR}_3$,
$\tilde{u}_2$, $\tilde{\mu}_2$ and $\tilde{\theta}_2$. We only
obtain explicitly the correction $\overline{H}^{(2)}$. In
\ref{app_gyro_2} we evaluate equation \eq{Hbar2_1} in detail. The
final result is
\begin{eqnarray}
\fl \overline{H}^{(2)} = \Lambda^2 \Psi^{(2)}_\phi
(\bR_\bot/\lambda \epsilon, \bR, \mu/\lambda^2, t/\tau, \lambda) +
\Lambda \Psi^{(2)}_{\phi B} (\bR_\bot/\lambda \epsilon, \bR, u,
\mu, \mu/\lambda^2, t/\tau, \lambda) \nonumber \\ + \Psi^{(2)}_B
(\bR, u, \mu), \label{hamiltonian_o2_2}
\end{eqnarray}
with
\begin{eqnarray}
\fl \Psi^{(2)}_\phi = \frac{1}{2 \lambda^2 B^2} \left \langle
\nabla_{(\bR_\bot/\lambda \epsilon)} \Phiwig \cdot \left ( \bun
\times \nabla_{(\bR_\bot/\lambda \epsilon)} \phiwig \right )
\right \rangle - \frac{1}{2 \lambda^2 B} \frac{\partial \langle
\phiwig^2 \rangle}{\partial (\mu/\lambda^2)}, \label{Psi2_phi}
\end{eqnarray}
\begin{eqnarray}
\fl \Psi^{(2)}_{\phi B} = & - \frac{u}{\lambda B} \left \langle
\left ( \nabla_{(\bR_\bot/\lambda \epsilon)} \phiwig \times \bun
\right ) \cdot \nabla_\bR \bun \cdot \rhobf \right \rangle -
\frac{\mu}{2 \lambda B^2} \nabla_\bR B \cdot
\nabla_{(\bR_\bot/\lambda \epsilon)} \phiave \nonumber \\ \fl & -
\frac{1}{4 \lambda B} \left \langle \nabla_{(\bR_\bot/ \lambda
\epsilon)} \phiwig \cdot \left [ \rhobf \rhobf - ( \rhobf \times
\bun ) (\rhobf \times \bun) \right ] \cdot \nabla_\bR B \right
\rangle - \frac{1}{B} \nabla_\bR B \cdot \langle \phiwig\, \rhobf
\rangle \nonumber \\ \fl &  - \frac{u^2}{\lambda^2 B} \bun \cdot
\nabla_\bR \bun \cdot \left \langle \frac{\partial
\phiwig}{\partial (\mu/\lambda^2)}\, \rhobf \right \rangle -
\frac{u^2}{2 \mu B} \bun \cdot \nabla_\bR \bun \cdot \langle
\phiwig\, \rhobf \rangle \nonumber\\ \fl & + \frac{u}{4 \lambda^2}
\nabla_\bR \bun : \left \langle \frac{\partial \phiwig}{\partial
(\mu/\lambda^2)}\, \left [ \rhobf ( \rhobf \times \bun ) + (\rhobf
\times \bun) \rhobf \right ] \right \rangle \nonumber \\ \fl & +
\frac{u}{4 \mu} \nabla_\bR \bun : \left \langle \phiwig\, \left [
\rhobf ( \rhobf \times \bun ) + (\rhobf \times \bun) \rhobf \right
] \right \rangle \label{Psi2_phiB}
\end{eqnarray}
and
\begin{eqnarray}
\fl \Psi^{(2)}_B = & - \frac{3u^2 \mu}{2B^2} \bun \cdot \nabla_\bR
\bun \cdot \nabla_\bR B + \frac{\mu^2}{4B} (\matI - \bun \bun) :
\nabla_\bR \nabla_\bR \bB \cdot \bun - \frac{3\mu^2}{4B^2}
|\nabla_{\bR\bot} B|^2 \nonumber \\ \fl & + \frac{u^2 \mu}{2B}
\nabla_\bR \bun : \nabla_\bR \bun + \left(\frac{\mu^2}{8} -
\frac{u^2 \mu}{4B}\right) \nabla_{\bR\bot} \bun : (\nabla_{\bR\bot}
\bun)^\mathrm{T} - \left(\frac{3 u^2 \mu}{8B} +
\frac{\mu^2}{16}\right) (\nabla_\bR \cdot \bun)^2 \nonumber \\ \fl
& + \left(\frac{3 u^2 \mu}{2B}-\frac{u^4}{2B^2}\right) |\bun \cdot
\nabla_\bR \bun|^2 + \left(\frac{ u^2 \mu}{8B} -
\frac{\mu^2}{16}\right) (\bun \cdot \nabla_\bR \times \bun)^2,
\label{Psi2_B}
\end{eqnarray}
where $\matrixtop{\mathbf{M}}^\mathrm{T}$ is the transpose of the
matrix $\matrixtop{\mathbf{M}}$. The final phase-space Lagrangian
is given then by \eq{eq:lagrangiangyroorderepsilon2}. We can write
it explicitly as
\begin{equation} \label{finalL}
\fl \overline{\mathcal{L}} = \left [ \frac{1}{\epsilon} \bA (\bR)
+ u \bun (\bR) - \epsilon \mu \bK (\bR) \right ] \cdot
\frac{d\bR}{dt} - \epsilon \mu \frac{d\theta}{dt} - \overline{H}+
O(\epsilon^3, \epsilon^4),
\end{equation}
where
\begin{eqnarray} \label{finalH}
\fl \overline{H} = \frac{1}{2} u^2 + \mu B(\bR) + \Lambda \epsilon
\phiave ( \bR_\bot/\lambda \epsilon, R_{||}, \mu/\lambda^2,
t/\tau) + \Lambda^2 \epsilon^2 \Psi^{(2)}_\phi (\bR_\bot/\lambda
\epsilon, \bR, \mu/\lambda^2, t/\tau, \lambda) \nonumber \\ +
\Lambda \epsilon^2 \Psi^{(2)}_{\phi B} (\bR_\bot/\lambda \epsilon,
\bR, u, \mu, \mu/\lambda^2, t/\tau, \lambda) + \epsilon^2
\Psi^{(2)}_B (\bR, u, \mu)
\end{eqnarray}
and
\begin{equation}
\bK (\bR) = \frac{1}{2} \bun(\bR) \bun(\bR) \cdot \nabla_\bR
\times \bun(\bR) - \nabla_\bR \eun_2 (\bR) \cdot \eun_1 (\bR).
\end{equation}
In previous work \cite{hahm88, brizard07}, only the contribution
$\Psi^{(2)}_\phi$ was kept because the terms that contained the
function $\phi$ were assumed to be larger. With the more natural
ordering \eq{orderings}, we find the new contributions
$\Psi^{(2)}_{\phi B}$ and $\Psi^{(2)}_B$, demonstrating that
magnetic geometry and electrostatic potential appear together and
cannot be separated. In Sections~\ref{sec:GyroEqsMotion},
\ref{sec:GyroPoissonEq} and \ref{sec:fieldtheory} we show that
$\Psi^{(2)}_{\phi B}$ and $\Psi^{(2)}_B$ modify both the equations
of motion and Poisson's equation.

We end this section pointing out that we chose the final form of
the Lagrangian \eq{finalL}, with $\overline{\Gammabf}^{(2)}_\bR =
0$, to have the same Poisson brackets as previous authors
\cite{brizard07}. There are other possible choices, e.g., making
the second order correction of the Hamiltonian independent of the
parallel velocity, condition that can be achieved by defining the
appropriate $\overline{\Gammabf}_\bR^{(2)}$.

\section{Gyrokinetic equations of motion and Vlasov equation}
\label{sec:GyroEqsMotion}

The equations of motion are given by \eq{ddt_poibra}. To obtain
them explicitly we need to find the Poisson bracket \eq{poi_bra}
that corresponds to the Lagrangian \eq{finalL}. Employing
\ref{app:poissonbrackets} we find the Poisson bracket to be
\begin{eqnarray} \label{eq:poissonbracket}
\fl \{F,G\} = \frac{1}{\epsilon}\left(\frac{\partial
F}{\partial\mu} \frac{\partial G}{\partial\theta} - \frac{\partial
F}{\partial\theta} \frac{\partial G}{\partial\mu}\right) +
\frac{{\bf B}^*}{B_{||}^*} \cdot
\left(\nabla^*_{\bR}F\frac{\partial G}{\partial u}- \frac{\partial
F}{\partial u}\nabla_{\bR}^*G\right) \nonumber \\ + \frac{\epsilon
}{B_{||}^*} \nabla_{\bR}^*F \cdot ( \bun \times \nabla_{\bR}^*G ),
\end{eqnarray}
where
\begin{equation} \label{Bstar}
\bB^* (\bR, u, \mu) := \bB(\bR) + \epsilon u \nabla_\bR \times
\bun (\bR) - \epsilon^2 \mu \nabla_\bR \times \bK (\bR),
\end{equation}
\begin{eqnarray} \label{Bstar_par}
\fl B^*_{||} (\bR, u, \mu) := \bB^* (\bR, u, \mu) \cdot \bun (\bR)
\nonumber \\ = B(\bR) + \epsilon u \bun(\bR) \cdot \nabla_\bR
\times \bun (\bR) - \epsilon^2 \mu \bun(\bR) \cdot \nabla_\bR
\times \bK (\bR)
\end{eqnarray}
and
\begin{equation}
\nabla_{\bR}^* := \nabla_{\bR} - {\bf
K}(\bR)\frac{\partial}{\partial\theta}.
\end{equation}

Employing the Poisson bracket in \eq{eq:poissonbracket} and the
Hamiltonian in \eq{finalH}, we find
\begin{eqnarray} \label{dRdt}
\fl \dot{\bR} \equiv \frac{d\bR}{dt} & = \left ( u + \Lambda
\epsilon^2 \frac{\partial \Psi^{(2)}_{\phi B}}{\partial u} +
\epsilon^2 \frac{\partial \Psi^{(2)}_B}{\partial u} \right )
\frac{\bB^*}{B_{||}^*} + \frac{1}{B_{||}^*} \bun \times \Bigg (
\epsilon \mu \nabla_\bR B + \frac{\Lambda \epsilon}{\lambda}
\nabla_{(\bR_\perp/\lambda \epsilon)} \phiave \nonumber \\ \fl & +
\frac{\Lambda^2 \epsilon^2}{\lambda} \nabla_{(\bR_\bot/\lambda
\epsilon)} \Psi^{(2)}_\phi + \frac{\Lambda \epsilon^2}{\lambda}
\nabla_{(\bR_\bot/\lambda \epsilon)} \Psi^{(2)}_{\phi B} +
\Lambda^2 \epsilon^3 \nabla_\bR \Psi^{(2)}_\phi
+ \Lambda \epsilon^3 \nabla_\bR \Psi^{(2)}_{\phi B} \nonumber \\
\fl & + \epsilon^3 \nabla_\bR \Psi^{(2)}_B \Bigg ),
\end{eqnarray}
\begin{eqnarray} \label{dudt}
\fl \dot{u} \equiv \frac{d u}{d t} &= - \frac{\mu}{B^*_{||}} \bB^*
\cdot \nabla_\bR B - \Lambda \epsilon \bun \cdot \nabla_\bR
\phiave - \Lambda^2 \epsilon^2 \bun \cdot \nabla_\bR
\Psi^{(2)}_\phi - \Lambda \epsilon^2 \bun \cdot \nabla_\bR
\Psi^{(2)}_{\phi B} \nonumber \\ \fl & - \epsilon^2 \bun \cdot
\nabla_\bR \Psi^{(2)}_B - \frac{1}{B_{||}^*} [ u \bun \times (\bun
\cdot \nabla_\bR \bun) - \epsilon \mu (\nabla_\bR \times \bK)_\bot
] \cdot \Bigg ( \frac{\Lambda \epsilon}{\lambda}
\nabla_{(\bR_\bot/\lambda \epsilon)} \phiave \nonumber \\ \fl & +
\frac{\Lambda^2 \epsilon^2}{\lambda} \nabla_{(\bR_\bot/\lambda
\epsilon)} \Psi^{(2)}_\phi + \frac{\Lambda \epsilon^2}{\lambda}
\nabla_{(\bR_\bot/\lambda \epsilon)} \Psi^{(2)}_{\phi B} +
\Lambda^2 \epsilon^3 \nabla_\bR \Psi^{(2)}_\phi + \Lambda
\epsilon^3 \nabla_\bR \Psi^{(2)}_{\phi B} \nonumber \\ \fl & +
\epsilon^3 \nabla_\bR \Psi^{(2)}_B \Bigg) ,
\end{eqnarray}
\begin{eqnarray} \label{dmudt}
\fl \dot{\mu} \equiv \frac{d \mu}{d t}=0
\end{eqnarray}
and
\begin{eqnarray} \label{dthetadt}
\fl \dot{\theta} \equiv \frac{d \theta}{d t} & = -
\frac{1}{\epsilon} B - \frac{\Lambda}{\lambda^2} \frac{\partial
\phiave}{\partial (\mu/\lambda^2)} - \frac{\Lambda^2
\epsilon}{\lambda^2} \frac{\partial \Psi^{(2)}_\phi}{\partial
(\mu/\lambda^2)} - \Lambda \epsilon \frac{\partial
\Psi^{(2)}_{\phi B}}{\partial \mu} - \frac{\Lambda
\epsilon}{\lambda^2} \frac{\partial \Psi^{(2)}_{\phi B}}{\partial
(\mu/\lambda^2)} - \epsilon \frac{\partial
\Psi^{(2)}_B}{\partial\mu} \nonumber \\ \fl & - \frac{\bB^* \cdot
\bK}{B_{||}^*} \Bigg( u + \Lambda \epsilon^2 \frac{\partial
\Psi^{(2)}_{\phi B}}{\partial u} + \epsilon^2 \frac{\partial
\Psi^{(2)}_B}{\partial u} \Bigg) - \frac{1}{B_{||}^*} (\bK \times
\bun) \cdot \Big ( \epsilon \mu \nabla_\bR B \nonumber \\ \fl & +
\frac{\Lambda \epsilon}{\lambda} \nabla_{(\bR_\perp/\lambda
\epsilon)} \phiave + \frac{\Lambda^2 \epsilon^2}{\lambda}
\nabla_{(\bR_\bot/\lambda \epsilon)} \Psi^{(2)}_\phi +
\frac{\Lambda \epsilon^2}{\lambda} \nabla_{(\bR_\bot/\lambda
\epsilon)} \Psi^{(2)}_{\phi B} + \Lambda^2 \epsilon^3 \nabla_\bR
\Psi^{(2)}_\phi \nonumber \\ \fl & + \Lambda \epsilon^3 \nabla_\bR
\Psi^{(2)}_{\phi B} + \epsilon^3 \nabla_\bR \Psi^{(2)}_B \Big ).
\end{eqnarray}
Note that we have emphasized the fact that the dependence of the
functions $\phiave$, $\Psi^{(2)}_{\phi}$ and $\Psi^{(2)}_{\phi B}$
on $\bR$ and $\mu$ (recall equation \eq{finalH}) can be fast or
slow. For this reason we distinguish between derivatives with
respect to the argument $\bR_\bot/\lambda \epsilon$ and
derivatives with respect to the argument $\bR$, and between
derivatives with respect to the argument $\mu$ and derivatives
with respect to the argument $\mu/\lambda^2$.

The new correction to the Hamiltonian $\Psi^{(2)}_{\phi B}$ gives
a contribution of order $\epsilon^2$ to the perpendicular and
parallel motion of the gyrocenter, comparable to the contribution
from $\Psi^{(2)}_\phi$, the term that is usually kept. The
correction to the Hamiltonian $\Psi^{(2)}_B$ gives a negligible
contribution to the perpendicular drift, but is needed for the
parallel motion. Thus, both corrections must be kept to obtain the
equations of motion to order $\epsilon^2$.

It is worth mentioning that equation \eq{dRdt} contains the
Ba\~nos drift \cite{banos67} in the definition of $u$. To make it
clear, instead of choosing the first order Lagrangian as in
\eq{lagr_n1_2_final} with $\overline{\Gammabf}^{(1)}_\bR$ given in
\eq{GammaRbar1_final}, we can choose it to have
\begin{equation}
\overline{\Gammabf}_\bR^{(1)\prime} = \mu \nabla_\bR \eun_2 \cdot
\eun_1 + \frac{\mu}{2} \bun \cdot \nabla_\bR \times \bun =
\overline{\Gammabf}_\bR^{(1)} + \mu \bun \cdot \nabla_\bR \times
\bun.
\end{equation}
This choice gives a different parallel velocity $u^\prime = u -
\epsilon \mu \bun \cdot \nabla_\bR \times \bun$ and a different
first order Hamiltonian $\overline{H}^{(1)\prime} =
\overline{H}^{(1)} + u \mu \bun \cdot \nabla_\bR \times \bun$.
With this new choice, the equation for $\dot{\bR}$ to first order
is
\begin{eqnarray}
\fl \dot{\bR} = (u^\prime + \epsilon \mu \bun \cdot \nabla_\bR
\times \bun) \bun + \frac{\epsilon \mu}{B} \bun \times \nabla_\bR
B + \frac{\epsilon (u^\prime)^2}{B} \bun \times (\bun \cdot
\nabla_\bR \bun) \nonumber \\ - \frac{\Lambda \epsilon}{\lambda B}
\nabla_{(\bR_\bot/\lambda \epsilon)} \phiave \times \bun +
O(\epsilon^2).
\end{eqnarray}
Note that the Ba\~nos drift has been made explicit. From here on,
we work only with our equations of motion \eq{dRdt}, \eq{dudt},
\eq{dmudt} and \eq{dthetadt} that are equivalent to the equations
obtained with this alternative choice that makes the Ba\~nos drift
manifest.

The gyrokinetic Vlasov equation is readily written for the
phase-space distribution $F (\bR, u, \mu, \theta, t)$ in
gyrokinetic coordinates, giving
\begin{equation}\label{eq:Vlasov}
\frac{\partial F}{\partial t} + \dot \bR \cdot \nabla_\bR F + \dot
u \frac{\partial F}{\partial u} + \dot \theta \frac{\partial
F}{\partial \theta} = 0,
\end{equation}
or employing the Poisson bracket,
\begin{equation}
\frac{\partial F}{\partial t} + \{F, \overline{H} \} = 0.
\end{equation}
In the absence of collisions and making use of the fact that $\dot
\bR$, $\dot u$ and $\dot \theta$ are independent of gyrophase, the
gyrophase independent piece of the distribution function $\langle
F \rangle$ and the gyrophase dependent piece $\tilde{F} = F -
\langle F \rangle$ are determined by two decoupled
equations~\cite{dubin83}, namely,
\begin{equation} \label{Vlasov_ave}
\frac{\partial \langle F \rangle}{\partial t} + \dot \bR \cdot
\nabla_\bR \langle F \rangle + \dot u \frac{\partial \langle F
\rangle}{\partial u} = 0
\end{equation}
and
\begin{equation}
\frac{\partial \tilde{F}}{\partial t} + \dot \bR \cdot \nabla_\bR
\tilde{F} + \dot u \frac{\partial \tilde{F}}{\partial u} + \dot
\theta \frac{\partial \tilde{F}}{\partial \theta} = 0.
\end{equation}
Since $\dot \theta \sim 1/\epsilon \gg
\partial/\partial t,\, \dot \bR \cdot
\nabla_\bR,\, \dot u (\partial/\partial u)$, it is usually assumed
that any dependence on the gyrophase disappears in a very short
time and $\tilde{F}$ can be neglected. As a result, we assume that
$F = \langle F \rangle$. By zeroing the gyrophase-dependent piece
of the distribution function we are eliminating the gyrofrequency
time scales, a crucial step to obtain the gyrokinetic equation. In
the presence of collisions it is possible to estimate the size of
$\tilde{F}$ and argue that it is small \cite{parra08,
Brizard2004}.

It is possible to write the gyrokinetic equation in conservative
form. Indeed, the equations of motion obtained from a phase-space
Lagrangian conserve phase-space volume. Therefore, the determinant
of the Jacobian matrix of the gyrokinetic transformation,
$J_{T_\epsilon}$, found in \ref{app:jacobians} to be
$\mbox{det}(J_{T_\epsilon}) = B_{||}^*$, satisfies the condition
\begin{equation} \label{consps}
\nabla_\bR \cdot \left ( B_{||}^* \dot \bR \right ) +
\frac{\partial}{\partial u} \left ( B_{||}^* \dot u \right ) +
\frac{\partial}{\partial \theta} \left ( B_{||}^* \dot \theta
\right ) = 0.
\end{equation}
For completeness, we prove this equation in \ref{app_consps}.
Since $\dot \theta$ and $B_{||}^*$ do not depend on $\theta$, this
equation reduces to
\begin{equation} \label{consps_2}
\nabla_\bR \cdot \left ( B_{||}^* \dot \bR \right ) +
\frac{\partial}{\partial u} \left ( B_{||}^* \dot u \right ) = 0.
\end{equation}
Using this expression and equation \eq{Vlasov_ave}, and taking
into account that our choice of Lagrangian \eq{finalL} implies
that $\partial B^*_{||}/\partial t = 0$, we find the Vlasov
equation in conservative form, i.e.,
\begin{equation} \label{Vlasov_ave_cons}
\frac{\partial}{\partial t} \left ( B_{||}^*  \langle F \rangle
\right ) +  \nabla_\bR \cdot \left ( B_{||}^* \dot \bR \langle F
\rangle \right ) + \frac{\partial}{\partial u} \left ( B_{||}^*
\dot u \langle F \rangle \right ) = 0.
\end{equation}

\section{Gyrokinetic Poisson's equation}
\label{sec:GyroPoissonEq}

In Sections~\ref{sect_theory} and \ref{sec:GyroEqsMotion} we have
obtained the gyrokinetic equations of motion and the gyrokinetic
Vlasov equation keeping the electrostatic potential as an unspecified
function. The system of equations of electrostatic gyrokinetic theory
is closed by coupling the gyrokinetic Vlasov equation to Poisson's
equation, which is the subject of study of this section.

First, in subsection~\ref{normalization_PoissonEq} we present the
normalization that we employ for Poisson's equation.  Since
several species enter in Poisson's equation, the normalization
given in Section~\ref{sect_order} must be modified. With this new
normalization, we obtain the gyrokinetic Poisson's equation in
subsection~\ref{sec:gyroPoissonChangeVar} by simply changing from
the coordinates $\boldr$ and $\bv$ to their gyrokinetic
counterparts $\bR$, $u$, $\mu$ and $\theta$. In
Section~\ref{sec:fieldtheory} we provide another way of obtaining
Poisson's equation that is based on field theory.

\subsection{Normalized Lagrangian in a system with several species}
\label{normalization_PoissonEq}

Poisson's equation in Gaussian units reads
\begin{eqnarray}\label{eq:Poisson}
\fl \nabla^2 \varphi (\boldr,t)= -4\pi \Bigg [ e\int d ^3 v_i\,
f_i(\boldr,\bv_i,t) + \sum_I Z_I e\int d ^3 v_I\,
f_I(\boldr,\bv_I,t) \nonumber \\ - e \int d^3v_e\,
f_e(\boldr,\bv_e,t) \Bigg ],
\end{eqnarray}
where $f_i(\boldr,\bv_i,t)$ is the particle distribution of the
dominant ions, $Z_I e$ and $f_I (\boldr, \bv_I, t)$ are the charge
and the distribution function of the impurity $I$, and
$f_e(\boldr,\bv_e,t)$ is the distribution function of the
electrons. Using the species-independent normalization
\begin{equation} \label{norm_spindep}
\underline{t} = \frac{c_s t}{L}, \underline{\boldr} =
\frac{\boldr}{L}, \underline{\bA} = \frac{\bA}{B_0 L},
\underline{\varphi} = \frac{e \varphi}{\epsilon_s T_{e0}}
\end{equation}
\noindent for time, space, vector potential and electrostatic
potential, and the species-dependent normalization
\begin{equation} \label{norm_spdep}
\underline{\bv_i} = \frac{\bv_i}{v_{ti}}, \underline{\bv_I} =
\frac{\bv_I}{v_{tI}}, \underline{\bv_e} = \frac{\bv_e}{v_{te}},
\underline{f_i} = \frac{v_{ti}^3}{n_{e0}} f_i, \underline{f_I} =
\frac{v_{tI}^3}{n_{e0}} f_I, \underline{f_e} =
\frac{v_{te}^3}{n_{e0}} f_e
\end{equation}
\noindent for the velocities and the distribution functions, we
get
\begin{eqnarray}\label{eq:Poissonnondim}
\fl - \frac{\epsilon_s \lambda_{De}^2}{L^2} \underline{\nabla}^2
\underline{\varphi} (\underline{\boldr}, \underline{t}) = \int d
^3 \underline{v_i}\, \underline{f_i}(\underline{\boldr},
\underline{\bv}_i, \underline{t}) + \sum_I Z_I \int d ^3
\underline{v_I}\, \underline{f_I} (\underline{\boldr},
\underline{\bv_I}, \underline{t}) \nonumber \\ - \int d
^3\underline{v_e}\, \underline{f_e}(\underline{\boldr},
\underline{\bv_e}, \underline{t}).
\end{eqnarray}
\noindent Here, $n_{e0}$ is a characteristic equilibrium value for
the electron density and
\begin{equation}
\lambda_{De} = \sqrt{\frac{T_{e0}}{4\pi e^2 n_{e0}}}
\end{equation}
is the electron Debye length. Recall that $\epsilon_s = \rho_s/L$.
This new normalization is somewhat different from the
normalization in Section~\ref{sect_order}. We have that $\check
\boldr = \underline{\boldr}$, $\check \bA = \underline{\bA}$,
$\check{\varphi} = \underline{\varphi}$ and $\check \bv_p =
\underline{\bv_p}$ (the subindex $p$ can take the value $i$, $e$,
and any of the values of $I$), but the normalization for the time
is different, giving $\check t = \tau \underline{t}$. The
gyrokinetic variables obtained in the previous sections can be
employed here without further changes because both the
normalization used in Section~\ref{sect_order} and the
normalization in equations \eq{norm_spindep} and \eq{norm_spdep}
give the same normalization for $\bR_p$, $u_p$, $\mu_p$ and
$\theta_p$. This normalization is species-dependent and for this
reason we use the subscripts $p = i, I, e$ to distinguish the
gyrokinetic variables for ions, impurities and electrons. In the
results of Sections~\ref{sect_order} and \ref{sect_theory}, the
constants $\epsilon$, $\Lambda$, $\lambda$ and $\tau$ depend on
the species, and it will be useful to use subscripts for them as
well. Their values for ions, the impurity species $I$ and
electrons are given in Table~\ref{table_lambdas}. Note the sign in
the definition of $\epsilon_e$ and $\lambda_e$. In
Table~\ref{table_lambdas}, $T_{i0}$ and $T_{I0}$ are the
characteristic temperatures of the ion species and the impurity
species $I$, respectively, and $M_I$ is the mass of the impurity
species $I$.

The function $\phi$ in \eq{defphi} is now species-dependent and
given by
\begin{eqnarray}\label{eq:PhicheckPhiunderline}
\fl \check \phi (\check \bR_{p\bot}/\lambda_p \epsilon_p, \check
R_{p ||}, \check \mu_p/\lambda_p^2, \check \theta_p + \pi \Theta
(-\lambda_p), \check t/\tau_p) \nonumber \\ = \underline{\phi}
(\underline{\bR_{p\bot}}/\epsilon_s, \underline{R_{p||}},
\underline{\mu_p}/\lambda_p^2, \underline{\theta_p} + \pi \Theta
(-\lambda_p), \underline{t}).
\end{eqnarray}
Later on we will use the the more compact notation
\begin{equation}\label{eq:phip}
\underline{\phi_p} (\underline{\bR_p} , \underline{\mu_p},
\underline{\theta_p}, \underline{t}) = \underline{\phi}
(\underline{\bR_{p\bot}}/\epsilon_s, \underline{R_{p||}},
\underline{\mu_p}/\lambda_p^2, \underline{\theta_p} + \pi \Theta(-
\lambda_p), \underline{t}).
\end{equation}

\begin{table} \label{table_lambdas}
\caption{Values of the species-dependent parameters $\epsilon$
$\Lambda$, $\lambda$ and $\tau$ for ions, the impurity species $I$
and electrons.}
\begin{indented}
\item[]
\begin{tabular}{l l l}
\br Ions & Impurity $I$ & Electrons \\
\mr  $\epsilon_i = \epsilon_s/\lambda_i$ & $\epsilon_I = \epsilon_s/\lambda_I$ & $-\epsilon_e = -\epsilon_s/\lambda_e \ll \epsilon_s$ \\
$\lambda_i = \sqrt{T_{e0}/T_{i0}}$ & $\lambda_I = Z_I \sqrt{T_{e0} m_i/T_{I0} M_I}$ & $-\lambda_e = \sqrt{m_i/m_e} \gg 1$ \\
$\tau_i = \sqrt{T_{i0}/T_{e0}}$ & $\tau_I = \sqrt{T_{I0} m_i/T_{e0} M_I}$ & $\tau_e = \sqrt{m_i/m_e} \gg 1$ \\
$\Lambda_i = (T_{e0}/T_{i0})^{3/2}$ & $\Lambda_I = Z_I^2 (T_{e0}/T_{I0})^{3/2} \sqrt{m_i/M_I}$ & $\Lambda_e = \sqrt{m_i/m_e} \gg 1$ \\
\br
\end{tabular}
\end{indented}
\end{table}

To write the total Lagrangian of the system, it is necessary to
normalize all the particle Lagrangians and Hamiltonians by the
same quantity. We choose that the normalized Lagrangian and
Hamiltonian be $\underline{\mathcal{L}} = \mathcal{L}/T_{e0}$ and
$\underline{H} = H/T_{e0}$. The gyrokinetic Lagrangian
\eq{finalL}, dependent on the species, becomes in the new
normalization
\begin{equation}\label{eq:Lnewnorm}
\fl \underline{\overline{\mathcal{L}}_p} = \frac{T_{p0}}{T_{e0}
\tau_p} \left [ \frac{\lambda_p}{\epsilon_s} \underline{\bA}
(\underline{\bR_p}) + \underline{u_p} \bun (\underline{\bR_p}) -
\frac{\epsilon_s}{\lambda_p} \underline{\mu_p}\, \underline{\bK}
(\underline{\bR_p}) \right ] \cdot
\frac{d\underline{\bR_p}}{d\underline{t}} - \frac{T_{p0}
\epsilon_s}{T_{e0} \lambda_p \tau_p} \underline{\mu_p}
\frac{d\underline{\theta_p}}{d\underline{t}} -
\underline{\overline{H}_p}
\end{equation}
where
\begin{eqnarray}
\fl \underline{H_p} =   \frac{T_{p0}}{T_{e0}} \left [ \frac{1}{2}
\underline{u_p}^2 + \underline{\mu_p}
\underline{B}(\underline{\bR_p}) \right ] + \frac{T_{p0}
\epsilon_s^2}{T_{e0} \lambda_p^2} \Psi^{(2)}_B (\underline{\bR_p},
\underline{u_p}, \underline{\mu_p}) + Z_p \epsilon_s \langle
\underline{\phi} \rangle ( \underline{\bR_{p\bot}}/\epsilon_s,
\underline{R_{p||}}, \underline{\mu_p}/\lambda_p^2, \underline{t}
) \nonumber \\ \fl + \frac{Z_p \Lambda_p \epsilon_s^2}{\lambda_p}
\Psi^{(2)}_\phi (\underline{\bR_{p\bot}}/\epsilon_s,
\underline{\bR_p}, \underline{\mu_p}/\lambda_p^2, \underline{t},
\lambda_p) + \frac{Z_p \epsilon_s^2}{\lambda_p} \Psi^{(2)}_{\phi
B} (\underline{\bR_{p\bot}}/\epsilon_s, \underline{\bR_p},
\underline{u_p}, \underline{\mu_p}, \underline{\mu_p}/\lambda_p^2,
\underline{t}, \lambda_p)
\end{eqnarray}
and
\begin{equation}
\underline{\bK} (\underline{\bR_p}) = \frac{1}{2}
\bun(\underline{\bR_p}) \bun(\underline{\bR_p}) \cdot
\nabla_{\underline{\bR_p}} \times \bun(\underline{\bR_p}) -
\nabla_{\underline{\bR_p}} \eun_2 (\underline{\bR_p}) \cdot \eun_1
(\underline{\bR_p}).
\end{equation}
Here, obviously, $Z_i = 1$ and $Z_e = -1$. Analogously to
\eq{eq:phip} we define
\begin{equation}
\Psi^{(2)}_{\phi, p} (\underline{\bR_p}, \underline{\mu_p},
\underline{t})=\Psi^{(2)}_\phi
(\underline{\bR_{p\bot}}/\epsilon_s, \underline{\bR_p},
\underline{\mu_p}/\lambda_p^2, \underline{t}, \lambda_p)
\end{equation}
and
\begin{equation}
\Psi^{(2)}_{\phi B, p} (\underline{\bR_p},
\underline{u_p},\underline{\mu_p}, \underline{t}) =
\Psi^{(2)}_{\phi B} (\underline{\bR_{p\bot}}/\epsilon_s,
\underline{\bR_p}, \underline{u_p}, \underline{\mu_p},
\underline{\mu_p}/\lambda_p^2, \underline{t}, \lambda_p).
\end{equation}

The equations of motion are those obtained in \eq{dRdt},
\eq{dudt}, \eq{dmudt} and \eq{dthetadt} multiplied by $\tau_p$. As
a result, the Vlasov equation for each species is
\begin{equation}\label{eq:Vlasovbar}
\frac{\partial \underline{F_p}}{\partial \underline{t}} + \tau_p
\dot{\bR} \cdot \nabla_{\underline{\bR_p}} \, \underline{F_p} +
\tau_p \dot{u} \frac{\partial \underline{F_p}}{\partial
\underline{u_p}} = 0.
\end{equation}
\noindent In what follows we work in non-dimensional variables but
do not underline them.

\subsection{Gyrokinetic Poisson's equation via the gyrokinetic change
of coordinates}
\label{sec:gyroPoissonChangeVar}

Our objective is to write Poisson's equation
(\ref{eq:Poissonnondim}) in terms of $F_p (\bR_p, u_p, \mu_p, t)$,
i.e. the solution of equation \eq{eq:Vlasovbar}. Since\footnote{We
stress that the transformation $T_\epsilon$ depends on the species
through the values of $\epsilon_p$, $\Lambda_p$ and $\lambda_p$
and denote it by $T_{\epsilon_p, p}$.} $(\boldr,\bv_p) =
T_{\epsilon_p, p} (\bR_p, u_p, \mu_p, \theta_p, t)$,
\begin{equation}
F_p (\bR_p, u_p, \mu_p, t)= f_p(T_{\epsilon_p, p} (\bR_p, u_p,
\mu_p, \theta_p, t), t).
\end{equation}
Using the obvious identity
\begin{equation}
\int d ^3 v_p\, f_p(\boldr,\bv_p,t) = \int d^3r^\prime\, d^3v_p\,
f_p(\boldr^\prime ,\bv_p,t) \delta(\boldr^\prime - \boldr)
\end{equation}
and the change of variables formula we can write Poisson's
equation as
\begin{eqnarray}\label{eq:Poissonpullback2}
\fl - \frac{\epsilon_s\lambda_{De}^2}{L^2} & \nabla^2 \varphi
(\boldr,t) \nonumber\\ \fl & = \int d ^3 R_i du_i d\mu_i d\theta_i
\left |\det\left(J_{T_{\epsilon_i, i}}\right) \right | F_i \,
\delta\Big(\pi^{\boldr}\Big(T_{\epsilon_i, i}(\bR_i, u_i, \mu_i,
\theta_i, t)\Big)-\boldr\Big) \nonumber \\ \fl & + \sum_I Z_I \int
d ^3 R_I du_I d\mu_I d\theta_I \left | \det
\left(J_{T_{\epsilon_I, I}}\right) \right | F_I \, \delta
\Big(\pi^{\boldr}\Big(T_{\epsilon_I, I}(\bR_I, u_I, \mu_I,
\theta_I, t)\Big)-\boldr\Big) \nonumber\\ \fl & - \int d ^3 R_e
du_e d\mu_e d\theta_e \left |\det\left(J_{T_{\epsilon_e,
e}}\right) \right | F_e\, \delta
\Big(\pi^{\boldr}\Big(T_{\epsilon_e, e}(\bR_e, u_e, \mu_e,
\theta_e, t)\Big)-\boldr\Big),
\end{eqnarray}
where $J_{T_{\epsilon_p, p}}(\bR_p, u_p, \mu_p)$ is the Jacobian
matrix of the transformation $T_{\epsilon_p, p}$, calculated in
\ref{app:jacobians}, and $\pi^{\boldr}(\boldr, \bv_p):=\boldr$ is
the projection onto the spatial part of the coordinates $(\boldr,
\bv_p)$. From an abstract viewpoint this is, perhaps, the simplest
way of writing (\ref{eq:Poissonnondim}) in terms of $F_p(\bR_p,
u_p, \mu_p, t)$. Observe that \eq{eq:Poissonpullback2} is an exact
relation. However, in practical terms, we have only computed the
explicit expression of $T_{\epsilon_p, p}$ up to a certain order
in $\epsilon_p$, or equivalently, in the species-independent
parameter $\epsilon_s$. Making use of the results of this paper we
can give an explicit expression for the change of variables up to
order $\epsilon_s^2$, namely
\begin{eqnarray}\label{eq:r_asfunctionofgyrokincoor}
\fl \pi^{\boldr} \Big(T_{\epsilon_p, p} (\bR_p, u_p, \mu_p,
\theta_p, t)\Big) = \bR_p + \frac{\epsilon_s}{\lambda_p}
\rhobf(\bR_p, \mu_p, \theta_p) \nonumber \\ +
\frac{\epsilon_s^2}{\lambda_p^2} \Bigg [ \bR_{p, 2} + \mu_{p, 1}
\frac{\partial \rhobf}{\partial \mu_p}+ \theta_{p, 1}
\frac{\partial \rhobf }{\partial \theta_p} \Bigg ] +
O(\epsilon_s^3),
\end{eqnarray}
which allows us to write the identity
\begin{eqnarray}\label{eq:transdelta}
\fl\delta\Big(\pi^{\boldr} & \Big(T_{\epsilon_p, p}(\bR_p, u_p,
\mu_p, \theta_p, t)\Big) - \boldr\Big)= \delta\Big( \bR_p +
\frac{\epsilon_s}{\lambda_p}\rhobf(\bR_p,\mu_p,\theta_p) - \boldr
\Big) \nonumber \\ \fl & + \frac{\epsilon_s^2}{\lambda_p^2} \Bigg
( \bR_{p, 2} + \mu_{p, 1} \frac{\partial \rhobf}{\partial \mu_p} +
\theta_{p, 1} \frac{\partial \rhobf}{\partial \theta_p} \Bigg )
\cdot\nabla \delta\Big( \bR_p + \frac{\epsilon_s}{\lambda_p}
\rhobf(\bR_p,\mu_p,\theta_p)-\boldr \Big) + O(\epsilon_s^3),
\end{eqnarray}
where $\nabla\delta$ denotes the gradient of the Dirac delta
function with respect to its natural arguments and the subscript
$p = i, I, e$ in the corrections $\bR_{p, 2}$, $\mu_{p, 2}$ and
$\theta_{p, 2}$ indicates that these corrections depend on the
species through $\Lambda_p$, $\lambda_p$ and $\phi_p (\bR_p,
\mu_p, \theta_p, t)$ in \eq{eq:phip}. Substituting
\eq{eq:transdelta} into \eq{eq:Poissonpullback2} finally gives
\begin{eqnarray}\label{eq:Poissonpullbackfinal}
\fl - \frac{\epsilon_s\lambda_{De}^2}{L^2} & \nabla^2 \varphi
(\boldr, t) = \sum_p Z_p \int d^3R_p du_p d\mu_p d\theta_p \,
B_{||, p}^* F_p \Bigg[\delta\Big( \bR_p +
\frac{\epsilon_s}{\lambda_p} \rhobf(\bR_p,\mu_p,\theta_p) - \boldr
\Big) \nonumber\\ \fl & + \frac{\epsilon_s^2}{\lambda_p^2} \Bigg (
\bR_{p, 2} + \mu_{p, 1} \frac{\partial \rhobf}{\partial \mu_p} +
\theta_{p, 1} \frac{\partial \rhobf}{\partial \theta_p} \Bigg )
\cdot\nabla \delta \Big( \bR_p + \frac{\epsilon_s}{\lambda_p}
\rhobf(\bR_p,\mu_p,\theta_p)-\boldr \Big) \Bigg] + \dots
\end{eqnarray}
Here we have used \ref{app:jacobians} to write $\mbox{det} (
J_{T_{\epsilon_p, p}} ) = B_{||, p}^*$. We employ the subscripts
$p = i, I, e$ in the determinant of the Jacobian $B^*_{||,p}$
because it depends on the species through its dependence on
$\epsilon_p = \epsilon_s/\lambda_p$. Expression
\eq{eq:Poissonpullbackfinal} may seem accurate to order
$\epsilon_s^2$, i.e., that on the right side of
\eq{eq:Poissonpullbackfinal} we are dropping only terms which are
of order $\epsilon_s^3$ or higher. However, it is easy to see that
this is not true. Using
\begin{eqnarray}
\fl \nabla \delta \left (\bR_p + \frac{\epsilon_s}{\lambda_p}
\rhobf - \boldr \right ) = \left ( \matI +
\frac{\epsilon_s}{\lambda_p} \nabla_{\bR_p} \rhobf \right )^{-1}
\cdot \nabla_{\bR_p} \delta \left (\bR_p +
\frac{\epsilon_s}{\lambda_p} \rhobf - \boldr \right ) \nonumber \\
= \nabla_{\bR_p} \delta \left (\bR_p +
\frac{\epsilon_s}{\lambda_p} \rhobf - \boldr \right ) +
O(\epsilon_s),
\end{eqnarray}
\begin{equation}
\frac{\epsilon_s}{\lambda_p} \frac{\partial \rhobf}{\partial
\mu_p} \cdot \nabla \delta \left (\bR_p +
\frac{\epsilon_s}{\lambda_p} \rhobf - \boldr \right ) =
\frac{\partial}{\partial \mu_p} \left [ \delta \left (\bR_p +
\frac{\epsilon_s}{\lambda_p} \rhobf - \boldr \right ) \right ]
\end{equation}
and
\begin{equation}
\frac{\epsilon_s}{\lambda_p} \frac{\partial \rhobf}{\partial
\theta_p} \cdot \nabla \delta \left ( \bR_p +
\frac{\epsilon_s}{\lambda_p} \rhobf - \boldr \right ) =
\frac{\partial}{\partial \theta_p} \left [\delta \left (\bR_p +
\frac{\epsilon_s}{\lambda_p} \rhobf - \boldr \right ) \right ],
\end{equation}
and integrating by parts so that the delta function does not
appear differentiated, we find that \eq{eq:Poissonpullbackfinal}
becomes
\begin{eqnarray} \label{eq:GKPoisson_v1}
\fl - \frac{\epsilon_s\lambda_{De}^2}{L^2} \nabla^2 \varphi
(\boldr, t) = \sum_p Z_p \int d^3R_p du_p d\mu_p d\theta_p \,
\delta \Big( \bR_p + \frac{\epsilon_s}{\lambda_p}
\rhobf(\bR_p,\mu_p,\theta_p) - \boldr \Big) \Bigg\{ B_{||, p}^*
F_p \nonumber\\ - \frac{\epsilon_s}{\lambda_p} \Bigg
[\frac{1}{\lambda_p} \nabla_{(\bR_{p\bot}/\epsilon_s)} \cdot \Big
( B_{||,p}^* F_p \bR_{p,2} \Big ) + \frac{\partial}{\partial
\mu_p} \Big ( B_{||,p}^* F_p \mu_{p, 1} \Big ) \nonumber \\
+ \frac{\partial}{\partial \theta_p} \Big ( B_{||,p}^* F_p
\theta_{p, 1} \Big ) \Bigg ] \Bigg \} + O(\epsilon_s^2).
\end{eqnarray}
Note that terms that seemed to be of second order in $\epsilon_s$
are in reality first order contributions. Similarly, terms that
seem to be of third order and are neglected in
\eq{eq:Poissonpullbackfinal} are in reality of second order. To
obtain a gyrokinetic Poisson's equation correct to order
$\epsilon_s^2$ it is necessary to carry
\eq{eq:r_asfunctionofgyrokincoor} to an order higher
\begin{eqnarray}\label{eq:r_asfunctionofgyrokincoor2}
\fl \pi^{\boldr} \Big(T_{\epsilon_p, p} (\bR_p, u_p, \mu_p,
\theta_p, t)\Big) = \bR_p + \frac{\epsilon_s}{\lambda_p} \rhobf +
\frac{\epsilon_s^2}{\lambda_p^2} \left ( \bR_{p, 2} + \mu_{p, 1}
\frac{\partial \rhobf}{\partial \mu_p} + \theta_{p, 1}
\frac{\partial \rhobf}{\partial \theta_p} \right ) \nonumber\\ +
\frac{\epsilon_s^3}{\lambda_p^3} \Bigg ( \tilde\bR_{p, 3} +
\bR_{p, 2} \cdot \nabla_{\bR_p} \rhobf + \tilde \mu_{p, 2}
\frac{\partial \rhobf}{\partial \mu_p} + \tilde \theta_{p, 2}
\frac{\partial \rhobf}{\partial \theta_p} + \frac{1}{2} \mu_{p,
1}^2 \frac{\partial^2 \rhobf}{\partial \mu_p^2} \nonumber\\ +
\mu_{p, 1} \theta_{p, 1} \frac{\partial^2 \rhobf}{\partial \mu_p
\partial \theta_p} + \frac{1}{2} \theta_{p, 1}^2
\frac{\partial^2 \rhobf}{\partial \theta_p^2} \Bigg) +
O(\epsilon_s^4).
\end{eqnarray}
All the terms entering this equation are computable from the
results found in Section~\ref{sect_theory}. We leave this for
future work.

By integrating over the delta function in \eq{eq:GKPoisson_v1} we
make the gyrokinetic Poisson's equation more explicit,
\begin{eqnarray} \label{eq:GKPoisson}
\fl - \frac{\epsilon_s\lambda_{De}^2}{L^2} \nabla^2 \varphi
(\boldr, t) = \sum_p Z_p \int du_p d\mu_p d\theta_p \,
\frac{1}{\mbox{det}(\matI + (\epsilon_s/\lambda_p) \nabla_{\bR_p}
\rhobf)} \Bigg \{ B_{||, p}^* F_p \nonumber\\ -
\frac{\epsilon_s}{\lambda_p} \Bigg [\frac{1}{\lambda_p}
\nabla_{(\bR_{p\bot}/\epsilon_s)} \cdot \Big ( B_{||,p}^* F_p
\bR_{p,2} \Big ) + \frac{\partial}{\partial \mu_p} \Big (
B_{||,p}^* F_p \mu_{p, 1} \Big ) \nonumber \\ +
\frac{\partial}{\partial \theta_p} \Big ( B_{||,p}^* F_p
\theta_{p, 1} \Big ) \Bigg ] \Bigg \}_{\bR_p = \widehat{\bR}
(\boldr, \mu_p, \theta_p, \epsilon_s/\lambda_p)} +
O(\epsilon_s^2),
\end{eqnarray}
where $\widehat{\bR} (\boldr, \mu_p, \theta_p,
\epsilon_s/\lambda_p)$ is the the function defined by solving for
$\bR_p$ the equation $\bR_p + (\epsilon_s/\lambda_p) \rhobf
(\bR_p, \mu_p, \theta_p) = \boldr$.

Finally, we would like to point out that usually equation
\eq{eq:GKPoisson} can be simplified even more. The following
discussion is not meant to be an exhaustive review of all possible
orderings and simplifications, but a brief comment on the most
typical approach. Usually $F_p$ has contributions with different
characteristic scales of variation, ranging from the shortest
scale $\rho_s$ to the background profile variation scale $L$. At
the same time $\nabla_{\bR_p} F_p \sim 1$ because the saturation
amplitude of the short scale fluctuation $F_{p, \rho_s}$ is
sufficiently small, i.e., $F_{p, \rho_s} = \epsilon_s
\textsf{F}_{p, \rho_s} (\bR_{p\bot}/\epsilon_s, R_{p||}, u_p,
\mu_p, t) \sim \epsilon_s \varphi (\boldr_\bot/\epsilon_s, t)$,
where $\textsf{F}_{p, \rho_s}$ is a function of order unity with
derivatives with respect to its arguments of order unity. For
$\nabla_{\bR_p} F_p \sim 1$, we can use the approximate expression
$\widehat{\bR} (\boldr, \mu_p, \theta_p, \epsilon_s/\lambda_p) =
\boldr - (\epsilon_s/\lambda_p) \rhobf(\boldr, \mu_p, \theta_p) +
O(\epsilon_s^2)$ because $F_p(\widehat{\bR}(\boldr, \mu_p,
\theta_p, \epsilon_s/\lambda_p), u_p, \mu_p, \theta_p, t) =
F_p(\boldr - (\epsilon_s/\lambda_p) \rhobf(\boldr, \mu_p,
\theta_p), u_p, \mu_p, \theta_p, t) + O(\epsilon_s^2)$. In
addition, the term containing $\bR_{p,2}$ in \eq{eq:GKPoisson}
becomes of next order because $\nabla_{\bR_p} F_p \sim
\nabla_{\bR_p} B_{||,p}^* \sim \nabla_{\bR_p} \cdot \bR_{p,2} \sim
1$. To simplify even more, we use $\mbox{det} (\matI +
(\epsilon_s/\lambda_p) \nabla_{\bR_p} \rhobf) = 1 +
(\epsilon_s/\lambda_p) \nabla_{\bR_p} \cdot \rhobf +
O(\epsilon_s^2)$. As a result,
\begin{eqnarray}\label{eq:GKPoissonsimple}
\fl - \frac{\epsilon_s\lambda_{De}^2}{L^2} & \nabla^2 \varphi
(\boldr, t) = \sum_p Z_p \int du_p d\mu_p d\theta_p \, \Bigg \{
B_{||, p}^* F_p - \frac{\epsilon_s}{\lambda_p} \Bigg [B_{||, p}^*
F_p \Big ( \nabla_{\bR_p} \cdot \rhobf \Big ) \nonumber \\ \fl & +
\frac{\partial}{\partial \mu_p} \Big ( B_{||,p}^* F_p \mu_{p, 1}
\Big ) + \frac{\partial}{\partial \theta_p} \Big ( B_{||,p}^* F_p
\theta_{p, 1} \Big ) \Bigg ] \Bigg \}_{\bR_p = \boldr -
(\epsilon_s/\lambda_p) \scrhobf(\boldr, \mu_p, \theta_p)} +
O(\epsilon_s^2).
\end{eqnarray}
This is the simplest Poisson's equation correct to first order in
$\epsilon_s$. As we have already pointed out, it is necessary to
include the corrections $\tilde {\bR}_{p,3}$, $\tilde{\mu}_{p, 2}$
and $\tilde{\theta}_{p,2}$ to find the contributions of order
$\epsilon_s^2$. We finish by checking that the assumption
$\nabla_{\bR_p} F_p \sim 1$ is consistent with the gyrokinetic
system of equations. Consider the short scale pieces of the Vlasov
equation \eq{eq:Vlasovbar} and of the gyrokinetic Poisson's
equation \eq{eq:GKPoissonsimple}. The result will be a typical
$\delta f$ formulation \cite{dorland00, dannert05, candy03,
chen03, peeters04}, demonstrating that it is possible to find a
closed non-linear system of equations to determine the short scale
fluctuations $\textsf{F}_{p, \rho_s}$ and $\varphi$. This does not
mean that $\delta f$ formulations are always valid, but it
indicates that the assumption $\nabla_{\bR_p} F_p \sim 1$ is
consistent with the gyrokinetic system of equations.

\section{Gyrokinetic Field Theory} \label{sec:fieldtheory}

Gyrokinetic field theory is the formulation of gyrokinetics as a
classical field theory and is defined by an action functional
$\Sigma$ that depends on the trajectories of the particles and the
electromagnetic field. A closed system of equations coupling the
electromagnetic field and the gyrokinetic distribution function is
obtained by finding the stationary points of $\Sigma$. The first
application of field theory to plasma physics is the work by Low
in \cite{Low1958} that was later extended to gyrokinetic theory in
the seminal papers by Sugama \cite{Sugama2000} and Brizard
\cite{Brizard2000}. The development of gyrokinetic field theory in
the last decade has been motivated mostly by the fact that it
allows to identify in a systematic way conservation laws from
symmetries of the Lagrangian. For the sake of completeness we
briefly recast now our results in field theory language following
the Lagrangian formulation of Sugama~\cite{Sugama2000}.

In subsection~\ref{sub:variational} we show how to obtain the
gyrokinetic equations of motion and the gyrokinetic Poisson's
equation from a variational principle. The advantage of this
procedure is that it ensures that an energy-like invariant is
conserved. We obtain the invariant and prove that it is conserved
in subsection~\ref{sub:noethers}.

\subsection{Equations via variational principle}
\label{sub:variational}
Using the abbreviated notation $\bZ_p =
\{\bR_p,u_p,\mu_p,\theta_p\}$, the action functional is
\begin{eqnarray}\label{eq:actionfieldtheory}
\fl\Sigma & \Big [\bZ_p(\bullet, \cdot;
\cdot),\varphi(\bullet,\cdot) \Big ] =
\frac{\lambda_{De}^2\epsilon_s^2}{2L^2} \int_{t_0}^{t_1} dt \int
d^3 r \,|\nabla\varphi(\boldr,t)|^2 \nonumber \\ \fl & + \sum_p
\int_{t_0}^{t_1} dt \int d^6 Z_{p0} \, B_{||, p}^* (\bZ_{p0})
F_{p0}(\bZ_{p0}) \, \overline{\mathcal{L}}_p
\Big(\bZ_p(\bZ_{p0},t_0;t), \dot\bZ_p(\bZ_{p0},t_0;t),
\varphi(\bullet,t)\Big),
\end{eqnarray}
where $F_{p0}$ is the distribution function of species $p$ at time
$t_0$, and $\bZ_p(\bZ_{p0},t_0;t)$ is the trajectory in
phase-space of a particle of species $p$ satisfying the initial
condition $\bZ_p(\bZ_{p0},t_0;t_0)=\bZ_{p0}$ at $t = t_0$. The
first term on the right side of \eq{eq:actionfieldtheory} is the
action for the electric field (recall that we are considering a
static magnetic field) and the second term is the sum of the
actions of the particles. The gyrokinetic Lagrangian
${\overline{\cal L}}_p$ of the species $p$ is defined in
\eq{eq:Lnewnorm} (recall that we have dropped the underlining for
normalized variables). Its third argument stresses that
$\overline{\mathcal{L}}_p$ depends as a functional on the
electrostatic potential.

The Euler-Lagrange equations for $\Sigma$ are obtained by finding
its stationary points under infinitesimal variations of the maps
$\bZ_p(\bZ_{p0},t_0;t)$ and $\varphi(\boldr,t)$. The allowed
infinitesimal perturbations to $\varphi(\boldr,t)$ vanish at the
boundary of the spatial domain of interest, and the perturbations
to both $\bZ_p(\bZ_{p0},t_0;t)$ and $\varphi(\boldr,t)$ must be
zero at $t = t_0$ and $t = t_1$. The calculation of the variation
with respect to $\bZ_p(\bZ_{p0},t_0;t)$ gives the gyrokinetic
equations of motion and is a repetition of that leading to
equations \eq{dRdt}, \eq{dudt}, \eq{dmudt} and \eq{dthetadt}
(recall the discussion on the phase-space Lagrangian methodology
in subsection~\ref{sub_change}). The distribution function at time
$t$ is
\begin{equation} \label{FpFp0}
F_p (\bZ_p, t) := F_{p0}( \bZ_{p0} ( \bZ_p, t; t_0 ) ),
\end{equation}
where $\bZ_{p0}(\bZ_p, t; t_0)$ is the inverse of the map
$\bZ_p(\bZ_{p0}, t_0; t)$, i.e., $\bZ_{p0} ( \bZ_p (\bZ_{p0}, t_0;
t), t; t_0) \equiv \bZ_{p0}$. Note that $F_p$ automatically
satisfies \eq{eq:Vlasovbar}. Observe also that from condition
\eq{consps}, which is automatically satisfied by \eq{dRdt},
\eq{dudt}, \eq{dmudt} and \eq{dthetadt}, we obtain
\begin{equation} \label{eq:psconserv}
\fl d^6Z_{p0}\, B_{||,p}^*(\bZ_{p0}) = d^6Z_p \, \frac{
B_{||,p}^*(\bZ_{p0}(\bZ_{p}, t; t_0))}{|\mbox{det} (J_{\bZ_{p0}
\mapsto \bZ_p}(\bZ_{p0}(\bZ_{p}, t; t_0)))|} = d^6Z_p \,
B_{||,p}^*(\bZ_p),
\end{equation}
where $(J_{\bZ_{p0} \mapsto \bZ_p})_\alpha^\beta = \partial
Z_p^\beta/\partial Z_{p0}^\alpha$ is the Jacobian matrix of the
map $\bZ_p(\bZ_{p0}, t_0; t)$, and $\mbox{det} (J_{\bZ_{p0}
\mapsto \bZ_p}(\bZ_{p0})) = B_{||,p}^* (\bZ_{p0})/B_{||,p}^*
(\bZ_p(\bZ_{p0},t_0;t))$ is its determinant. This property is
proven in \ref{app:psconserv} for completeness.

The stationary points of $\Sigma$ under variations of
$\varphi(\boldr,t)$ are given by $\delta_\varphi \Sigma = 0$, with
\begin{eqnarray}\label{eq:auxPoissonvariational}
\fl\delta_\varphi {\Sigma} &=
\frac{\lambda_{De}^2\epsilon_s^2}{L^2} \int dt\, d^3 r\, \nabla
\delta \varphi (\boldr, t) \cdot \nabla \varphi(\boldr,t)
\nonumber\\ \fl & - \sum_p Z_p \epsilon_s \int dt\, d^6 Z_{p0} \,
B_{||, p}^* (\bZ_{p0}) F_{p0}(\bZ_{p0}) \delta_{\varphi} \langle
\phi_p (\bZ_p(\bZ_{p0}, t_0; t), t) \rangle \nonumber \\ \fl & -
\sum_p \frac{Z_p\Lambda_p\epsilon_s^2}{\lambda_p} \int dt\, d^6
Z_{p0}\, B_{||,p}^*(\bZ_{p0}) F_{p0}(\bZ_{p0}) \delta_{\varphi}
\Psi^{(2)}_{\phi, p} \Big ( \bZ_p(\bZ_{p0}, t_0; t), t \Big )
\nonumber \\ \fl & - \sum_p \frac{Z_p\epsilon_s^2}{\lambda_p} \int
dt\, d^6 Z_{p0}\, B_{||, p}^* (\bZ_{p0}) F_{p0}(\bZ_{p0})
\delta_{\varphi} \Psi^{(2)}_{\phi B, p} \Big ( \bZ_p(\bZ_{p0},
t_0; t), t \Big ).
\end{eqnarray}
In \ref{app:poissonvariationalprinciple} we have evaluated all the
terms. The final result is
\begin{equation}\label{eq:gyrokinPoissonfieldtheory1}
\delta_\varphi \Sigma = \epsilon_s \int dt\, d^3r\, \delta \varphi
(\boldr, t) \mathcal{P} (\boldr, t),
\end{equation}
where
\begin{eqnarray}\label{eq:gyrokinPoissonfieldtheory}
\fl \mathcal{P} (\boldr, t) & = -
\frac{\lambda_{De}^2\epsilon_s}{L^2} \nabla^2 \varphi(\boldr,t)
\nonumber \\ \fl & - \sum_p Z_p \int d^3 R_p du_p d\mu_p
d\theta_p\, B_{||, p}^* (\bR_p, u_p, \mu_p ) F_p(\bR_p, u_p,
\mu_p, t) \Bigg \{ \delta \left (\bR_p +
\frac{\epsilon_s}{\lambda_p} \rhobf - \boldr \right ) \nonumber \\
\fl & + \frac{\epsilon_s}{\lambda_p} \Bigg[
\frac{\epsilon_s}{\lambda_p} \bR_{p, 2\perp} \cdot \nabla_{\bR_p}
\delta \left (\bR_p + \frac{\epsilon_s}{\lambda_p} \rhobf - \boldr
\right ) + \mu_{p,1} \frac{\partial}{\partial \mu_p} \delta \left
(\bR_p + \frac{\epsilon_s}{\lambda_p} \rhobf - \boldr \right )
\nonumber \\ \fl & + \theta_{p,1} \frac{\partial}{\partial
\theta_p} \delta \left (\bR_p + \frac{\epsilon_s}{\lambda_p}
\rhobf - \boldr \right ) \Bigg] \Bigg \}.
\end{eqnarray}
By imposing that $\delta_\varphi\Sigma=0$ for any $\delta \varphi
(\boldr, t)$, we find that $\mathcal{P} (\boldr, t) = 0$. This is
Poisson's equation. By integrating by parts to leave the delta
function undifferentiated, and then integrating over the delta
function, we recover \eq{eq:GKPoisson}. Note that even though the
Hamiltonian is obtained to second order in $\epsilon_s$, and hence
the Vlasov equation is also known to second order, the gyrokinetic
Poisson's equation that we have found using the variational
principle is only correct to first order in $\epsilon_s$. It
coincides to first order with the first order equation that we
found independently in subsection~\ref{sec:gyroPoissonChangeVar}.

Observe that equation \eq{eq:gyrokinPoissonfieldtheory} keeps some
second order terms that could have been neglected, as done in
\eq{eq:GKPoissonsimple}. These terms are important if we want to
take advantage of the field theory formulation of gyrokinetics.
The action \eq{eq:actionfieldtheory} is invariant under time
translations, so Noether's theorem automatically provides a
conserved quantity, interpreted as the total energy of the system,
which we denote by ${\cal H}(t)$ because it is indeed the field
theory Hamiltonian. However, $\dot{{\cal H}}(t)=0$ {\it on the
equations of motion of} $\Sigma$, that is, on solutions of
\eq{dRdt}, \eq{dudt}, \eq{dmudt}, \eq{dthetadt} and
\eq{eq:gyrokinPoissonfieldtheory}, without neglecting any terms.
We show this in the next subsection.

\subsection{Conservation of energy} \label{sub:noethers}

In this subsection we prove that if the equations of motion are
obtained via a variational principle of the action
\eq{eq:actionfieldtheory}, there is an energy-like invariant
$\mathcal{H} (t)$. This is an application of Noether's theorem,
and it is based on the fact that in the action
\eq{eq:actionfieldtheory} the only time dependence is through the
functions $\bZ_p (\bZ_{p0}, t_0; t)$ and $\varphi (\boldr, t)$. To
obtain the energy-like invariant, we use the equations of motion
derived from the variational principle in
subsection~\ref{sub:variational}.

In order to find the conserved quantity we first perform the
change of variable $t = t' + \delta t$ in every integral on the
right side of \eq{eq:actionfieldtheory}, which, of course, does
not change the value of the action, giving
\begin{eqnarray}\label{eq:actiontimetrans}
\fl \Sigma \Big [ \bZ_p(\bullet,
\cdot;\cdot),\varphi(\bullet,\cdot) \Big ] = \int_{t_0 - \delta
t}^{t_1 - \delta t} dt' \Bigg [
\frac{\lambda_{De}^2\epsilon_s^2}{2L^2} \int d^3 r
\,|\nabla\varphi(\boldr,t' + \delta t)|^2 \nonumber \\ + \sum_p
\int d^6 Z_{p0} B_{||, p}^* (\bZ_{p0}) F_{p0}(\bZ_{p0}) \left.
\overline{\mathcal{L}}_p \right |_{t = t^\prime + \delta t} \Bigg
],
\end{eqnarray}
where
\begin{equation}
\fl \left. \overline{\mathcal{L}}_p \right |_{t = t^\prime +
\delta t} = \overline{\mathcal{L}}_p
\Big(\bZ_p(\bZ_{p0},t_0;t'+\delta t),
\dot\bZ_p(\bZ_{p0},t_0;t'+\delta t), \varphi(\bullet, t' + \delta
t)\Big).
\end{equation}
We expand the right side of the previous equation up to first
order in $\delta t$ to find
\begin{eqnarray}\label{eq:actiontimetrans2}
\fl \Sigma \Big[ \bZ_p & (\bullet,
\cdot;\cdot),\varphi(\bullet,\cdot) \Big ] = \Sigma \Big[
\bZ_p(\bullet, \cdot; \cdot), \varphi(\bullet, \cdot) \Big ]
\nonumber\\ \fl & + \delta_\varphi \left. \Sigma\Big[
\bZ_p(\bullet, \cdot;\cdot),\varphi(\bullet,\cdot) \Big ] \right
|_{\delta \varphi = \delta t (\partial \varphi/\partial t)} +
\sum_p \delta_{\bZ_p} \left. \Sigma\Big[ \bZ_p(\bullet,
\cdot;\cdot),\varphi(\bullet,\cdot) \Big ] \right |_{\delta \bZ_p
= \delta t \dot \bZ_p} \nonumber\\ \fl & - \delta t\Bigg[
\frac{\lambda_{De}^2\epsilon_s^2}{2L^2}  \int d^3 r
\,|\nabla\varphi(\boldr,t')|^2 + \sum_p \int d^6 Z_{p0} \, B_{||,
p}^* (\bZ_{p0}) F_{p0}(\bZ_{p0}) \left. \overline{\mathcal{L}}_p
\right |_{t = t'} \Bigg]_{t'=t_0}^{t'=t_1},
\end{eqnarray}
where $\delta_\varphi \Sigma |_{\delta \varphi = \delta t
(\partial \varphi/\partial t)}$ and $\delta_{\bZ_p} \Sigma
|_{\delta \bZ_p = \delta t \dot \bZ_p}$ are the variations of
$\Sigma$ under perturbations of both $\varphi(\boldr, t)$ and
$\bZ_p (\bZ_{p0}, t_0; t)$ with the specific form $\delta \varphi
= \delta t (\partial \varphi/\partial t)$ and $\delta Z_p = \delta
t \dot Z_p$. For the variations with respect to $\bZ_p$, we obtain
\begin{eqnarray}
\fl \delta_{\bZ_p} \Sigma & = \int_{t_0}^{t_1} dt^\prime \int
d^6Z_{p0}\, B_{||,p}^*(\bZ_{p0}) F_{p0} (\bZ_{p0}) \delta_{\bZ_p}
\overline{\mathcal{L}}_p \Big ( \bZ_p(\bZ_{p0}, t_0; t^\prime),
\dot\bZ_p (\bZ_{p0}, t_0; t^\prime) \Big ) \nonumber
\\ \fl & = \int_{t_0}^{t_1} dt^\prime \int d^6Z_{p0}\,
B_{||,p}^*(\bZ_{p0}) F_{p0} (\bZ_{p0}) \sum_{\alpha = 1}^6 \left.
\left ( \delta Z_p^\alpha \frac{\partial
\overline{\mathcal{L}}_p}{\partial Z_p^\alpha} + \delta \dot
Z_p^\alpha \frac{\partial \overline{\mathcal{L}}_p}{\partial \dot
Z_p^\alpha} \right ) \right |_{t = t^\prime} \nonumber \\ \fl & =
\left [ \int d^6Z_{p0}\, B_{||,p}^*(\bZ_{p0}) F_{p0} (\bZ_{p0})
\sum_{\alpha = 1}^6 \left. \delta Z_p^\alpha \frac{\partial
\overline{\mathcal{L}}_p}{\partial \dot Z_p^\alpha} \right |_{t =
t^\prime} \right ]_{t^\prime = t_0}^{t^\prime = t_1},
\end{eqnarray}
where to obtain this last equality we have integrated by parts in
$t$ and we have used the equations of motion
\begin{equation}
\frac{d}{dt} \left ( \frac{\partial
\overline{\mathcal{L}}_p}{\partial \dot{Z}_p^\alpha} \right ) =
\frac{\partial \overline{\mathcal{L}}_p}{\partial Z_p^\alpha},
\quad \alpha = 1, 2, \ldots, 6.
\end{equation}
Applying $\delta \bZ_p = \delta t \dot \bZ_p$ we find
\begin{equation} \label{varZpNoethers}
\fl \delta_{\bZ_p} \Sigma |_{\delta \bZ_p = \delta t \dot \bZ_p} =
\delta t \left [ \int d^6Z_{p0}\, B_{||,p}^*(\bZ_{p0}) F_{p0}
(\bZ_{p0}) \sum_{\alpha = 1}^6 \left. \dot Z_p^\alpha
\frac{\partial \overline{\mathcal{L}}_p}{\partial \dot Z_p^\alpha}
\right |_{t = t^\prime} \right ]_{t^\prime = t_0}^{t^\prime =
t_1}.
\end{equation}
For the variations with respect to $\delta \varphi$ we obtain
that
\begin{equation} \label{varvarphiNoethers}
\left. \delta_\varphi \Sigma \right |_{\delta \varphi = \delta t
(\partial \varphi/\partial t)} = 0,
\end{equation}
where we have employed \eq{eq:gyrokinPoissonfieldtheory1} and
Poisson's equation $\mathcal{P} (\boldr, t) = 0$. To obtain
\eq{eq:gyrokinPoissonfieldtheory1} we assumed that $\delta
\varphi$ vanishes at the boundaries of the domain, and this is not
necessarily the case for $\delta \varphi = \delta t(\partial
\varphi/\partial t)$. We avoid this problem by assuming either
that the domain extends to infinity, where $\varphi = 0$, or that
we are in a periodic box and the contribution from one half of the
boundary cancels with the contribution from the other half. In
either case, we are assuming that there is no net energy flux
through the boundary.

By substituting \eq{varZpNoethers} and \eq{varvarphiNoethers} into
\eq{eq:actiontimetrans2} and observing that the term linear in
$\delta t$ in \eq{eq:actiontimetrans2} has to be identically zero,
that is, the sum of the second, third and fourth terms in the
right side must vanish, we find
\begin{eqnarray}\label{eq:actiontimetrans3}
\fl \delta t\Bigg[ - & \frac{\lambda_{De}^2\epsilon_s^2}{2L^2}
\int d^3 r \,|\nabla\varphi(\boldr,t')|^2 \nonumber \\ \fl & +
\sum_p \int d^6 Z_{p0} \, B_{||, p}^* (\bZ_{p0}) F_{p0}(\bZ_{p0})
\left. \left ( \sum_{\alpha = 1}^6 \dot Z_p^\alpha \frac{\partial
\overline{\mathcal{L}}_p}{\partial \dot Z_p^\alpha} -
\overline{\mathcal{L}}_p \right ) \right |_{t = t^\prime}
\Bigg]_{t'=t_0}^{t'=t_1} = 0.
\end{eqnarray}
Then,
\begin{equation}\label{eq:actiontimetrans5}
\Big [ \mathcal{H} (t^\prime) \Big]_{t'=t_0}^{t'=t_1} = 0.
\end{equation}
for solutions of the equations of motion, where
\begin{eqnarray}\label{eq:actiontimetrans6}
\fl{\cal H}(t)= \sum_p \int d^6 Z_{p0} \, B_{||, p}^* (\bZ_{p0})
F_{p0}(\bZ_{p0}) \overline{H}_p \Big( \bZ_p(\bZ_{p0}, t_0;
t),\varphi(\bullet,t) \Big) \nonumber \\ -
\frac{\lambda_{De}^2\epsilon_s^2}{2L^2}  \int d^3 r
\,|\nabla\varphi(\boldr,t)|^2.
\end{eqnarray}
Since this must hold for every interval $[t_0,t_1]$, we deduce
that
\begin{equation}
\frac{d {\cal H}(t)}{d t} = 0
\end{equation}
on the equations of motion. Finally, using equations \eq{FpFp0}
and \eq{eq:psconserv}, we can write the integrals over $\bZ_{p0}$
in \eq{eq:actiontimetrans6} as integrals over $\bZ_p$, giving
\begin{eqnarray}\label{eq:actiontimetrans7}
\fl{\cal H}(t) = \sum_p \int d^6 Z_p \, B_{||,p}^* (\bZ_p)
F_p(\bZ_p, t) \overline{H}_p \Big (\bZ_p, \varphi(\bullet,t) \Big)
- \frac{\lambda_{De}^2\epsilon_s^2}{2L^2}  \int d^3
r\,|\nabla\varphi(\boldr,t)|^2.
\end{eqnarray}
The conservation of this energy-like invariant is only satisfied
on the equations of motion, that is, the variational equations of
motion and the variational Poisson's equation obtained in
subsection~\ref{sub:variational} must be used. It is important to
keep all the given terms, even if they are higher order than first
(the order to which Poisson's equation is correct). For example,
it is necessary to obtain the \emph{exact} function
$\widehat{\bR}(\boldr, \mu_p, \theta_p, \epsilon_s/\lambda_p)$ in
\eq{eq:GKPoisson} to have exact conservation of $\mathcal{H}$.

\section{Conclusions and further work}
\label{sec:conclusions}

The gyrokinetic ordering in a static magnetic field consists of
the ordering assumptions \eq{orderings}, defined by a single
parameter $\epsilon=\rho/L=\omega/\Omega$. In this paper we have
strictly implemented the gyrokinetic ordering in the phase-space
Lagrangian to obtain explicitly the gyrokinetic Lagrangian to
order $\epsilon^2$ for general magnetic geometry.

Our approach differs from previous phase-space Lagrangian (or
Hamiltonian) derivations of gyrokinetics. In previous work
\cite{brizard07} the calculation is performed in two steps. First,
with zero fluctuating electrostatic potential, an expansion in
powers of $\epsilon = \rho/L$ is performed and a gyrophase
independent guiding-center Lagrangian is determined to order
$\epsilon$. Then, the electrostatic fluctuations, whose size is
given by a new expansion parameter $\epsilon_\varphi =
Ze\varphi/Mv_t^2$, are switched-on, reintroducing a gyrophase
dependence that is removed order by order in $\epsilon_\varphi$
yielding the final gyrokinetic Lagrangian, usually computed up to
order $\epsilon_\varphi^2$. When the expansion in
$\epsilon_\varphi$ is performed, the fact that there has been a
previous expansion in $\epsilon$ is ignored and the terms of order
$\epsilon\epsilon_\varphi$ are never calculated. Thus, the final
Lagrangian is missing relevant terms of order
$\epsilon\epsilon_\varphi$ and $\epsilon^2$.

The novelty of our work can be easily understood by examining the
explicit expression of the second-order gyrokinetic Hamiltonian,
$\overline{H}^{(2)}$, given in equations~\eq{hamiltonian_o2_2},
\eq{Psi2_phi}, \eq{Psi2_phiB} and \eq{Psi2_B}. It shows in a
transparent way that gyrokinetic theory ties together geometry and
turbulence, so that no splitting between guiding-center and
gyrokinetic dynamics is possible. See, for example,
$\Psi^{(2)}_{\phi B}$ in \eq{Psi2_phiB} where magnetic geometry
and electrostatic potential appear together.  This is the first
time that the electrostatic gyrokinetic equations in general
geometry are fully computed to order $\epsilon^2$ and the
calculations are pursued to the point of reaching formulae, like
the one for $\overline{H}^{(2)}$, that can be straightforwardly
implemented in a computer code.

From the new phase-space Lagrangian in \eq{finalL} and the new
Hamiltonian in \eq{finalH} we obtain a new Vlasov equation and a
new gyrokinetic Poisson's equation. The Vlasov equation is correct
to second order in the expansion parameter $\epsilon$, and it is
to our knowledge the highest order full $f$ gyrokinetic equation
available in the literature for general geometry. In the limit
where the electrostatic potential has a scale of variation much
larger than the gyroradius of the species of interest, this
equation is also the highest order drift kinetic equation that we
are aware of.

The gyrokinetic Poisson's equation derived from the new Lagrangian
is, however, only correct to first order in $\epsilon$. We have
calculated Poisson's equation employing two methods. In
Section~\ref{sec:GyroPoissonEq}, in the integrals in velocity
space that enter in Poisson's equation we have simply changed from
the coordinates $\{ \boldr, \bv \}$ to the gyrokinetic variables.
In Section~\ref{sec:fieldtheory} we have used a variational
formalism. The equation obtained with the variational formalism is
set by the form of the Hamiltonian and it is only correct to first
order, although it contains terms that are higher order. These
higher order terms do not add accuracy, but they are necessary to
have an exact energy-like invariant. By directly changing from $\{
\boldr, \bv \}$ to the gyrokinetic variables in the integrals that
enter Poisson's equation, it is possible to obtain a higher order
Poisson's equation if the higher order corrections
$\tilde{\bR}_3$, $\tilde{\mu}_2$ and $\tilde{\theta}_2$ are
calculated. It is not necessary to calculate the Hamiltonian to
next order. The calculation of $\tilde{\bR}_3$, $\tilde{\mu}_2$
and $\tilde{\theta}_2$ is however very tedious and is left for
future work. If this procedure is followed and the third order
Hamiltonian is not obtained, the conservation of the energy-like
invariant will not be exact.

Some natural steps following the present work are the extension to
the electromagnetic case and the introduction of external flows.
We will also investigate the implications of the new terms in the
Hamiltonian for the transport of toroidal angular momentum in
tokamaks.

\ack  The authors are indebted to the programme of visits to TJ-II
at Laboratorio Nacional de Fusi\'on of CIEMAT (Spain) and to the
summer programme ``Gyrokinetics in Laboratory and Astrophysical
Plasmas" at the Isaac Newton Institute for Mathematical Sciences,
without which this work would not have been possible.

This research was partially supported by the Post-doctoral
programme of the Engineering and Physical Sciences Research
Council of the UK, by the Junior Research Fellowship programme of
Christ Church at University of Oxford, and by grant ENE2009-07247,
Ministerio de Ciencia e Innovaci\'on (Spain).

\appendix

\section{Proof that the inverse of the matrix
in \eq{lagr_bra} defines a Poisson bracket}
\label{app_PoissonBracket}

Let $L_{\alpha\beta}(\bZ)$ be an invertible, antisymmetric matrix
of dimension $2n$ whose components are functions defined on a
region ${\cal U} \subset {\mathbb R}^{2n}$, and such that
\begin{equation}\label{eq:ClosedForm}
\frac{\partial L_{\alpha \beta}}{\partial Z^\gamma} +
\frac{\partial L_{\gamma \alpha}}{\partial Z^\beta} +
\frac{\partial L_{\beta \gamma}}{\partial Z^\alpha} = 0,
\quad 1\leq \alpha,\beta,\gamma \leq 2n.
\end{equation}
Denote by $P^{\alpha\beta}(\bZ)$ the inverse of
$L_{\alpha\beta}(\bZ)$,
$P^{\alpha\beta}(\bZ)=(L^{-1}(\bZ))^{\alpha\beta}$. Then,
$P^{\alpha\beta}$ defines a Poisson bracket by contraction with
the differentials of pairs of functions on ${\cal U}$, i.e.,
\begin{equation}\label{poi_braAppendix}
\{ F, G \} = \sum_{\alpha,\beta = 1}^{2n} P^{\alpha\beta}
\frac{\partial F}{\partial Z^\alpha} \frac{\partial G}{\partial
Z^\beta}.
\end{equation}

We must prove that \eq{poi_braAppendix} satisfies skew-symmetry,
given in \eq{eq:PoissonProperty1}, the Leibniz rule, given in
\eq{eq:PoissonProperty2}, and the Jacobi identity, given in
\eq{eq:PoissonProperty3}. Skew-symmetry is satisfied because the
inverse of an antisymmetric matrix is antisymmetric, i.e.,
$P^{\alpha \beta} = - P^{\beta \alpha}$. The Leibniz rule is
trivial to check. As for the Jacobi identity, it is immediate to
see that it is equivalent to
\begin{equation}\label{eq:JacobiComponents}
\fl \sum_{\delta = 1}^{2n} P^{\alpha\delta}\frac{\partial
P^{\beta\gamma}}{\partial Z^\delta} + \sum_{\delta = 1}^{2n}
P^{\beta\delta}\frac{\partial P^{\gamma\alpha}}{\partial Z^\delta}
+ \sum_{\delta = 1}^{2n} P^{\gamma\delta}\frac{\partial
P^{\alpha\beta}}{\partial Z^\delta}=0, \quad 1\leq
\alpha,\beta,\gamma \leq 2n.
\end{equation}
Showing that \eq{eq:JacobiComponents} is equivalent to
\eq{eq:ClosedForm} is a simple exercise of application of the
formula of the derivative of the inverse of a matrix,
\begin{equation}\label{eq:formuladerivmatrix}
\frac{\partial (L^{-1})^{\alpha \beta}}{\partial Z^\gamma} = -
\sum_{\delta, \rho = 1}^{2n} (L^{-1})^{\alpha \delta}
\frac{\partial L_{\delta \rho}}{\partial Z^\gamma} (L^{-1})^{\rho
\beta}.
\end{equation}

\section{Calculation of the Lagrangian after the non-perturbative change of variables}
\label{app_DK}

In this Appendix we prove equations \eq{dS_NPdR} and
\eq{dS_NPdtheta}. For equation \eq{dS_NPdR}, we use that according
to \eq{S_NP}
\begin{eqnarray} \label{dSdR_app1}
\fl \nabla_{\bR_g} S_{NP} = - \int_0^{\mu_g}
\frac{d\mu_g^\prime}{2\mu_g^\prime} \, \big [ \nabla \bA (\bR_g +
\epsilon \rhobf^\prime ) \cdot \rhobf^\prime + \epsilon
\nabla_{\bR_g} \rhobf^\prime \cdot \nabla \bA (\bR_g + \epsilon
\rhobf^\prime) \cdot \rhobf^\prime \nonumber \\ + \nabla_{\bR_g}
\rhobf^\prime \cdot \bA (\bR_g + \epsilon \rhobf^\prime) \big ].
\end{eqnarray}
Recall that the prime $^\prime$ here indicates that the function
depends on $\mu_g^\prime$. Employing
\begin{eqnarray}
\fl \frac{1}{2\mu_g^\prime} \nabla \bA (\bR_g + \epsilon
\rhobf^\prime ) \cdot \rhobf^\prime = \frac{1}{2\mu_g^\prime}
\rhobf^\prime \cdot \nabla \bA (\bR_g + \epsilon \rhobf^\prime ) +
\frac{1}{2\mu_g^\prime} \rhobf^\prime \times \left [ \nabla \times
\bA (\bR_g + \epsilon \rhobf^\prime ) \right ] \nonumber \\ =
\frac{1}{\epsilon} \frac{\partial}{\partial \mu_g^\prime} [ \bA
(\bR_g + \epsilon \rhobf^\prime ) ] + \frac{1}{2 \mu_g^\prime}
\rhobf^\prime \times \bB (\bR_g + \epsilon \rhobf^\prime ),
\end{eqnarray}
\begin{eqnarray}
\fl \frac{\epsilon}{2\mu_g^\prime} \nabla_{\bR_g} \rhobf^\prime
\cdot \nabla \bA (\bR_g + \epsilon \rhobf^\prime) \cdot
\rhobf^\prime \nonumber \\ = \frac{\epsilon}{2\mu_g^\prime}
\nabla_{\bR_g} \rhobf^\prime \cdot [ \rhobf^\prime \cdot \nabla
\bA (\bR_g + \epsilon \rhobf^\prime) ] +
\frac{\epsilon}{2\mu_g^\prime} \nabla_{\bR_g} \rhobf^\prime \cdot
[ \rhobf^\prime \times \bB (\bR_g + \epsilon \rhobf^\prime) ]
\nonumber \\ = \nabla_{\bR_g} \rhobf^\prime \cdot
\frac{\partial}{\partial \mu_g^\prime} [ \bA (\bR_g + \epsilon
\rhobf^\prime) ] + \frac{\epsilon}{2\mu_g^\prime} \nabla_{\bR_g}
\rhobf^\prime \cdot [ \rhobf^\prime \times \bB (\bR_g + \epsilon
\rhobf^\prime) ]
\end{eqnarray}
and
\begin{equation}
\fl \frac{1}{2\mu_g^\prime} \nabla_{\bR_g} \rhobf^\prime \cdot \bA
(\bR_g + \epsilon \rhobf^\prime) = \nabla_{\bR_g} \left (
\frac{\partial \rhobf^\prime}{\partial \mu_g^\prime} \right )
\cdot \bA (\bR_g + \epsilon \rhobf^\prime),
\end{equation}
equation \eq{dSdR_app1} becomes
\begin{eqnarray} \label{dSdR_app2}
\fl \nabla_{\bR_g} S_{NP} = - \frac{1}{\epsilon} \bA (\bR_g +
\epsilon \rhobf) + \frac{1}{\epsilon} \bA_g - \nabla_{\bR_g}
\rhobf \cdot \bA (\bR_g + \epsilon \rhobf) \nonumber \\ -
\int_0^{\mu_g} \frac{d\mu_g^\prime}{2\mu_g^\prime} \, \left [
\rhobf^\prime \times \bB (\bR_g + \epsilon \rhobf^\prime ) +
\epsilon \nabla_{\bR_g} \rhobf^\prime \cdot ( \rhobf^\prime \times
\bB (\bR_g + \epsilon \rhobf^\prime) ) \right ].
\end{eqnarray}
To obtain equation \eq{dS_NPdR} from this equation we use
\eq{drhodmu} to write
\begin{equation}
\int_0^{\mu_g} \frac{d\mu_g^\prime}{2\mu_g^\prime} \,
\rhobf^\prime \times \bB_g = \rhobf \times \bB_g,
\end{equation}
and we employ \eq{gradrho} to get
\begin{eqnarray}
\fl \nabla_{\bR_g} \rhobf^\prime \cdot ( \rhobf^\prime \times \bB
(\bR_g + \epsilon \rhobf^\prime) ) = [ ( \rhobf^\prime \times
\bun_g) \cdot \bB (\bR_g + \epsilon \rhobf^\prime) ]\nabla_{\bR_g}
\bun_g \cdot \rhobf^\prime \nonumber \\ +
\frac{2\mu_g^\prime}{B_g} [ \bun_g \cdot \bB (\bR_g + \epsilon
\rhobf^\prime) ] \nabla_{\bR_g} \eun_{2g} \cdot \eun_{1g}.
\end{eqnarray}

For equation \eq{dS_NPdtheta}, we use that according to \eq{S_NP}
\begin{eqnarray} \label{dSdtheta_app1}
\fl \frac{\partial S_{NP}}{\partial \theta_g} = - \int_0^{\mu_g}
\frac{d\mu_g^\prime}{2\mu_g^\prime} \, \left [ \epsilon
\frac{\partial \rhobf^\prime}{\partial \theta_g} \cdot \nabla \bA
(\bR_g + \epsilon \rhobf^\prime) \cdot \rhobf^\prime +
\frac{\partial \rhobf^\prime}{\partial \theta_g} \cdot \bA (\bR_g
+ \epsilon \rhobf^\prime) \right ].
\end{eqnarray}
Employing
\begin{eqnarray}
\fl \frac{\epsilon}{2\mu_g^\prime} \frac{\partial
\rhobf^\prime}{\partial \theta_g} \cdot \nabla \bA (\bR_g +
\epsilon \rhobf^\prime) \cdot \rhobf^\prime \nonumber
\\ = \frac{\epsilon}{2\mu_g^\prime} \frac{\partial
\rhobf^\prime}{\partial \theta_g} \cdot [ \rhobf^\prime \cdot
\nabla \bA (\bR_g + \epsilon \rhobf^\prime) ] +
\frac{\epsilon}{2\mu_g^\prime} \frac{\partial
\rhobf^\prime}{\partial \theta_g} \cdot [ \rhobf^\prime \times \bB
(\bR_g + \epsilon \rhobf^\prime) ] \nonumber \\ = \frac{\partial
\rhobf^\prime}{\partial \theta_g} \cdot \frac{\partial}{\partial
\mu_g^\prime} [ \bA (\bR_g + \epsilon \rhobf^\prime) ] +
\frac{\epsilon}{2\mu_g^\prime} \frac{\partial
\rhobf^\prime}{\partial \theta_g} \cdot [ \rhobf^\prime \times \bB
(\bR_g + \epsilon \rhobf^\prime) ]
\end{eqnarray}
and
\begin{equation}
\frac{1}{2\mu_g^\prime} \frac{\partial \rhobf^\prime}{\partial
\theta_g} \cdot \bA (\bR_g + \epsilon \rhobf^\prime) =
\frac{\partial^2 \rhobf^\prime}{\partial \theta_g \partial
\mu_g^\prime} \cdot \bA (\bR_g + \epsilon \rhobf^\prime),
\end{equation}
equation \eq{dSdtheta_app1} becomes equation \eq{dS_NPdtheta}. To
obtain the final form of the equation we have also used
\eq{drhodtheta} to write
\begin{eqnarray}
\frac{\partial \rhobf^\prime}{\partial \theta_g} \cdot [
\rhobf^\prime \times \bB (\bR_g + \epsilon \rhobf^\prime) ] = -
\frac{2\mu_g^\prime}{B_g} \bun_g \cdot \bB (\bR_g + \epsilon
\rhobf^\prime).
\end{eqnarray}

\section{Comparison to first order with the results in reference \cite{parra08}}
\label{app_comp}
In this Appendix we compare the gyrokinetic variable
transformation obtained in this article with the variables found
in \cite{parra08}.

To be able to compare with the results in \cite{parra08}, given in
the form $\bR_{PC} (\boldr, \bv, t)$, $E_{PC} (\boldr, \bv, t)$,
$\mu_{PC} (\boldr, \bv, t)$ and $\theta_{PC} (\boldr, \bv, t)$, we
will use the transformation $(\boldr, \bv) = T_\epsilon (\bR, u,
\mu, \theta, t)$ to write them as $\bR_{PC} (\bR, u, \mu, t)$,
$E_{PC} (\bR, u, \mu, t)$, $\mu_{PC} (\bR, u, \mu, t)$ and
$\theta_{PC} (\bR, u, \mu, \theta, t)$. After doing so, we will
see that the gyrokinetic variables $\bR_{PC}$, $E_{PC}$ and
$\mu_{PC}$ in \cite{parra08} are gyrophase independent quantities
up to the order that they are defined. This is a property that
must be satisfied because otherwise the new gyrokinetic variables
would have fast time dependence through the gyrophase. The
variable $\mu_{PC}$ must be a function of $\mu$ only because there
is only one adiabatic invariant associated with the gyromotion.

The gyrokinetic variables of \cite{parra08} are the gyrocenter
position
\begin{eqnarray}
\fl \bR_{PC} = & \boldr + \frac{\epsilon}{B} \bv \times \bun +
\frac{\epsilon^2}{B} \left [ \left ( v_{||} \bun + \frac{1}{4}
\bv_\bot \right ) (\bv \times \bun) + (\bv \times \bun) \left (
v_{||} \bun + \frac{1}{4} \bv_\bot \right ) \right ]
\dotcross \nabla \left ( \frac{ \bun }{ B } \right ) \nonumber \\
\fl & + \frac{\epsilon^2 v_{||}}{B^2} \bv_\bot \cdot \nabla \bun +
\frac{\epsilon^2 v_{||}}{B^2} \bun \bun \cdot \nabla \bun \cdot
\bv_\bot  + \frac{\epsilon^2}{8B^2} \bun [ \bv_\bot \bv_\bot -
(\bv \times \bun) (\bv \times \bun) ] : \nabla \bun \nonumber\\
\fl & - \frac{\Lambda \epsilon^2}{\lambda B^2}
\nabla_{(\bR_\bot/\lambda \epsilon)} \Phiwig \times \bun +
O(\epsilon^3),
\end{eqnarray}
where $\mathbf{a} \mathbf{b} \dotcross \matrixtop{\mathbf{M}} =
\mathbf{a} \times ( \mathbf{b} \cdot \matrixtop{\mathbf{M}} )$;
the gyrokinetic kinetic energy
\begin{equation}
E_{PC} = \frac{v^2}{2} + \Lambda \epsilon \phiwig + O(\epsilon^2);
\end{equation}
the magnetic moment
\begin{eqnarray}
\fl \mu_{PC} = \frac{v_\bot^2}{2B} - \frac{\epsilon
v_\bot^2}{2B^3} (\bv \times \bun) \cdot \nabla B - \frac{\epsilon
v_{||}^2}{B^2} \bun \cdot \nabla \bun \cdot (\bv \times \bun) -
\frac{\epsilon v_{||} v_\bot^2}{2B^2} \bun \cdot \nabla \times
\bun \nonumber \\ - \frac{\epsilon v_{||}}{4B^2} [ \bv_\bot (\bv
\times \bun) + (\bv \times \bun) \bv_\bot ] : \nabla \bun +
\frac{\Lambda \epsilon}{B} \phiwig + O(\epsilon^2);
\end{eqnarray}
and the gyrophase
\begin{eqnarray}
\fl \theta_{PC} = \arctan \left ( \frac{\bv \cdot \eun_2}{\bv
\cdot \eun_1} \right ) - \frac{\epsilon}{B^2} \bv_\bot \cdot
\nabla B - \frac{\epsilon v_{||}^2}{v_\bot^2 B} \bun \cdot \nabla
\bun \cdot \bv_\bot + \frac{\epsilon}{B} (\bv \times \bun) \cdot
\nabla \eun_2 \cdot \eun_1 \nonumber \\ - \frac{\epsilon v_{||}}{4
v_\bot^2 B} [ \bv_\bot \bv_\bot - (\bv \times \bun) (\bv \times
\bun) ]: \nabla \bun - \frac{\Lambda \epsilon}{\lambda^2 B}
\frac{\partial \Phiwig}{\partial (\mu/\lambda^2)} + O(\epsilon^2).
\end{eqnarray}

To these variables we apply the gyrokinetic transformation. First
we apply the non-perturbative change of variables
$T_{NP,\epsilon}$. For the variable $\bR_{PC}$ we find
\begin{eqnarray} \label{Rp_1}
\fl \bR_{PC} = & \bR_g - \epsilon^2 \rhobf \cdot \nabla_{\bR_g}
\left ( \frac{\bun_g}{B_g} \right ) \times \left ( v_{||g} \bun_g
+ \rhobf \times \bB_g \right )  + \frac{\epsilon^2 v_{||g}}{B_g}
(\rhobf \times \bun_g) \cdot \nabla_{\bR_g} \bun_g \nonumber \\\fl
& - \epsilon^2 \left [ \left ( v_{||g} \bun_g + \frac{1}{4} \rhobf
\times \bB_g \right ) \rhobf + \rhobf \left ( v_{||g} \bun_g +
\frac{1}{4} \rhobf \times \bB_g \right ) \right ] \dotcross
\nabla_{\bR_g} \left ( \frac{ \bun_g }{ B_g } \right ) \nonumber
\\ \fl & + \frac{\epsilon^2 v_{||g}}{B_g} \bun \bun \cdot
\nabla_{\bR_g} \bun \cdot (\rhobf \times \bun_g) -
\frac{\epsilon^2}{8} \bun_g [ \rhobf \rhobf - (\rhobf \times
\bun_g) (\rhobf \times \bun_g) ] : \nabla_{\bR_g} \bun_g
\nonumber\\ \fl & - \frac{\Lambda \epsilon^2}{\lambda B_g^2}
\nabla_{(\bR_\bot/\lambda \epsilon)} \Phiwig \times \bun_g +
O(\epsilon^3),
\end{eqnarray}
where we have used $\boldr = \bR_g + \epsilon \rhobf$, $B(\boldr)
= B_g + O(\epsilon)$, $\bun(\boldr) = \bun_g + O(\epsilon)$,
$v_{||} = v_{||g} + O(\epsilon)$, $\bv_\bot = \rhobf \times \bB_g
+ O(\epsilon)$ and
\begin{eqnarray}
\fl \frac{1}{B(\boldr)} \bv \times \bun(\boldr) = \frac{1}{B_g}
\bv \times \bun_g - \epsilon \rhobf \cdot \nabla_{\bR_g} \left (
\frac{\bun_g}{B_g} \right ) \times \left ( v_{||g} \bun_g + \rhobf
\times \bB_g \right ) + O(\epsilon^2) \nonumber \\ = - \rhobf  -
\epsilon \rhobf \cdot \nabla_{\bR_g} \left ( \frac{\bun_g}{B_g}
\right ) \times \left ( v_{||g} \bun_g + \rhobf \times \bB_g
\right ) + O(\epsilon^2).
\end{eqnarray}
We then apply the transformation $(\bR_g, v_{||g}, \mu_g,
\theta_g) = T_{P, \epsilon} (\bR, u, \mu, \theta, t)$ by using
$\bR_g = \bR + \epsilon^2 \bR_2 + O(\epsilon^3)$, with $\bR_2$
given in \eq{R2}, $v_{||g} = u + O(\epsilon)$, $\mu_g = \mu +
O(\epsilon)$ and $\theta_g = \theta + O(\epsilon)$. As a result we
find the expression
\begin{eqnarray} \label{Rp_2}
\fl \bR_{PC} = & \bR + \frac{\epsilon^2}{2} \bun [ \rhobf \rhobf +
(\rhobf \times \bun) (\rhobf \times \bun) ] : \nabla_\bR \bun +
\frac{\epsilon^2}{4B} [ \rhobf \rhobf + (\rhobf \times \bun)
(\rhobf \times \bun) ] \cdot \nabla_\bR B \nonumber
\\ \fl & + \frac{\epsilon^2 u}{B} (\rhobf \times \bun) \cdot
\nabla_\bR \bun + \epsilon^2 u \bun \cdot \nabla_\bR \left (
\frac{\bun}{B} \right ) \times \rhobf - \frac{\epsilon^2 u}{B}
\bun \bun \cdot \nabla_\bR \bun \cdot (\rhobf \times \bun)
\nonumber \\ \fl & - \frac{\epsilon^2 u}{B} \bun \times \nabla_\bR
\bun \cdot \rhobf + O(\epsilon^3),
\end{eqnarray}
where $B$ and $\bun$ depend on $\bR$. To simplify expression
\eq{Rp_2} we use
\begin{equation} \label{matI}
\rhobf \rhobf + (\rhobf \times \bun) (\rhobf \times \bun) =
\frac{2\mu}{B} ( \matI - \bun \bun )
\end{equation}
and
\begin{equation}
\fl (\rhobf \times \bun) \cdot \nabla_\bR \bun - \bun \times
\nabla_\bR \bun \cdot \rhobf = [ (\bun \times \nabla_\bR) \times
\bun ] \times \rhobf = (\nabla_\bR \cdot \bun) (\rhobf \times
\bun).
\end{equation}
We finally obtain
\begin{eqnarray} \label{Rp_final}
\bR_{PC} = & \bR - \frac{\epsilon^2 \mu}{B^2} \bun \bun \cdot
\nabla_\bR B + \frac{\epsilon^2 \mu}{2B^2} \nabla_{\bR\bot} B +
O(\epsilon^3).
\end{eqnarray}
To find this result we have used $\nabla_\bR \cdot \bun = - B^{-1}
\bun \cdot \nabla_\bR B$.

Applying the non-perturbative change of variables to $E_{PC}$ we
find
\begin{equation}
\fl E_{PC} = \frac{1}{2} v_{||g}^2 + \frac{1}{2} | \rhobf \times
\bB_g |^2 + \Lambda \epsilon \phiwig + O(\epsilon^2) = \frac{1}{2}
v_{||g}^2 + \mu_g B_g + \Lambda \epsilon \phiwig + O(\epsilon^2).
\end{equation}
To transform to the variables $\bR$, $u$, $\mu$ and $\theta$ we
use $v_{||g} = u + \epsilon u_1 + O(\epsilon^2)$, $\mu_g = \mu +
\epsilon \mu_1 + O(\epsilon^2)$, with $u_1$ and $\mu_1$ given in
\eq{u1} and \eq{mu1}, $\bR_g = \bR + O(\epsilon^2)$ and $\theta_g
= \theta + O(\epsilon)$, giving
\begin{equation}\label{Ep_final}
E_{PC} = \frac{1}{2} u^2 + \mu B (\bR) + O(\epsilon^2).
\end{equation}

For $\mu_{PC}$, we find that the transformation $(\boldr, \bv) =
T_{NP, \epsilon} ( \bR_g, v_{||g}, \mu_g, \theta_g)$ gives
\begin{eqnarray}
\fl \mu_{PC} = \mu_g - \epsilon v_{||g} \rhobf \cdot
\nabla_{\bR_g} \bun_g \cdot (\rhobf \times \bun_g) +
\frac{\epsilon v_{||g}^2}{B_g} \bun_g \cdot \nabla_{\bR_g} \bun_g
\cdot \rhobf - \frac{\epsilon v_{||g} \mu_g}{B_g} \bun_g \cdot
\nabla_{\bR_g} \times \bun_g \nonumber \\ + \frac{\epsilon
v_{||g}}{4} [ (\rhobf \times \bun_g) \rhobf + \rhobf (\rhobf
\times \bun_g) ] : \nabla_{\bR_g} \bun_g + \frac{\Lambda
\epsilon}{B_g} \phiwig + O(\epsilon^2),
\end{eqnarray}
where we have used $v_{||} = v_{||g} + O(\epsilon)$, $\bv_\bot =
\rhobf \times \bB_g + O(\epsilon)$,
\begin{eqnarray}
\fl \frac{v_\bot^2}{2B(\boldr)} = \frac{1}{2B_g} [ v^2 - (\bv
\cdot \bun(\boldr))^2 ] - \frac{\epsilon \mu_g}{B_g} \rhobf \cdot
\nabla_{\bR_g} B_g + O(\epsilon^2)
\end{eqnarray}
and
\begin{eqnarray}
\fl v^2 - (\bv \cdot \bun(\boldr))^2 = v_{||g}^2 + 2 \mu_g B_g -
\left [ \left ( v_{||g} \bun_g + \rhobf \times \bB_g \right )
\cdot \bun (\boldr) \right ]^2 + O(\epsilon^2) \nonumber \\ =
2\mu_g B_g - 2 \epsilon v_{||g} \rhobf \cdot \nabla_{\bR_g} \bun_g
\cdot (\rhobf \times \bB_g) + O(\epsilon^2).
\end{eqnarray}
Then employing $\mu_g = \mu + \epsilon \mu_1 + O(\epsilon^2)$,
with $\mu_1$ given in \eq{mu1}, $\bR_g = \bR + O(\epsilon^2)$,
$v_{||g} = u + O(\epsilon)$ and $\theta_g = \theta + O(\epsilon)$,
we find
\begin{eqnarray}\label{mup_final}
\fl \mu_{PC} = \mu - \frac{\epsilon v_{||g} \mu_g}{B_g} \bun_g
\cdot \nabla_{\bR_g} \times \bun_g + \frac{\epsilon v_{||g}}{2} [
\rhobf (\rhobf \times \bun_g) - (\rhobf \times \bun_g) \rhobf ] :
\nabla_{\bR_g} \bun_g + O(\epsilon^2) \nonumber
\\ = \mu + O(\epsilon^2),
\end{eqnarray}
where to obtain the final equality we have used \eq{matI}.

Finally, for $\theta_{PC}$ we obtain that the non-perturbative
transformation gives
\begin{eqnarray} \label{thetap_1}
\fl \theta_{PC} = \theta_g  - \frac{\epsilon v_{||g}}{2\mu_g}
\rhobf \cdot \nabla_{\bR_g} \bun_g \cdot \rhobf -
\frac{\epsilon}{B_g} (\rhobf \times \bun_g) \cdot \nabla_{\bR_g}
B_g - \frac{\epsilon v_{||g}^2}{2 \mu_g B_g} \bun_g \cdot
\nabla_{\bR_g} \bun_g \cdot (\rhobf \times \bun_g) \nonumber\\ +
\frac{\epsilon v_{||g}}{8 \mu_g} [ \rhobf \rhobf - (\rhobf \times
\bun_g) (\rhobf \times \bun_g) ]: \nabla_{\bR_g} \bun_g -
\frac{\Lambda \epsilon}{\lambda^2 B_g} \frac{\partial
\Phiwig}{\partial (\mu/\lambda^2)} + O(\epsilon^2),
\end{eqnarray}
where we have used
\begin{eqnarray}
\fl \arctan \left ( \frac{\bv \cdot \eun_2(\boldr)}{\bv \cdot
\eun_1(\boldr)} \right ) = \arctan \left [ \frac{( v_{||g} \bun_g
+ \rhobf \times \bB_g) \cdot \eun_2(\boldr)}{(v_{||g} \bun_g +
\rhobf \times \bB_g) \cdot \eun_1(\boldr)} \right ] \nonumber \\ =
\theta_g + \frac{\epsilon}{2\mu_g B_g} [ (\rhobf \times \bB_g)
\cdot \eun_{1g} ] \rhobf \cdot \nabla_{\bR_g} \eun_{2g} \cdot
(v_{||g} \bun_g + \rhobf \times \bB_g) \nonumber
\\ - \frac{\epsilon}{2\mu_g B_g} [ (\rhobf \times \bB_g) \cdot
\eun_{2g} ] \rhobf \cdot \nabla_{\bR_g} \eun_{1g} \cdot (v_{||g}
\bun_g + \rhobf \times \bB_g) + O(\epsilon^2) \nonumber \\ =
\theta_g - \frac{\epsilon v_{||g}}{2\mu_g} \rhobf \cdot
\nabla_{\bR_g} \bun_g \cdot \rhobf + \epsilon \rhobf \cdot
\nabla_{\bR_g} \eun_{2g} \cdot \eun_{1g} + O(\epsilon^2).
\end{eqnarray}
Here we have employed $\nabla_{\bR_g} \eun_{1g} = -
(\nabla_{\bR_g} \bun_g \cdot \eun_{1g}) \bun_g - (\nabla_{\bR_g}
\eun_{2g} \cdot \eun_{1g}) \eun_{2g}$ and $\nabla_{\bR_g}
\eun_{2g} = - (\nabla_{\bR_g} \bun_g \cdot \eun_{2g}) \bun_g +
(\nabla_{\bR_g} \eun_{2g} \cdot \eun_{1g}) \eun_{1g}$ to write the
last equality. Equation \eq{thetap_1} is now rewritten by using
$\theta_g = \theta + \epsilon \theta_1 + O(\epsilon^2)$, with
$\theta_1$ given in \eq{theta1}, $\bR_g = \bR + O(\epsilon^2)$,
$v_{||g} = u + O(\epsilon)$ and $\mu_g = \mu + O(\epsilon)$,
giving
\begin{eqnarray} \label{thetap_final}
\fl \theta_{PC} = \theta - \frac{\epsilon u}{4 \mu} [ \rhobf
\rhobf + (\rhobf \times \bun) (\rhobf \times \bun) ]: \nabla_\bR
\bun + O(\epsilon^2) = \theta + \frac{\epsilon u}{2B^2} \bun \cdot
\nabla_\bR B + O(\epsilon^2),
\end{eqnarray}
where we have employed \eq{matI} and $\nabla_\bR \cdot \bun = -
B^{-1} \bun \cdot \nabla_\bR B$.

From equation \eq{Rp_final}, \eq{Ep_final}, \eq{mup_final} and
\eq{thetap_final} we see that the gyrokinetic variables defined in
\cite{parra08} are simple functions of the gyrokinetic variables
in this article. Notice that $\bR_{PC}$ and $E_{PC}$ do not depend
on the gyrophase and that $\mu_{PC}$ is equal to $\mu$ to the
order of interest, as expected.

\section{Calculation of the second order gyrokinetic Hamiltonian}
\label{app_gyro_2}

In this Appendix we show how to obtain \eq{hamiltonian_o2_2} from
equation \eq{Hbar2_1}.

First, by employing equations \eq{GammaR_o2}, \eq{Gammatheta_o2},
\eq{S2}, \eq{R2} and \eq{nablaS2_slow}, we find that $\langle
\Gammabf^{(2)}_\bR \rangle = 0$, $\langle \nabla_\bR S^{(2)}_P
\rangle = 0$, $\langle \partial S^{(2)}_P/\partial t \rangle = 0$,
\begin{equation}
\left \langle \bR_2 \right \rangle = - \frac{\mu}{2B^2}
\nabla_{\bR\bot} B
\end{equation}
and
\begin{equation}
\langle \Gamma_\theta^{(2)} \rangle = \frac{\mu^2}{4B^2} (\matI -
\bun \bun) : \nabla_\bR \nabla_\bR \bB \cdot \bun.
\end{equation}
Here we have used equation \eq{rhorhoave} repeatedly. These
results give that equation \eq{Hbar2_1} is equal to
\begin{eqnarray}
\fl \overline{H}^{(2)} & = - \frac{u^2 \mu}{2B^2} \bun \cdot
\nabla_\bR \bun \cdot \nabla_\bR B + \frac{\mu^2}{4B} (\matI -
\bun \bun) : \nabla_\bR \nabla_\bR \bB \cdot \bun -
\frac{\mu^2}{2B^2} |\nabla_{\bR\bot} B|^2 + \frac{\langle u_1^2
\rangle }{2} \nonumber \\ \fl & + \langle \bR_2 \cdot
\nabla_{(\bR_\bot/\epsilon)} H^{(1)} \rangle + \mathcal{T}_1 +
\mathcal{T}_2 + \mathcal{T}_3 + \mathcal{T}_4,
\label{lagrangian_gyro_app1}
\end{eqnarray}
where we have used $(\nabla_\bR \times \bun) \times \bun = \bun
\cdot \nabla_\bR \bun$ to write
\begin{eqnarray}
\fl \bun \cdot [ (\nabla_\bR \times \bun) \times \langle \bR_2
\rangle ] = \frac{\mu}{2B^2} [(\nabla_\bR \times \bun) \times
\bun] \cdot \nabla_\bR B = \frac{\mu}{2B^2} \bun \cdot \nabla_\bR
\bun \cdot \nabla_\bR B.
\end{eqnarray}
The terms $\mathcal{T}_1$, $\mathcal{T}_2$, $\mathcal{T}_3$ and
$\mathcal{T}_4$ in equation \eq{lagrangian_gyro_app1} are
\begin{equation}
\mathcal{T}_1 = B \Bigg \langle \Bigg ( \Gammabf_{\bR\bot}^{(1)} +
\bR_2 \cdot \nabla_\bR \bA \Bigg ) \cdot \frac{\partial
\bR_2}{\partial \theta} \Bigg \rangle, \label{T1app2_1}
\end{equation}
\begin{equation}
\mathcal{T}_2 = B \Bigg \langle \left ( \Gammabf_\bR^{(1)} \cdot
\bun + u_1 \right ) \bun \cdot \frac{\partial \bR_2}{\partial
\theta} \Bigg \rangle, \label{T2app2_1}
\end{equation}
\begin{equation}
\mathcal{T}_3 = B \Bigg \langle \left (\Gamma_\theta^{(1)} - \mu_1
\right ) \frac{\partial \theta_1}{\partial \theta} \Bigg \rangle
\label{T3app2_1}
\end{equation}
and
\begin{eqnarray}
\fl \mathcal{T}_4 = - u \bun \cdot \Bigg \langle u_1
\frac{\partial \Gammabf_\bR^{(1)}}{\partial u} + \mu_1
\frac{\partial \Gammabf_\bR^{(1)}}{\partial \mu} + \theta_1
\frac{\partial \Gammabf_\bR^{(1)}}{\partial \theta} \Bigg \rangle
+ B \Bigg \langle \mu_1 \frac{\partial
\Gamma_\theta^{(1)}}{\partial \mu} + \theta_1 \frac{\partial
\Gamma_\theta^{(1)}}{\partial \theta} \Bigg \rangle \nonumber \\ +
\Bigg \langle \mu_1 \frac{\partial H^{(1)}}{\partial \mu} +
\theta_1 \frac{\partial H^{(1)}}{\partial \theta} \Bigg \rangle.
\label{T4app2_1}
\end{eqnarray}
We proceed to write these terms in more convenient forms.

The term $\mathcal{T}_1$ in \eq{T1app2_1} is rewritten using
\begin{eqnarray}
\fl \bR_2 \cdot \nabla_\bR \bA \cdot \frac{\partial
\bR_2}{\partial \theta} = \frac{1}{2} \left ( \bR_2 \cdot
\nabla_\bR \bA \cdot \frac{\partial \bR_2}{\partial \theta} -
\frac{\partial \bR_2}{\partial \theta} \cdot \nabla_\bR \bA \cdot
\bR_2 \right ) + \frac{1}{2} \frac{\partial}{\partial \theta}
\left ( \bR_2 \cdot \nabla_\bR \bA \cdot \bR_2 \right ) \nonumber
\\ = \frac{1}{2} ( \bB \times \bR_2 ) \cdot \frac{\partial
\bR_2}{\partial \theta} + \frac{1}{2} \frac{\partial}{\partial
\theta} \left ( \bR_2 \cdot \nabla_\bR \bA \cdot \bR_2 \right )
\end{eqnarray}
to obtain
\begin{equation}
\mathcal{T}_1 = B \Bigg \langle \Bigg ( \Gammabf_{\bR\bot}^{(1)} +
\frac{1}{2} \bB \times \bR_2 \Bigg ) \cdot \frac{\partial
\bR_2}{\partial \theta} \Bigg \rangle.
\end{equation}
Employing \eq{R2_1}, this last result becomes
\begin{equation}
\fl\mathcal{T}_1 = \frac{1}{2} \left \langle \left (
\Gammabf_\bR^{(1)} - \nabla_{(\bR_\bot/\epsilon)} S^{(2)}_P \right
) \cdot \left \{ \bun \times \left [ \frac{\partial
\Gammabf_\bR^{(1)}}{\partial \theta} +
\nabla_{(\bR_\bot/\epsilon)} \left ( \frac{\partial
S^{(2)}_P}{\partial \theta} \right ) \right ] \right \} \right
\rangle.
\end{equation}
Realizing that
\begin{eqnarray}
\fl \Gammabf_\bR^{(1)} \cdot \left [ \bun \times
\nabla_{(\bR_\bot/\epsilon)} \left ( \frac{\partial
S_P^{(2)}}{\partial \theta} \right ) \right ] -
\nabla_{(\bR_\bot/\epsilon)} S_P^{(2)} \cdot \left [ \bun \times
\frac{\partial \Gammabf_\bR^{(1)}}{\partial \theta} \right ]
\nonumber \\ = \frac{\partial}{\partial \theta} \left [
\Gammabf_\bR^{(1)} \cdot \left ( \bun \times
\nabla_{(\bR_\bot/\epsilon)} S_P^{(2)} \right ) \right ],
\end{eqnarray}
we finally obtain
\begin{equation}
\fl \mathcal{T}_1 = \frac{1}{2} \left \langle \Gammabf_\bR^{(1)}
\cdot \left ( \bun \times \frac{\partial
\Gammabf_\bR^{(1)}}{\partial \theta} \right ) \right \rangle -
\frac{1}{2} \left \langle \nabla_{(\bR_\bot/\epsilon)} S^{(2)}_P
\cdot \left [ \bun \times \nabla_{(\bR_\bot/\epsilon)} \left (
\frac{\partial S^{(2)}_P}{\partial \theta} \right ) \right ]
\right \rangle. \label{T1app2_2}
\end{equation}
These two terms can be expressed in terms of magnetic and electric
field gradients. Using equation \eq{GammaR_o1} and $\langle \rhobf
\rhobf \rangle$ from \eq{rhorhoave}, we find
\begin{eqnarray}
\fl \left \langle \Gammabf_\bR^{(1)} \cdot \left ( \bun \times
\frac{\partial \Gammabf_\bR^{(1)}}{\partial \theta} \right )
\right \rangle = \frac{u^2 \mu}{B} (\bun \times \nabla_\bR \bun
\times \bun) : (\nabla_\bR \bun)^\mathrm{T} - \frac{\mu^2}{2 B^2}
|\nabla_{\bR\bot} B|^2,
\end{eqnarray}
where $\matrixtop{\mathbf{M}}^\mathrm{T}$ is the transpose of the
matrix $\matrixtop{\mathbf{M}}$. Employing that $\bun \times
\nabla_\bR \bun \times \bun = (\nabla_{\bR\bot} \bun)^\mathrm{T} -
(\nabla_\bR \cdot \bun) ( \matI - \bun \bun )$, we finally find
\begin{eqnarray}
\fl \left \langle \Gammabf_\bR^{(1)} \cdot \left ( \bun \times
\frac{\partial \Gammabf_\bR^{(1)}}{\partial \theta} \right )
\right \rangle = \frac{u^2 \mu}{B} \nabla_\bR \bun : \nabla_\bR
\bun - \frac{u^2 \mu}{B} (\nabla_\bR \cdot \bun)^2 -
\frac{\mu^2}{2 B^2} |\nabla_{\bR\bot} B|^2.
\end{eqnarray}
Substituting this expression and the definition of $S^{(2)}_P$ in
\eq{S2} into equation \eq{T1app2_2} gives
\begin{eqnarray}
\fl \mathcal{T}_1 = \frac{u^2 \mu}{2B} \nabla_\bR \bun :
\nabla_\bR \bun - \frac{u^2 \mu}{2B} (\nabla_\bR \cdot \bun)^2 -
\frac{\mu^2}{4 B^2} |\nabla_{\bR\bot} B|^2 \nonumber \\ -
\frac{\Lambda^2}{2 \lambda^2 B^2} \left \langle
\nabla_{(\bR_\bot/\lambda \epsilon)} \Phiwig \cdot \left ( \bun
\times \nabla_{(\bR_\bot/\lambda \epsilon)} \phiwig \right )
\right \rangle. \label{T1app2_final}
\end{eqnarray}

Using equations \eq{R2_1} and \eq{u1_1}, the term $\mathcal{T}_2$
in \eq{T2app2_1} becomes
\begin{equation}
\mathcal{T}_2 = \bB \cdot \overline{\Gammabf}_\bR^{(1)} \left
\langle \frac{\partial^2 S^{(2)}_P}{\partial u \partial \theta}
\right \rangle = 0. \label{T2app2_final}
\end{equation}

Employing equations \eq{mu1_1} and \eq{theta1_1} the term
$\mathcal{T}_3$ in \eq{T3app2_1} is written as
\begin{equation}
\mathcal{T}_3 = B \left \langle \frac{\partial S^{(2)}_P}{\partial
\theta} \frac{\partial^2 S^{(2)}_P}{\partial \mu
\partial \theta} \right \rangle = \frac{B}{2}
\frac{\partial}{\partial \mu} \left \langle \left ( \frac{\partial
S^{(2)}_P}{\partial \theta} \right )^2 \right \rangle.
\label{T3app2_final}
\end{equation}

Finally, the term $\mathcal{T}_4$ in \eq{T4app2_1} is manipulated
to yield
\begin{eqnarray}
\fl \mathcal{T}_4 = - \left \langle u u_1 \bun \cdot
\frac{\partial \Gammabf_\bR^{(1)}}{\partial u} \right \rangle +
\left \langle \left ( \mu_1 \frac{\partial}{\partial \mu} +
\theta_1 \frac{\partial}{\partial \theta} \right ) \left ( H^{(1)}
- u \bun \cdot \Gammabf_\bR^{(1)} + B \Gamma^{(1)}_\theta \right )
\right \rangle.
\end{eqnarray}
Using equation \eq{eq_S1GK} to write $H^{(1)} - u \bun \cdot
\Gammabf_\bR^{(1)} + B \Gamma^{(1)}_\theta = - B (
\partial S^{(2)}_P/\partial \theta) + \langle H^{(1)} - u \bun
\cdot \Gammabf_\bR^{(1)} + B \Gamma^{(1)}_\theta \rangle$ and
equation \eq{u1_1} to write $\bun \cdot \Gammabf_\bR^{(1)} = - u_1
+ \bun \cdot \overline{\Gammabf}_\bR^{(1)}$, and employing that
$\langle u_1 \rangle = 0$ and $\langle \mu_1 \rangle = 0$, we
obtain
\begin{equation}
\mathcal{T}_4 = \frac{u}{2} \frac{\partial \langle u_1^2
\rangle}{\partial u} - B \left \langle \mu_1 \frac{\partial^2
S^{(2)}_P}{\partial \mu
\partial \theta} + \theta_1 \frac{\partial^2 S^{(2)}_P}{\partial
\theta^2} \right \rangle.
\end{equation}
Using equations \eq{mu1_1} and \eq{theta1_1}, we find
\begin{equation}
\mathcal{T}_4 = \frac{u}{2} \frac{\partial \langle u_1^2
\rangle}{\partial u} - B \left \langle \Gamma_\theta^{(1)}
\frac{\partial^2 S^{(2)}_P}{\partial \mu
\partial \theta} \right \rangle - B \frac{\partial}{\partial
\mu} \left \langle \left ( \frac{\partial S^{(2)}_P}{\partial
\theta} \right )^2 \right \rangle. \label{T4app2_final}
\end{equation}
Here we have used
\begin{equation}
\fl \left \langle \frac{\partial S^{(2)}_P}{\partial \theta}
\frac{\partial^2 S^{(2)}_P}{\partial \mu \partial \theta} -
\frac{\partial S^{(2)}_P}{\partial \mu} \frac{\partial^2
S^{(2)}_P}{\partial \theta^2} \right \rangle = 2 \left \langle
\frac{\partial S^{(2)}_P}{\partial \theta} \frac{\partial^2
S^{(2)}_P}{\partial \mu \partial \theta} \right \rangle
=\frac{\partial}{\partial \mu} \left \langle \left (
\frac{\partial S^{(2)}_P}{\partial \theta} \right )^2 \right
\rangle,
\end{equation}
where we have integrated by parts in $\theta$ to obtain the first
equality.

Substituting the results $H^{(1)} = \Lambda \phiave + \Lambda
\phiwig$, \eq{T1app2_final}, \eq{T2app2_final}, \eq{T3app2_final}
and \eq{T4app2_final} into equation \eq{lagrangian_gyro_app1}
gives
\begin{eqnarray}
\fl \overline{H}^{(2)} & =  - \frac{u^2 \mu}{2B^2} \bun \cdot
\nabla_\bR \bun \cdot \nabla_\bR B + \frac{\mu^2}{4B} (\matI -
\bun \bun) : \nabla_\bR \nabla_\bR \bB \cdot \bun -
\frac{3\mu^2}{4B^2} |\nabla_{\bR\bot} B|^2 \nonumber \\ \fl & +
\frac{u^2 \mu}{2B} \nabla_\bR \bun : \nabla_\bR \bun - \frac{u^2
\mu}{2B} (\nabla_\bR \cdot \bun)^2 + \frac{\Lambda}{\lambda}
\langle \bR_2 \cdot \nabla_{(\bR_\bot/\lambda \epsilon)} \phiwig
\rangle \nonumber\\ \fl & - \frac{\Lambda \mu}{2 \lambda B^2}
\nabla_\bR B \cdot \nabla_{(\bR_\bot/\lambda \epsilon)} \phiave -
\frac{\Lambda^2}{2 \lambda^2 B^2} \left \langle
\nabla_{(\bR_\bot/\lambda \epsilon)} \Phiwig \cdot \left ( \bun
\times \nabla_{(\bR_\bot/\lambda \epsilon)} \phiwig \right )
\right \rangle \nonumber \\ \fl & + \frac{\langle u_1^2 \rangle
}{2} + \frac{u}{2} \frac{\partial \langle u_1^2 \rangle}{\partial
u} - B \left \langle \Gamma_\theta^{(1)} \frac{\partial^2
S^{(2)}_P}{\partial \mu \partial \theta} \right \rangle -
\frac{B}{2} \frac{\partial}{\partial \mu} \left \langle \left (
\frac{\partial S^{(2)}_P}{\partial \theta} \right )^2 \right
\rangle. \label{lagrangian_gyro_app2}
\end{eqnarray}
Employing equations \eq{mu1_1}, \eq{eq_H1GK}, \eq{eq_Hbar1GK} and
\eq{H1} to write
\begin{equation}
B \frac{\partial S^{(2)}_P}{\partial \theta} = - \Lambda \phiwig -
u u_1 - B \Gamma_\theta^{(1)}, \label{dSdtheta}
\end{equation}
we find
\begin{eqnarray}
\fl - B \Bigg \langle & \Gamma_\theta^{(1)} \frac{\partial^2
S^{(2)}_P}{\partial \mu \partial \theta} \Bigg \rangle -
\frac{B}{2} \frac{\partial}{\partial \mu} \left \langle \left (
\frac{\partial S^{(2)}_P}{\partial \theta} \right )^2 \right
\rangle = - \frac{\Lambda^2}{2 \lambda^2 B} \frac{\partial \langle
\phiwig^2 \rangle}{\partial (\mu/\lambda^2)} - \frac{u^2}{2B}
\frac{\partial \langle u_1^2 \rangle}{\partial \mu} \nonumber\\
\fl & - \frac{\Lambda u}{\lambda^2 B} \left \langle \frac{\partial
\phiwig}{\partial (\mu/\lambda^2)} \, u_1 \right \rangle -
\frac{\Lambda u}{B} \left \langle \phiwig\, \frac{\partial
u_1}{\partial \mu} \right \rangle - \Lambda \left \langle
\phiwig\, \frac{\partial \Gamma_\theta^{(1)}}{\partial \mu} \right
\rangle - u \left \langle u_1 \frac{\partial
\Gamma_\theta^{(1)}}{\partial \mu} \right \rangle.
\end{eqnarray}
Substituting this result into \eq{lagrangian_gyro_app2} gives
equation \eq{hamiltonian_o2_2} with $\Psi^{(2)}_\phi$ and
$\Psi^{(2)}_{\phi B}$ as in \eq{Psi2_phi} and \eq{Psi2_phiB}, and
$\Psi^{(2)}_B$ given by
\begin{eqnarray}
\fl \Psi^{(2)}_B & = - \frac{u^2 \mu}{2B^2} \bun \cdot \nabla_\bR
\bun \cdot \nabla_\bR B + \frac{\mu^2}{4B} (\matI - \bun \bun) :
\nabla_\bR \nabla_\bR \bB \cdot \bun - \frac{3\mu^2}{4B^2}
|\nabla_{\bR\bot} B|^2 \nonumber \\ \fl & + \frac{u^2 \mu}{2B}
\nabla_\bR \bun : \nabla_\bR \bun - \frac{u^2 \mu}{2B} (\nabla_\bR
\cdot \bun)^2  + \frac{\langle u_1^2 \rangle }{2} + \frac{u}{2}
\frac{\partial \langle u_1^2 \rangle}{\partial u} - \frac{u^2}{2B}
\frac{\partial \langle u_1^2 \rangle}{\partial \mu} \nonumber \\
\fl & - u \left \langle u_1 \frac{\partial
\Gamma_\theta^{(1)}}{\partial \mu} \right \rangle.
\label{lagrangian_gyro_app3}
\end{eqnarray}

To obtain equation \eq{Psi2_B} from equation
\eq{lagrangian_gyro_app3} we only need to calculate $\langle u_1^2
\rangle$ and $\langle u_1 (\partial \Gamma_\theta^{(1)}/\partial
\mu) \rangle$. The gyroaverage of $u_1^2$ is
\begin{equation}
\fl \langle u_1^2 \rangle = \frac{u^2 \mu}{B} |\bun \cdot
\nabla_\bR \bun|^2 + \frac{B^2}{4} \left \langle \left [ (\rhobf
\times \bun) \cdot \nabla_\bR \bun \cdot \rhobf \right ]^2 \right
\rangle - \frac{\mu^2}{4} (\bun \cdot \nabla_\bR \times \bun)^2 ,
\label{u12ave}
\end{equation}
where we have used the definition of $u_1$ in \eq{u1}, we have
taken the gyroaverage $\langle \rhobf \rhobf \rangle$ from
\eq{rhorhoave}, and we have employed equation \eq{rhorhowig} to
write $[ \rhobf (\rhobf \times \bun) + (\rhobf \times \bun) \rhobf
] : \nabla_\bR \bun = 2 (\rhobf \times \bun ) \cdot \nabla_\bR
\bun \cdot \rhobf - 2 (\mu/B) \bun \cdot \nabla_\bR \times \bun$
and hence obtain
\begin{eqnarray}
\fl \left \langle \left \{ \left [ \rhobf ( \rhobf \times \bun ) +
(\rhobf \times \bun) \rhobf \right ]: \nabla_\bR \bun \right \}^2
\right \rangle = 4 \left \langle \left [ ( \rhobf \times \bun )
\cdot \nabla_\bR \bun \cdot \rhobf \right ]^2 \right \rangle
\nonumber \\ - \frac{4\mu^2}{B^2} (\bun \cdot \nabla_\bR \times
\bun)^2.
\end{eqnarray}
The gyroaverage of the second term in \eq{u12ave} is given by
\begin{eqnarray}
\fl \left \langle \left [ ( \rhobf \times \bun ) \cdot \nabla_\bR
\bun \cdot \rhobf \right ]^2 \right \rangle = \frac{\mu^2}{2B^2}
(\bun \cdot \nabla_\bR \times \bun)^2 + \frac{\mu^2}{2B^2}
\nabla_{\bR\bot} \bun : (\nabla_{\bR\bot} \bun)^\mathrm{T}
\nonumber \\ + \frac{\mu^2}{2B^2} (\bun \times \nabla_\bR \bun
\times \bun) : \nabla_\bR \bun = \frac{\mu^2}{2B^2} (\bun \cdot
\nabla_\bR \times \bun)^2 \nonumber \\ + \frac{\mu^2}{B^2}
\nabla_{\bR\bot} \bun : (\nabla_{\bR\bot} \bun)^\mathrm{T} -
\frac{\mu^2}{2B^2} (\nabla_\bR \cdot \bun)^2, \label{trick2}
\end{eqnarray}
\noindent where we have used
\begin{eqnarray}
\fl \langle \rho_i \rho_j \rho_k \rho_l \rangle =
\frac{\mu^2}{2B^2} [ (\delta_{ij} - \hat{b}_i \hat{b}_j)
(\delta_{kl} - \hat{b}_k \hat{b}_l) + (\delta_{ik} - \hat{b}_i
\hat{b}_k) (\delta_{jl} - \hat{b}_j \hat{b}_l) \nonumber \\ +
(\delta_{il} - \hat{b}_i \hat{b}_l) (\delta_{jk} - \hat{b}_j
\hat{b}_k) ].
\end{eqnarray}
Here $\delta_{ij}$ is the Kronecker delta. We have employed $\bun
\times \nabla_\bR \bun \times \bun = (\nabla_{\bR\bot}
\bun)^\mathrm{T} - (\nabla_\bR \cdot \bun) ( \matI - \bun \bun)$
to obtain the second equality in \eq{trick2}. Substituting
equation \eq{trick2} into equation \eq{u12ave} gives
\begin{eqnarray}
\fl \langle u_1^2 \rangle = \frac{u^2 \mu}{B} |\bun \cdot
\nabla_\bR \bun|^2 + \frac{\mu^2}{4} \nabla_{\bR\bot} \bun :
(\nabla_{\bR\bot} \bun)^\mathrm{T} - \frac{\mu^2}{8} (\nabla_\bR
\cdot \bun)^2 \nonumber \\ - \frac{\mu^2}{8} (\bun \cdot
\nabla_\bR \times \bun)^2. \label{u12ave_final}
\end{eqnarray}
The gyroaverage of $u_1 (\partial \Gamma_\theta^{(1)}/\partial
\mu)$ is
\begin{equation}
\left \langle u_1 \frac{\partial \Gamma_\theta^{(1)}}{\partial
\mu} \right \rangle = \frac{u\mu}{B^2} \bun \cdot \nabla_\bR \bun
\cdot \nabla_\bR B. \label{u1Gammatheta1_final}
\end{equation}
Finally, substituting equations \eq{u12ave_final} and
\eq{u1Gammatheta1_final} into \eq{lagrangian_gyro_app3} gives
\eq{Psi2_B}.

\section{Poisson bracket}
\label{app:poissonbrackets}

In this Appendix we prove that the Poisson bracket that
corresponds to the gyrokinetic Lagrangian in \eq{finalL} is
\eq{eq:poissonbracket}. Since the symplectic part\footnote{It is
common to call {\it the symplectic part} of a phase-space
Lagrangian to the piece linear in the time derivatives of the
coordinates.} of the gyrokinetic Lagrangian \eq{finalL} is exactly
the same as in \cite{brizard07}, the Poisson bracket in
gyrokinetic coordinates will also be.

As explained in subsection \ref{sub_change}, to obtain the Poisson
bracket, given in \eq{poi_bra}, we need to calculate the inverse
of the matrix $L$ in \eq{lagr_bra}. We explicitly write this
matrix by writing the gyrokinetic coordinates as $\{ Z^\alpha
\}_{\alpha = 1}^6$, with $(Z^1, Z^2, Z^3) = \bR$, $Z^4 = u$, $Z^5
= \mu$ and $Z^6 = \theta$. The gyrokinetic Lagrangian \eq{finalL}
is written as in \eq{gyroL_s}, with $(\overline{\Gamma}_1,
\overline{\Gamma}_2, \overline{\Gamma}_3) = \epsilon^{-1} \bA
(\bR) + u \bun (\bR) - \epsilon \mu \bK (\bR)$,
$\overline{\Gamma}_4 = 0$, $\overline{\Gamma}_5 = 0$ and
$\overline{\Gamma}_6 = - \epsilon \mu$. Then, using \eq{lagr_bra}
for the gyrokinetic Lagrangian, we find that the matrix $L$ is
given by
\begin{equation} \label{gyromatrixL}
L_{\alpha\beta}:=\frac{\partial \overline{\Gamma}_\beta}{\partial
Z^\alpha}- \frac{\partial \overline{\Gamma}_\alpha}{\partial
Z^\beta},
\end{equation}
or in matrix form
\begin{eqnarray}\label{eq:matrixL}
L = \left(
\begin{array}{ccc:c:c:c}
\ddots & & &\vdots &\vdots &\vdots \\
 &-\epsilon^{-1}\bB^*\times \matI& &-\bun&\epsilon\bK&\mathbf{0}\\
 & &\ddots &\vdots &\vdots &\vdots \\
\hdashline
\dots &\bun& \dots &0 &0 &0 \\
\hdashline
\dots&-\epsilon \bK&\dots&0&0&-\epsilon\\
\hdashline \dots&\mathbf{0}&\dots&0&\epsilon&0
\end{array}
\right),
\end{eqnarray}
where $\bB^*$ is defined in \eq{Bstar}. Its inverse is given by
\begin{eqnarray}
\fl P = L^{-1} = \frac{1}{B_{||}^*} \left(
\begin{array}{ccc:c:c:c}
\ddots & & &\vdots&\vdots&\vdots\\
 &\epsilon \bun \times \matI& &\bB^*&\mathbf{0}&\epsilon \bK \times \bun\\
 & &\ddots&\vdots &\vdots&\vdots\\
\hdashline \dots&-\bB^*&\dots&0&0& \bB^* \cdot \bK \\
\hdashline
\dots&\mathbf{0}&\dots&0&0&\epsilon^{-1} B^*_{||}\\
\hdashline \dots&-\epsilon \bK \times \bun&\dots&- \bB^* \cdot
\bK&-\epsilon^{-1} B^*_{||}&0
\end{array}
\right),
\end{eqnarray}
where $B_{||}^*$ is defined in \eq{Bstar_par}. It is easy to check
by direct calculation that $P$ is the inverse of $L$.

The Poisson bracket of two functions $F(\bZ)$ and $G(\bZ)$ is then
given by equation \eq{poi_bra} that can be compactly rewritten as
\eq{eq:poissonbracket}.

\section{Calculation of the Jacobian}
\label{app:jacobians}

In this Appendix we show that the determinant of the Jacobian
matrix of the gyrokinetic transformation is $B^*_{||}$, defined in
\eq{Bstar_par}. This result coincides with the results in
\cite{brizard07} because of our choice for the final form of the
Lagrangian \eq{finalL}.

To obtain the Jacobian of the gyrokinetic transformation, we use
the matrix $L$, defined in \eq{lagr_bra}. This matrix is defined
for both the original coordinates $\{ \boldr, \bv \}$ and the new
gyrokinetic coordinates $\{ \bR, u, \mu, \theta \}$. The matrix
$L$ in the original phase space and the matrix $L$ in the new
gyrokinetic phase space are related by the Jacobian matrix of the
gyrokinetic transformation $T_\epsilon$. It is possible to use
this relation to calculate the determinant of the Jacobian matrix
by calculating the matrix $L$ in both coordinate systems.

We denote the original coordinates by
$\{X^\alpha\}_{\alpha=1}^{6}$, with $(X^1, X^2, X^3) = \boldr$ and
$(X^4, X^5, X^6) = \bv$, and the gyrokinetic coordinates by $\{
Z^\alpha \}_{\alpha = 1}^6$, with $(Z^1, Z^2, Z^3) = \bR$, $Z^4 =
u$, $Z^5 = \mu$ and $Z^6 = \theta$. The Jacobian matrix of the
transformation is given by
\begin{equation}
(J_{T_\epsilon})^\alpha_{\beta}(\bZ) =\frac{\partial
X^\alpha(\bZ)}{\partial Z^\beta}.
\end{equation}
We write the Lagrangian in the coordinates $\bX$ as
\begin{equation}
{\cal L}^\bX = \sum_{\alpha=1}^6\gamma_\alpha(\bX)\frac{d
X^\alpha}{d t} - H^\bX(\bX,t),
\end{equation}
with $(\gamma_1, \gamma_2, \gamma_3) = \epsilon^{-1} \bA (\boldr)
+ \bv$, $\gamma_4 = 0$, $\gamma_5 = 0$ and $\gamma_6 = 0$. We
write the Lagrangian $\overline{\mathcal{L}}$ in gyrokinetic
coordinates $\bZ$ as in \eq{gyroL_s}, with $(\overline{\Gamma}_1,
\overline{\Gamma}_2, \overline{\Gamma}_3) = \epsilon^{-1} \bA
(\bR) + u \bun (\bR) - \epsilon \mu \bK (\bR)$,
$\overline{\Gamma}_4 = 0$, $\overline{\Gamma}_5 = 0$ and
$\overline{\Gamma}_6 = - \epsilon \mu$.

From the Lagrangians $\mathcal{L}^\bX$ and
$\overline{\mathcal{L}}$, we obtain the matrix $L$, defined in
\eq{lagr_bra}, in both coordinate systems, given by
\eq{gyromatrixL} for the gyrokinetic coordinates, and by
\begin{equation}
l_{\alpha\beta}:=\frac{\partial\gamma_\beta}{\partial X^\alpha}-
\frac{\partial\gamma_\alpha}{\partial X^\beta}
\end{equation}
for the original coordinates. It is immediate to check that $L =
J_{T_\epsilon}^\mathrm{T}{l}J_{T_\epsilon}$, with the superscript
$^\mathrm{T}$ standing for matrix transposition. It is then
obvious that
\begin{equation} \label{detJ}
\det(J_{T_\epsilon}) = \sqrt{\frac{\mbox{det}({L})}{\mbox{det}({
l})}},
\end{equation}
where we have used that the Jacobian of $T_\epsilon$ is positive
at $\epsilon = 0$ to determine the sign in front of the square
root. Then, to calculate the Jacobian is enough to calculate the
determinants of the matrices $L$ and $l$.

The matrix $l$ is
\begin{eqnarray}
l= \left(
\begin{array}{ccc:ccc}
\ddots & & &\ddots& & \\
 &- \epsilon^{-1} \bB \times \matI & & &- \matI& \\
 & &\ddots& & &\ddots\\
\hdashline
\ddots& & &\ddots& & \\
 &\matI& & &\mathbf{0}& \\
 & &\ddots& & &\ddots\\
\end{array}
\right),
\end{eqnarray}
and the matrix $L$ was given in \eq{eq:matrixL}. The determinant
of $l$ is
\begin{equation} \label{detl}
\mbox{det}({l})=1.
\end{equation}
As for $L$, given in \eq{eq:matrixL}, we have that
\begin{equation} \label{detL}
\mbox{det}({L}) = \epsilon^2 \left|
\begin{array}{ccc:c}
\ddots& & &\vdots\\
 &- \epsilon^{-1} \bB^* \times \matI& &-\bun\\
 & &\ddots&\vdots\\
\hdashline \dots&\bun&\dots&0
\end{array}
\right|.
\end{equation}
Writing the matrix in the reference system $\{ \eun_1, \eun_2,
\bun \}$, where $\bun = (0, 0, 1)$, we find that
\begin{eqnarray}
\fl \mbox{det} ({L}) = \epsilon^2 \left |
\begin{array}{ccc:c}
0&\epsilon^{-1} \bB^* \cdot \bun& -\epsilon^{-1} \bB^* \cdot \eun_2 &0\\
-\epsilon^{-1} \bB^* \cdot \bun&0&\epsilon^{-1} \bB^* \cdot \eun_1 &0\\
\epsilon^{-1} \bB^* \cdot \eun_2& - \epsilon^{-1} \bB^* \cdot \eun_1&0&-1\\
\hdashline 0&0&1&0
\end{array}
\right | = \epsilon^2 \left |
\begin{array}{cc}
0&\epsilon^{-1} B_{||}^*\\
- \epsilon^{-1} B^*_{||}&0
\end{array}
\right | \nonumber \\ = (B^*_{||})^2.
\end{eqnarray}
Substituting this result and \eq{detl} into \eq{detJ}, we finally
obtain
\begin{equation}\label{eq:Jacobianxi}
\det(J_{T_\epsilon} ) = B_{||}^*.
\end{equation}

\section{Conservation of phase-space volume}
\label{app_consps}

In this Appendix we prove that
\begin{equation} \label{consps_app}
\sum_{\alpha = 1}^6 \frac{\partial}{\partial Z^\alpha} \left [
\mbox{det} (J_{T_\epsilon}) \frac{dZ^\alpha}{dt} \right ] = 0.
\end{equation}
This equation is satisfied by any gyrokinetic Lagrangian
$\overline{\mathcal{L}}$ with a symplectic part
$\overline{\Gamma}_\alpha$ that is independent of time, as is in
our Lagrangian \eq{finalL}. Relation \eq{consps_app} gives
equation \eq{consps}.

To prove \eq{consps_app} we use equations \eq{eqmotion}, \eq{detJ}
and \eq{detl}. From \eq{detJ} and \eq{detl} we find
\begin{eqnarray} \label{app_ps_aux1}
\fl \sum_{\alpha = 1}^6 \frac{\partial}{\partial Z^\alpha} \left [
\mbox{det} (J_{T_\epsilon}) \frac{dZ^\alpha}{dt} \right ] =
\frac{1}{2 \sqrt{\mbox{det} (L)}} \Bigg \{ \sum_{\alpha = 1}^6
\frac{dZ^\alpha}{dt} \frac{\partial}{\partial Z^\alpha} \left [
\mbox{det} (L) \right ] \nonumber \\ + 2\, \mbox{det} (L)
\sum_{\alpha = 1}^6 \frac{\partial}{\partial Z^\alpha} \left (
\frac{dZ^\alpha}{dt} \right ) \Bigg \},
\end{eqnarray}
and using \eq{eqmotion} for the gyrokinetic Lagrangian
$\overline{\mathcal{L}}$, we obtain
\begin{equation}
\frac{dZ^\alpha}{dt} = \sum_{\beta = 1}^6 (L^{-1})^{\alpha \beta}
\frac{\partial \overline{H}}{\partial Z^\beta}, \quad \alpha =
1,2, \ldots, 6 .
\end{equation}
Employing that $(L^{-1})^{\alpha \beta} = - (L^{-1})^{\beta
\alpha}$, we find that this equation leads to
\begin{equation} \label{div_dZdt}
\sum_{\alpha = 1}^6 \frac{\partial}{\partial Z^\alpha} \left (
\frac{dZ^\alpha}{dt} \right ) = \sum_{\alpha, \beta = 1}^6
\frac{\partial (L^{-1})^{\alpha \beta}}{\partial Z^\alpha}
\frac{\partial \overline{H}}{\partial Z^\beta}.
\end{equation}
Equation \eq{div_dZdt} can be further simplified by using
\eq{eq:formuladerivmatrix}, giving
\begin{eqnarray} \label{div_dZdt_2}
\fl \sum_{\alpha = 1}^6 \frac{\partial}{\partial Z^\alpha} \left (
\frac{dZ^\alpha}{dt} \right ) = - \sum_{\alpha, \beta, \gamma =
1}^6 (L^{-1})^{\alpha \beta} \frac{\partial L_{\beta
\gamma}}{\partial Z^\alpha} \frac{dZ^\gamma}{dt} \nonumber \\ = -
\frac{1}{2} \sum_{\alpha, \beta, \gamma = 1}^6 (L^{-1})^{\alpha
\beta} \left ( \frac{\partial L_{\beta \gamma}}{\partial Z^\alpha}
+ \frac{\partial L_{\gamma \alpha}}{\partial Z^\beta} \right )
\frac{dZ^\gamma}{dt},
\end{eqnarray}
where to obtain the last equality we have used that $L_{\alpha
\beta} = - L_{\beta \alpha}$ and $(L^{-1})^{\alpha \beta} = -
(L^{-1})^{\beta \alpha}$. Substituting \eq{div_dZdt_2} into
\eq{app_ps_aux1} and using that the derivatives of a determinant
are
\begin{equation} \label{eq:ddet}
\frac{\partial}{\partial Z^\alpha}\,  \mbox{det} ( L ) =
\mbox{det} (L) \sum_{\beta, \gamma = 1}^6 (L^{-1})^{\beta \gamma}
\frac{\partial L_{\gamma \beta}}{\partial Z^\alpha},
\end{equation}
finally gives \eq{consps_app}. To prove that all the terms cancel we
just need to use that, as trivially deduced from its definition
\eq{lagr_bra}, $L_{\alpha\beta}$ satisfies \eq{eq:ClosedForm} with
$n=3$.

\section{Proof of equation \eq{eq:psconserv}}
\label{app:psconserv}
In this Appendix we prove \eq{eq:psconserv} by showing that
$\mbox{det} (J_{\bZ_{p0} \mapsto \bZ_p}(\bZ_{p0})) = B_{||,p}^*
(\bZ_{p0})/B_{||,p}^* (\bZ_p(\bZ_{p0},t_0;t))$. First we evaluate
the time derivative of $\mbox{det}(J_{\bZ_{p0} \mapsto \bZ_p})$,
given by
\begin{equation}
\fl \frac{d}{dt} [ \mbox{det}(J_{\bZ_{p0} \mapsto \bZ_p}) ] =
\mbox{det} (J_{\bZ_{p0} \mapsto \bZ_p}) \sum_{\beta, \gamma = 1}^6
(J_{\bZ_{p0} \mapsto \bZ_p}^{-1})_\beta^\gamma \frac{\partial \dot
Z_p^\beta}{\partial Z_{p0}^\gamma } = \mbox{det} (J_{\bZ_{p0}
\mapsto \bZ_p}) \sum_{\beta = 1}^6 \frac{\partial \dot
Z_p^\beta}{\partial Z_p^\beta },
\end{equation}
where we have employed the formula for the derivative of a
determinant in \eq{eq:ddet} and omitted the arguments hoping that
no confusion will be caused. Using \eq{consps}, or its equivalent
\eq{consps_app}, we obtain that
\begin{equation}
\fl \frac{1}{\mbox{det}(J_{\bZ_{p0} \mapsto \bZ_p})} \frac{d}{dt}
[ \mbox{det}(J_{\bZ_{p0} \mapsto \bZ_p}) ] = - \frac{1}{B_{||,p}^*
(\bZ_p(Z_{p0}, t_0; t))} \frac{d}{dt} [ B_{||,p}^* (\bZ_p(Z_{p0},
t_0; t)) ],
\end{equation}
that is, the product $B_{||,p}^* (\bZ_p(Z_{p0}, t_0; t))
\mbox{det}(J_{\bZ_{p0} \mapsto \bZ_p})$ is constant in time. Since
for $t = t_0$ the map $\bZ_p (\bZ_{p0}, t_0; t)$ is the identity,
giving $\mbox{det}(J_{\bZ_{p0} \mapsto \bZ_p}) = 1$, we find that
the constant is $B_{||,p}^* (Z_{p0})$, implying that $\mbox{det}
(J_{\bZ_{p0} \mapsto \bZ_p}(\bZ_{p0})) = B_{||,p}^*
(\bZ_{p0})/B_{||,p}^* (\bZ_p(\bZ_{p0},t_0;t))$.

\section{Manipulations leading to equation~\eq{eq:gyrokinPoissonfieldtheory1}}
\label{app:poissonvariationalprinciple}

In this Appendix we obtain equation
\eq{eq:gyrokinPoissonfieldtheory1} from
\eq{eq:auxPoissonvariational}.

First, we evaluate the variations with respect to $\varphi(\boldr,
t)$ in \eq{eq:auxPoissonvariational} term by term. For the first
term in \eq{eq:auxPoissonvariational}, we find
\begin{equation} \label{term1variational}
\int dt\, d^3 r\, \nabla \delta \varphi (\boldr, t) \cdot \nabla
\varphi(\boldr,t) = - \int dt\, d^3 r\, \delta \varphi (\boldr, t)
\nabla^2 \varphi(\boldr,t),
\end{equation}
where we have integrated by parts and we have taken into account
that $\delta \varphi = 0$ at the boundary. Using relation
\eq{defphi} and equation \eq{eq:phip} to write
\begin{equation} \label{eq:phipdelta}
\phi_p (\bR_p, \mu_p, \theta_p, t) = \int d^3r\, \delta \left (
\bR_p + \frac{\epsilon_s}{\lambda_p} \rhobf(\bR_p, \mu_p,
\theta_p) - \boldr \right ) \varphi (\boldr, t),
\end{equation}
we find
\begin{equation} \label{term2variational}
\delta_{\varphi} \langle \phi_p (\bR_p, \mu_p, \theta_p, t)
\rangle = \int d^3r\, \delta \varphi (\boldr, t) \left \langle
\delta \left ( \bR_p + \frac{\epsilon_s}{\lambda_p} \rhobf -
\boldr \right ) \right \rangle.
\end{equation}
Using relation \eq{eq:phipdelta} again and the identity
\begin{eqnarray}
\fl \frac{1}{2} \delta_\varphi \langle
\nabla_{(\bR_{p\bot}/\epsilon_s)} \Phiwig_p \cdot ( \bun \times
\nabla_{(\bR_{p\bot}/\epsilon_s)} \phiwig_p ) \rangle =
\frac{1}{2} \langle \nabla_{(\bR_{p\bot}/\epsilon_s)}
(\delta_\varphi \Phiwig_p) \cdot ( \bun \times
\nabla_{(\bR_{p\bot}/\epsilon_s)} \phiwig_p ) \rangle \nonumber \\
- \frac{1}{2} \langle \nabla_{(\bR_{p\bot}/\epsilon_s)}
(\delta_\varphi \phiwig_p) \cdot ( \bun \times
\nabla_{(\bR_{p\bot}/\epsilon_s)} \Phiwig_p ) \rangle \nonumber \\
= - \langle \nabla_{(\bR_{p\bot}/\epsilon_s)} (\delta_\varphi
\phiwig_p) \cdot ( \bun \times \nabla_{(\bR_{p\bot}/\epsilon_s)}
\Phiwig_p ) \rangle,
\end{eqnarray}
where to obtain the last equality we have integrated by parts in
$\theta_p$ and we have used that $\partial \Phiwig_p/\partial
\theta_p = \phiwig_p$ and that $\partial (\delta_\varphi
\Phiwig_p)/\partial \theta_p = \delta_\varphi \phiwig_p$, we
obtain
\begin{eqnarray} \label{term3variational}
\fl \Lambda_p \delta_{\varphi} \Psi^{(2)}_{\phi, p} (\bR_p, \mu_p,
\theta_p, t) + \delta_{\varphi} \Psi^{(2)}_{\phi B, p} (\bR_p,
u_p, \mu_p, \theta_p, t) \nonumber \\ = \int d^3r\, \delta \varphi
(\boldr, t) \Bigg \langle \frac{1}{\lambda_p} \bV_{\bR_p} (\bR_p,
u_p, \mu_p, \theta_p, t) \cdot \nabla_{(\bR_{p\bot}/\epsilon_s)}
\delta \left ( \bR_p + \frac{\epsilon_s}{\lambda_p} \rhobf -
\boldr \right ) \nonumber \\ + V_{\mu_p} (\bR_p, u_p, \mu_p,
\theta_p, t) \frac{\partial}{\partial \mu_p} \left [ \delta \left
( \bR_p + \frac{\epsilon_s}{\lambda_p} \rhobf - \boldr \right )
\right ] \nonumber \\ + V_{\theta_p} (\bR_p, u_p, \mu_p, \theta_p,
t) \delta \left ( \bR_p + \frac{\epsilon_s}{\lambda_p} \rhobf -
\boldr \right ) \Bigg \rangle.
\end{eqnarray}
In this equation we have separated the different terms into three
types: the terms that contain the gradient of the delta function,
the terms that contain the derivative with respect to $\mu_p$ of
the delta function, and the terms that contain the delta function.
The coefficient multiplying the gradient of the delta function is
\begin{eqnarray}
\fl \bV_{\bR_p} (\bR_p, u_p, \mu_p, \theta_p, t) = -
\frac{\Lambda_p}{\lambda_pB^2} \bun \times
\nabla_{(\bR_{p\perp}/\epsilon_s)} \tilde \Phi_p - \frac{u_p}{B}
\bun \times \nabla_{\bR_p} \bun \cdot \rhobf - \frac{\mu_p}{2B^2}
\nabla_{\bR_{p\bot}} B \nonumber \\ - \frac{1}{4B} \Big (
\rhobf\rhobf - (\rhobf\times\bun)(\rhobf\times\bun) \Big ) \cdot
\nabla_{\bR_p} B;
\end{eqnarray}
the coefficient multiplying the derivative with respect to $\mu_p$
of the delta function is
\begin{eqnarray}
\fl V_{\mu_p} (\bR_p, u_p, \mu_p, \theta_p, t) =
-\frac{\Lambda_p}{B}\tilde\phi_p - \frac{u_p^2}{B} \bun \cdot
\nabla_{\bR_p} \bun \cdot \rhobf \nonumber \\ + \frac{u_p}{4}
\nabla_{\bR_p} \bun: \Big ( \rhobf (\rhobf\times\bun) +
(\rhobf\times\bun) \rhobf \Big );
\end{eqnarray}
and the coefficient multiplying the delta function is
\begin{eqnarray}
\fl V_{\theta_p} (\bR_p, u_p, \mu_p, \theta_p, t) = -
\frac{\Lambda_p}{\lambda_p^2B} \frac{\partial
\tilde\phi_p}{\partial (\mu_p/\lambda_p^2)} - \frac{1}{B}
\nabla_{\bR_p} B \cdot \rhobf - \frac{u_p^2}{2\mu_p B} \bun \cdot
\nabla_{\bR_p} \bun \cdot \rhobf \nonumber \\ + \frac{u_p}{4
\mu_p}\nabla_{\bR_p} \bun: \Big (\rhobf (\rhobf\times\bun) +
(\rhobf\times\bun)\rhobf \Big).
\end{eqnarray}

Substituting \eq{term1variational}, \eq{term2variational} and
\eq{term3variational} into \eq{eq:auxPoissonvariational}, and
employing \eq{FpFp0} and \eq{eq:psconserv} to write the integrals
as integrals over $\bZ_p$ and not over $\bZ_{p0}$, we find that
the variations are of the form given in
\eq{eq:gyrokinPoissonfieldtheory1} with
\begin{eqnarray} \label{Poisson_app}
\fl \mathcal{P} (\boldr, t) & = -
\frac{\lambda_{De}^2\epsilon_s}{L^2} \nabla^2 \varphi(\boldr,t)
\nonumber \\ \fl & - \sum_p Z_p \int d^3 R_p du_p d\mu_p
d\theta_p\, B_{||, p}^* (\bR_p, u_p, \mu_p ) F_p(\bR_p, u_p,
\mu_p, t) \Bigg \{ \delta \left (\bR_p +
\frac{\epsilon_s}{\lambda_p} \rhobf - \boldr \right ) \nonumber\\
\fl & + \frac{\epsilon_s}{\lambda_p} \Bigg [
\frac{\epsilon_s}{\lambda_p} \bV_{\bR_p} \cdot \nabla_{\bR_p}
\delta \left (\bR_p + \frac{\epsilon_s}{\lambda_p} \rhobf - \boldr
\right ) + V_{\mu_p} \frac{\partial}{\partial \mu_p} \delta \left
(\bR_p + \frac{\epsilon_s}{\lambda_p} \rhobf - \boldr \right )
\nonumber \\ \fl & + V_{\theta_p} \delta \left (\bR_p +
\frac{\epsilon_s}{\lambda_p} \rhobf - \boldr \right ) \Bigg] \Bigg
\}.
\end{eqnarray}
Note that we did not need to keep the gyroaverages in
\eq{term2variational} and \eq{term3variational} because there is
an overall integral in $\theta_p$ and neither $B_{||,p}^*$ nor
$F_p$ depend on $\theta_p$.

We have only left to prove that \eq{Poisson_app} is equal to
\eq{eq:gyrokinPoissonfieldtheory}. Using the result in \eq{R2} and
employing \eq{rhorhoave} and \eq{rhorhowig} to write
\begin{equation}
\frac{\mu_p}{B} \nabla_{\bR_{p\bot}} B + \frac{1}{2} \Big [ \rhobf
\rhobf - (\rhobf \times \bun) (\rhobf \times \bun) \Big ] \cdot
\nabla_{\bR_p} B = \rhobf \rhobf \cdot \nabla_{\bR_p} B,
\end{equation}
it is clear that $\bV_{\bR_p}$ is exactly the perpendicular
component of $\bR_{p, 2}$. Using \eq{mu1}, we find that $V_{\mu_p}
= \mu_{p, 1}$. It only remains to use that $V_{\theta_p} =
-\partial \theta_{p,1}/\partial \theta_p$, with $\theta_{p,1}$
given in \eq{theta1}. To obtain this identity we have used $\rhobf
=\partial (\rhobf \times \bun) /\partial \theta_p$, $ \rhobf
(\rhobf  \times \bun ) + (\rhobf  \times \bun ) \rhobf = -
(1/2)\partial[\rhobf \rhobf - (\rhobf  \times \bun ) (\rhobf
\times \bun ) ]/\partial \theta_p$ and $\phiwig_p =\partial
\tilde\Phi_p/\partial \theta_p$. Using $V_{\theta_p} = -
\partial \theta_{p,1}/\partial \theta_p$ and integrating by parts in
$\theta_p$ in \eq{Poisson_app} gives the final form
\eq{eq:gyrokinPoissonfieldtheory}.

\end{document}